\documentclass[iop]{emulateapj}
\usepackage{apjfonts}
\usepackage{epsfig}
\usepackage{graphicx}
\usepackage{url}
\usepackage{natbib}

\defcitealias{cmg+06}{CD06}
\defcitealias{ccg+08}{CD08}
\defcitealias{lm10a}{LM10}
\defcitealias{mlp+06}{MLPP}

\shorttitle{Kinematics of Sgr Tidal Debris in Kapteyn's Selected Areas}
\shortauthors{Carlin et al.}
\journalinfo{Accepted for publication in The Astrophysical Journal}

\begin{document}

\title{Kinematics and Chemistry of Stars Along the Sagittarius Trailing Tidal Tail and Constraints on the Milky Way Mass Distribution}

\author{Jeffrey L. Carlin\altaffilmark{1,2,5}, Steven R. Majewski\altaffilmark{1}, Dana I. Casetti-Dinescu\altaffilmark{3}, David R. Law\altaffilmark{4}, Terrence M. Girard\altaffilmark{3}, and Richard J. Patterson\altaffilmark{1}}

\altaffiltext{1}{Department of Astronomy, University of Virginia,  P.O. Box 400325, Charlottesville, VA 22904-4325, USA (jc4qn@mail.astro.virginia.edu)}
\altaffiltext{2}{Department of Physics, Applied Physics, and Astronomy, Rensselaer Polytechnic Institute, 110 8th Street, Troy, NY 12180, USA (carlij@rpi.edu)}
\altaffiltext{3}{Astronomy Department, Yale University, P.O. Box 208101, New Haven, CT 06520-8101, USA}
\altaffiltext{4}{Department of Physics and Astronomy, University of California, Los Angeles, CA 90095, USA ; Hubble Fellow}
\altaffiltext{5}{Visiting Astronomer, Kitt Peak National Observatory, National Optical Astronomy Observatory, which is operated by the Association of Universities for Research in Astronomy (AURA) under cooperative agreement with the National Science Foundation.}

\begin{abstract}

We present three-dimensional kinematics of Sagittarius (Sgr) trailing
tidal debris in six fields located 70-130$\arcdeg$ along the stream
from the Sgr dwarf galaxy core. The data are from our proper-motion
(PM) survey of Kapteyn's Selected Areas, in which we have measured
accurate PMs to faint magnitudes in $\sim40\arcmin\times40\arcmin$
fields evenly spaced across the sky. The radial velocity (RV)
signature of Sgr has been identified among our follow-up spectroscopic
data in four of the six fields and combined with mean PMs of
spectroscopically-confirmed members to derive space motions of Sgr
debris based on $\sim$15-64 confirmed stream members per field. These
kinematics are compared to predictions of the
\citet{lm10a} model of Sgr disruption; we find reasonable agreement
with model predictions in RVs and PMs along Galactic
latitude. However, an upward adjustment of the Local Standard of Rest
velocity ($\Theta_{\rm LSR}$) from its standard 220 km s$^{-1}$ to at
least $232\pm14$ km~s$^{-1}$ (and possibly as high as $264\pm23$
km~s$^{-1}$) is necessary to bring 3-D model debris kinematics and our
measurements into agreement. Satisfactory model fits that
simultaneously reproduce known position, distance, and radial velocity
trends of the Sgr tidal streams, while significantly increasing
$\Theta_{\rm LSR}$, could only be achieved by increasing the Galactic
bulge and disk mass while leaving the dark matter halo fixed to the
best-fit values from
\citet{lm10a}. We derive low-resolution spectroscopic abundances along
this stretch of the Sgr stream and find a constant [Fe/H] $\sim$ -1.15
(with $\sim0.5$ dex scatter in each field -- typical for dwarf
galaxy populations) among the four fields with reliable
measurements. A constant metallicity suggests that debris along the
$\sim60\arcdeg$ span of this study was all stripped from Sgr on the
same orbital passage.

\end{abstract}


\section{Introduction}

With the profusion of data provided in recent years by deep,
large-area photometric surveys such as the Two Micron All Sky Survey
(2MASS) and Sloan Digital Sky Survey (SDSS), a wealth of stellar
substructure has been uncovered in the Milky Way (MW) halo. The
finding and subsequent mapping of numerous stellar tidal streams and
overdensities (e.g., Sagittarius --- \citealt{ili+01,msw+03,bze+06}; Monoceros ---
\citealt{nyr+02,iil+03,yng+03}; other SDSS streams ---
\citealt{g09,bei+07,g06,g06a,gd06}) has borne out the idea
\citep{sz78,m93,mmh96} that remnants of accreted dwarf galaxies make up much
of the stellar halo of the Milky Way. The direct confirmation of the 
accretion of late-infalling subhalos via discovery of ubiquitous
long-lived, coherent tidal debris streams has provided strong
constraints on models of small-scale hierarchical structure formation
under the prevailing $\Lambda$-Cold Dark Matter ($\Lambda$CDM)
cosmology (e.g., \citealt{ans+03a,bj05,fjb+06}). Furthermore, because
the tidal streams retain the kinematical signatures of the orbits of their
progenitors (i.e., angular momentum and energy), stellar debris in the
streams can be used as sensitive probes of the underlying Galactic
gravitational potential (e.g., \citealt{jzs+99,ili+01,h04,mga+04,jlm05,ljm05,mlp+06}).

\begin{figure*}[!t]
\begin{center}
\includegraphics[width=4.25in]{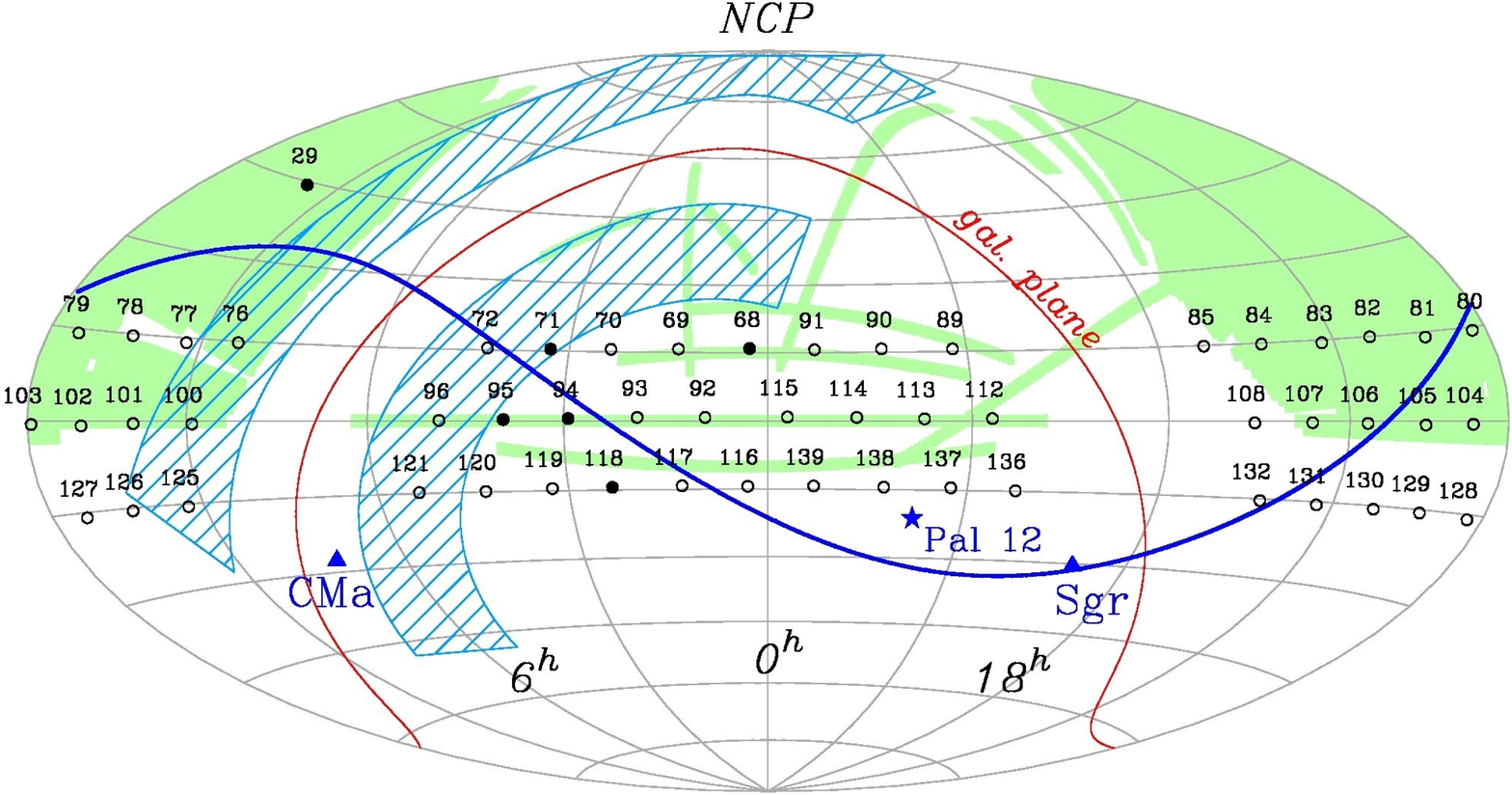}
\includegraphics[width=2.4in,trim=0.2in 0.2in 0.2in 0.2in,clip]{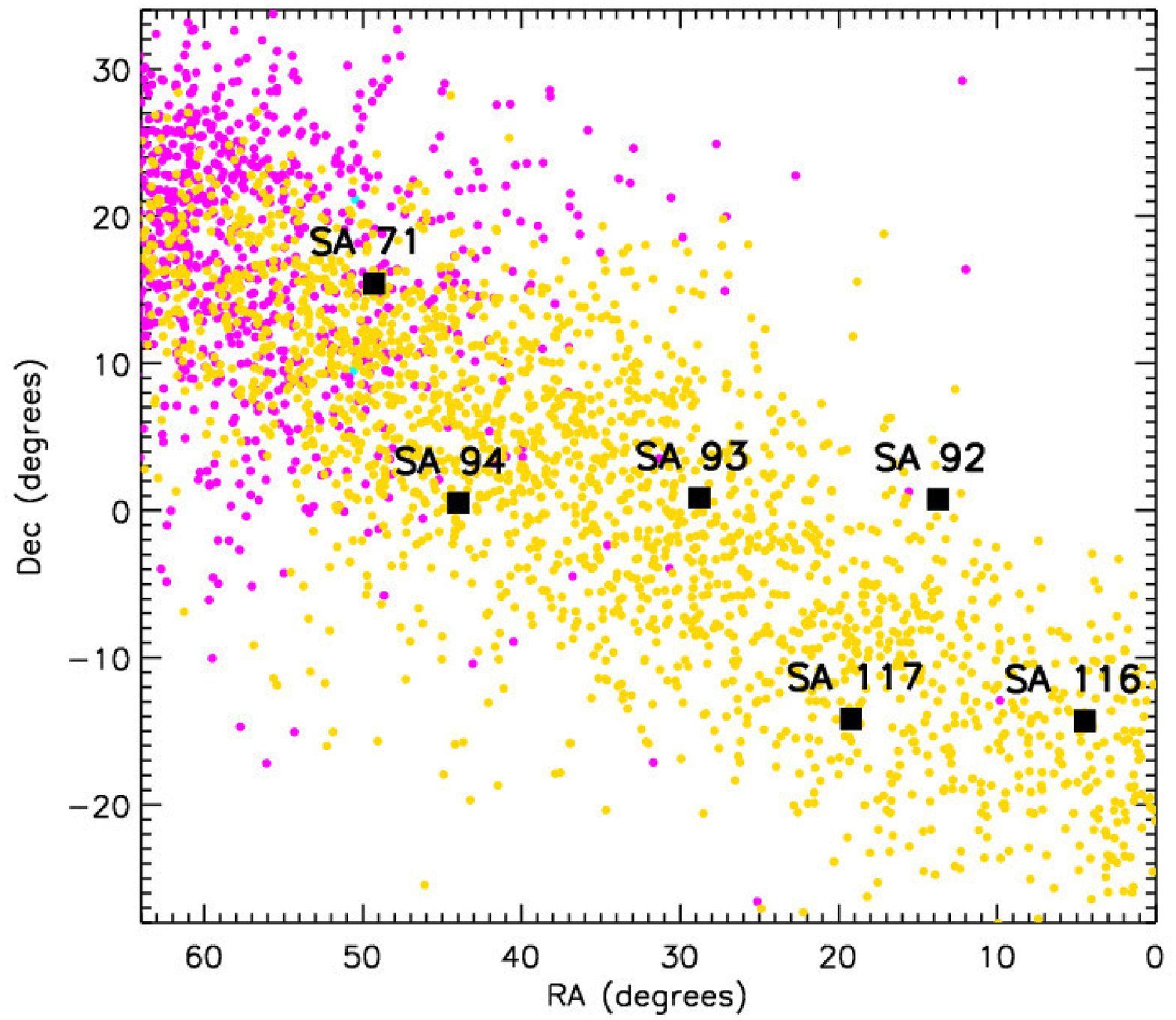}
\caption{{\it Left panel: } Distribution of Kapteyn's Selected Areas (in equatorial coordinates) for which we have derived proper motions, shown in an Aitoff projection. Solid points are those fields for which we have additional deep, 4-meter plates (see text). The current orbital plane of Sagittarius is overlaid as a solid blue line, and the shaded (light green) areas represent the sky coverage of SDSS (as of DR5). Regions containing stellar overdensities suggested in the literature to be part of the "Monoceros ring" are denoted by the blue hatched areas on either side of the disk. {\it Right panel: } Spatial distribution of the Kapteyn Selected Areas used in this study overlaid on the predicted distribution of Sagittarius tidal debris from the best-fit triaxial halo model of \citet{lm10a}.  Gold colored points represent debris stripped from the Sgr progenitor on the past two perigalactic passages (0-1.3 Gyr ago), and magenta points the previous two passages (1.3-3.2 Gyr ago). Note that all of the fields are sampling predominantly debris stripped on the same orbital passage (i.e., the gold points), with only SA 71 slightly sampling earlier-stripped (magenta) debris.} 
\label{fig:radec}
\end{center}
\end{figure*}

The best-known and {\it only} widely agreed-upon case of a presently
visible dwarf galaxy undergoing tidal disruption in the Milky Way halo
is the Sagittarius (Sgr) dwarf spheroidal (dSph)\footnote{Though we
note that there is now evidence for extended tidal debris populations
around the Carina \citep{mmz+06,mmj08} and Leo I \citep{smm+07}
dSphs. Also, some debate still exists over whether the HI Magellanic
Stream derives from tidal stripping of Small or Large Magellanic Cloud
gas versus from ram pressure stripping.}.  The core of this galaxy was
first discovered by
\citet{igi94} in a kinematical study of the outer Galactic bulge, with
the first large-scale mapping of the Sgr leading and trailing tidal arms
done by \citet{msw+03} using 2MASS M-giant stars.  Various studies
have reported the discovery of stars (e.g.,
\citealt{msw+03,mga+04,bze+06,ynj+09,cbi+10}; a comprehensive summary of the
earlier detections appears in \citealt{msw+03}) or star clusters
(e.g., Pal 12: \citealt{dmg+00}; Whiting 1: \citealt{czm07}; many
clusters: \citealt{bfi03}; a summary of Sgr clusters appears in
\citealt{lm10b}) plausibly associated with debris from Sgr, either
trailing or leading it along its orbit.  Line-of-sight velocities
(i.e., {\it radial velocities}, or RVs) of Sgr members have been
determined at a few positions along the stream (e.g.,
\citealt{dhm+01,mkl+04,mbb+07}), and, along with the spatial
distribution of these stars, provide constraints on models of the
Sgr-Milky Way interaction (e.g.,
\citealt{jsh95,hw01,ili+01,h04,mga+04}). A comprehensive effort at
modeling the Sgr disruption constrained by all observations available
after about a decade of study was done by \citet{ljm05}, who were able
to reproduce most extant data, but were unable to completely reconcile
the apparent need for a prolate MW halo potential to produce the
leading arm {\it radial velocities} on the one hand, and an oblate
halo to match the {\it positions} of leading debris on the other.
This contradiction has apparently been recently resolved by
\citet{lmj09}, who propose that the Milky Way might have a triaxial
halo; a comprehensive $N$-body model based on the best-fitting
triaxial halo \citep[hereafter LM10]{lm10a} reasonably matches nearly
all existing constraints (spatial {\it and} kinematical) of Sgr tidal
debris.\footnote{Further complications have arisen due to an apparent
bifurcation of the leading stream \citep{bze+06}; the
\citetalias{lm10a} model was not designed to address this
issue. Several attempts to explain this detail invoke overlapping
debris from multiple orbital wraps (\citealt{fbe+06}; though
\citealt{ynj+09} find similar stellar populations in both arms, likely
ruling out this scenario) or a disk-galaxy progenitor for Sgr
(\citealt{pbe+10}; but cf. \citealt{lkm+10}). The highly-elliptical
shape of the Sgr dwarf has recently been reproduced by \citep{lkm+10},
who model Sgr as a disk galaxy embedded in an extended dark
halo. Tidal stirring transforms the initially disk dwarf into an
extended elliptical shape over two pericentric passages; the rotation
of the progenitor may also explain the bifurcation of the Sgr leading
arm.} The Sagittarius dwarf and its tidal debris are thus proving to
be an excellent laboratory for studying both the dynamics of tidally
disrupting dwarf galaxies and star stream formation, as well as the
shape and strength of the Galactic gravitational potential that is the
cause of this disruption. It is this model of \citetalias{lm10a} to
which we shall compare our data throughout this work.

\subsection{Our Proper Motion Survey}

To date, no systematic survey has addressed the tangential velocities
(derived from proper motions) of the identified major Galactic tidal
streams.  Only a few studies (e.g.,
\citealt{dmg+02,ccg+08,cgm+09a,ccg+10,krh10}) have published {\it any}
proper motion results for major Galactic substructures, and typically
not at a level of precision that is useful for constraining dynamical
models of tidal stream production and evolution.  In an effort to
detect and characterize halo substructures, we have been working on a
project to obtain full phase-space information (positions and full 3-D
space motions) for individual stars in Kapteyn's Selected Areas (SAs;
see \citealt{cmg+06} for an overview of this project).  The sky
positions of the Selected Areas were chosen by Jacobus \citet{k06} to
provide evenly spaced coverage for a systematic exploration of Milky
Way structure. We have attempted to carry on at least part of this
legacy by taking advantage of Mt.~Wilson 60-inch telescope
photographic plate material taken by Kapteyn and collaborators
\citep{skv+30} for their survey to make up the first-epoch data of our
survey (in particular, near-equatorial fields at $\delta = 0\arcdeg,
+15\arcdeg$, and $-15\arcdeg$).  The distribution on the sky of those
SAs that make up our survey is shown in an Aitoff projection in
Figure~\ref{fig:radec}. Some of the equatorial SA fields lie along the
orbit of the Sagittarius dwarf galaxy (which is approximated by the
blue, solid curve in Figure~\ref{fig:radec}) -- it is a subset of
these fields (in particular, six fields along the trailing tidal tail;
see the right panel of Figure~\ref{fig:radec}) that are the focus of
the present work.

\subsection{Constraints on the Local Standard of Rest Velocity}

The positions of tidal debris that have been found over a large
stretch of the Sagittarius orbit place fairly strong constraints on
the three-dimensional motions of the Sgr dwarf. It is, however,
important to confirm and refine the models by measuring space
velocities of stream stars (especially proper motions, which are
difficult to measure for stars in distant Galactic substructures). In
the case of the Sgr trailing tail, however, proper motions measured
for debris stars are also rather sensitive to the Sun's motion through
the Galaxy.  This arises because much of the Sagittarius trailing
tidal tail is positioned at a roughly constant distance below the
Galactic plane, with the Sgr orbital plane nearly coincidental with
the Galactic $X_{\rm GC}-Z_{\rm GC}$ plane.\footnote{Throughout this
paper, when we refer to Galactic Cartesian ($X,Y,Z)_{\rm GC}$
coordinates, we are specifically referring to a right-handed Cartesian
frame centered on the Galactic center, with $X_{\rm GC}$ positive in
the direction from the Sun to the Galactic center, $Y_{\rm GC}$ in the
direction of the Sun's motion through the Galaxy, and $Z_{\rm GC}$
upward out of the plane. Assuming the Sun is at $R_0 = 8.0$ kpc from
the Galactic center, this places the Sun at $(X,Y,Z)_{\rm
GC}$=(-8.0,0,0) kpc. The corresponding velocity components will be
denoted ($U,V,W)_{\rm GC}$, where the ``GC'' denotes velocities
relative to the Galactic rest frame.} \citet[hereafter
``MLPP'']{mlp+06} noted that because of this orientation, longitudinal
proper motions of Sgr trailing debris located sufficiently far away
from the South Galactic Pole contain virtually no contribution from
Sagittarius motions, and almost entirely reflect the solar motion.

Efforts to measure fundamental dynamical properties of the Milky Way,
such as its rotation curve, $\Theta(R)$, are complicated by our Sun's
own (poorly known) motion within the Galaxy.  Measurements of
$\Theta_{\rm LSR}$, the Galactic rotation speed at the solar circle
(the {\it Local Standard of Rest}, ``LSR''), vary by 25\%, despite
many efforts at its determination. The value adopted by the IAU in
1985 of $\Theta_{\rm LSR}$ = 220 km s$^{-1}$ (see \citealt{kl86}) has
long represented a reasonable approximation to existing measurements
(note, however, that prior to the 1985 IAU adoption of $\Theta_{\rm
LSR}$ = 220 km s$^{-1}$, the 1964 IAU general assembly adopted 250 km
s$^{-1}$; see a listing of pre-1985 measurements of $\Theta_{\rm LSR}$
in \citealt{kl86}). Constraints taking into account the ellipticity of
the disk have suggested the LSR velocity could be as low as $\sim$180
km s$^{-1}$ \citep{kt94}. A similarly low value of $184 \pm 8$ km
s$^{-1}$ was found by \citet{om98}, who modified previous methods of
determining the Oort constants by including radial variations of gas
density in their mass modeling of the Galactic rotation curve. Proper
motions of Galactic Cepheids from $Hipparcos$ \citep{fw97} yield a
result of $\Theta_{\rm LSR} =(217.5 \pm 7.0) (R_0/8)$ km s$^{-1}$
(where $R_0$ is the distance from the Sun to the Galactic center; the
IAU adopted value is $R_0 = 8.5$ kpc), in line with the IAU
standard. Using re-reduced $Hipparcos$ data \citep{v07} with improved
systematic errors, \citet{yzk08} found $\Theta_{\rm LSR} =(243 \pm 9)
(R_0/8)$ km s$^{-1}$ based on thin-disk O-B5 stars. Estimates based on
absolute PMs of Galactic bulge stars in the field of view of globular
cluster M4 using the {\it Hubble Space Telescope (HST)} yield
$(202.4\pm20.8) (R_0/8)$ km s$^{-1}$ \citep{krh+04} and $(220.8 \pm
13.6) (R_0/8)$ km s$^{-1}$ \citep{bpk+03} (with both studies using
data from the same $HST$ observations). Long-term $VLBA$ monitoring of
Sgr A*, the radio source at the Galactic center, led to a proper
motion of Sgr A* from which \citet{rb04} revised the LSR velocity
upward to $(235.6\pm1.2) (R_0/8)$ km s$^{-1}$. \citet{gsw+08} combined
stellar kinematics near the Galactic center with the proper motion of
Sgr A* to derive $(229
\pm 18) (R_0/8.4)$ km s$^{-1}$. More recently, $\Theta_{\rm LSR}$ has
been suggested to be even higher, $(254\pm16) (R_0/8.4)$ km s$^{-1}$,
based on trigonometric parallaxes of Galactic star-forming regions
\citep{rmz+09}. Reanalysis of these same data, including the Sgr A*
proper motion, by \citet{bhr09} found a similar $(244\pm13)$ km
s$^{-1}$. \citet{krh10} provided constraints on the MW halo potential
by analysing the GD-1 \citep{gd06} stellar stream, combining SDSS
photometry, USNO-B+SDSS proper motions (see \citealt{mml+04,mml+08}),
and spectroscopy to obtain 6-D phase-space data over a large stretch
of the stream, which they used to estimate $\Theta_{\rm LSR} =
(224\pm13) (R_0/8.4)$ km s$^{-1}$ (though this result is made somewhat
more uncertain due to a systematic dependence on the flattening of the
disk+halo potential). Finally, a combined estimate including many of
the above results as priors finds a value of $(236\pm11) (R_0/8.2)$ km
s$^{-1}$
\citep{bhr09}. Most of the estimates discussed here rely on
the Oort constants, and thus are dependent on our incomplete knowledge
of $R_0$. Despite numerous attempts at determining the circular
velocity at the solar circle, this constant remains poorly
constrained. It is clear that independent methods would be valuable to
obtain alternative estimates of $\Theta_{\rm LSR}$.

Here we use a new, independent method for ascertaining $\Theta_{\rm
LSR}$ that has the advantage over most of the previously mentioned
methods in that the results have virtually complete decoupling from an
assumed value of $R_0$.  As discussed in \citetalias{mlp+06}, the
trailing arm of the Sagittarius tidal stellar stream is ideally placed
to serve as an absolute velocity reference for the LSR.  With an
orbital pole of $(l_p,b_p) = (274,-14)\arcdeg$, Sgr is almost on a
polar orbit, and the line of nodes of the intersection of the Galactic
midplane and the Sgr debris plane is almost coincident with the
Galactic $X_{\rm GC}$ axis (the axis containing the Sun and Galactic
center). This is illustrated in Figure~\ref{fig:sgr_xyz}, which shows
the projection of Sgr debris from the \citetalias{lm10a} model onto
the Galactic $X_{\rm GC}-Z_{\rm GC}$, $Y_{\rm GC}-Z_{\rm GC}$, and
$X_{\rm GC}-Y_{\rm GC}$ planes. In the upper left panel (the $X_{\rm
GC}-Z_{\rm GC}$ plane), the Sgr orbital plane is nearly face-on, while
in the other two panels, few Sgr debris points are seen more than
$\sim5$ kpc on either side of the $X_{\rm GC}-Z_{\rm GC}$ plane (i.e.,
$|Y_{\rm GC}| \lesssim 5$ kpc for nearly all Sgr debris). The motions
of Sgr stars {\it within} its (virtually non-precessing;
\citealt{jlm05}) debris plane, as observed from the LSR, are therefore
almost entirely in the Galactic $U$ and $W$ velocity components (i.e.,
in the $X_{GC}$-$Z_{GC}$ plane), whereas $V$ motions of Sgr tidal tail
stars almost entirely reflect {\it solar motion} --- i.e.,
$\Theta_{\rm LSR}$ (plus the Sun's peculiar motion in $V$, established
to be in the range $\sim$ +5 to +12 km s$^{-1}$; e.g.,
\citealt{db98}).  The Sgr trailing tail is positioned fairly
equidistantly from the Galactic disk for a substantial fraction of its
stretch across the Southern MW hemisphere \citep{msw+03}.  This band
of stars arcing almost directly ``beneath'' us within the
$X_{GC}$-$Z_{GC}$ plane (see the upper panel of
Figure~\ref{fig:sgr_xyz}) provides a remarkable, stationary zero-point
reference against which to make direct measurement of the solar motion
{\it almost completely independent of the Sun's distance from the GC.}

Because of the fortuitous orientation of the Sgr debris, the majority
of $\Theta_{\rm LSR}$ motion (i.e., $V$) is seen in the proper motions
of these stars, with the reflex solar motion almost entirely contained
in the $\mu_l \cos($b) component for Sgr trailing arm stars (at least
for those stream stars away from the South Galactic Pole (SGP)
coordinate ``discontinuity", where the $\mu_l \cos($b) of Sgr stream
stars switches sign).  Fig.~4 of
\citetalias{mlp+06} shows the
essence of the proposed experiment via measurement of $\mu_l \cos($b)
for Sgr trailing arm stars, which shows a trend with debris longitude,
$\Lambda_{\sun}$\footnote{$\Lambda_{\sun}$ was defined by
\citealt{msw+03} as longitude in the Sgr debris plane as seen from the
Sun; $\Lambda_{\sun}=0^{\circ}$ at the present Sgr position, and
increases along the trailing tail.}, that reflects the solar motion.
In the region from $100\arcdeg \lesssim \Lambda_{\sun}
\lesssim 200\arcdeg$, $\mu_l \cos($b) is nearly constant, because the
motion of Sgr debris contributes little to the $V$-component of
velocity. Thus accurate measurement of $\mu_l \cos($b) for Sgr
trailing tail stars along this stretch of the stream will provide a
means of estimating $\Theta_{\rm LSR}$ with almost no dependence on
$R_0$.

Five of the Kapteyn fields for which we have precise ($\sim$1 mas
yr$^{-1}$) proper motions lie squarely on the Sgr trailing arm in this
$\Lambda_{\sun}$ range, and one other (SA 92) is on the periphery of
the stream. In Section~\ref{nbodyresults.sec}, we will use the mean
Sgr debris proper motions derived in four of these six fields to
derive constraints on $\Theta_{\rm LSR}$.

\subsection{Metallicities and Detailed Abundances of the Sagittarius System}

\citet{cmc+07} presented one of the first studies of high-resolution spectroscopic metallicities derived for Sgr debris. Their work showed that M-giants along the
Sagittarius leading stream exhibit a significant metallicity gradient
(which had previously been suggested to be present over smaller
separations from the Sgr core based on photometric techniques; e.g.,
\citealt{a01,mga+04,bnc+06}), decreasing from a mean [Fe/H] = -0.4 in
the core to $\sim -0.7$ between $\sim 60-120\arcdeg$ from the core
(i.e., between $300 > \Lambda_{\sun} > 240\arcdeg$) , and to $\sim
-1.1$ at $\gtrsim300\arcdeg$ from the main Sgr body. Such a population
gradient along the stream likely arose due a strong metallicity
gradient being present in the dSph before its tidal disruption; thus
the outer, more metal-poor populations were preferentially lost as
tidal stripping progressed at earlier times relative to the more
intermediate-age (and higher metallicity) populations remaining in the
core (a mechanism for this process has been demonstrated in the
context of an $N-$body model by \citetalias{lm10a}). In addition,
apparently some younger populations were formed even after Sgr began
disrupting. The existence of a population gradient has also been seen
by \citet{bnc+06}, who found that the relative numbers of blue
horizontal branch (BHB) stars to red clump (RC, or red horizontal
branch) stars are much higher in a leading stream field than in the
Sgr core.  Since BHB stars arise in older, more metal-poor populations
than the RC stars, this must indicate that the stripped population was
made up of predominantly older, less-enriched stars than remain in the
core today. \citet{kyd10} extended the search for chemical
evolutionary signatures to the trailing tail of Sgr, observing a
handful of stars selected from the 2MASS M-giant catalogs of
\citet{msw+03} at high resolution in each of two fields at distances
of $66\arcdeg$ and $132\arcdeg$ from the core. Keller et al. combined
the mean metallicities in these two fields with the [Fe/H] = -0.4
result for the Sgr core from
\citet{mbb+05}, and derived a metallicity fit as a function of
$\Lambda_{\odot}$ of $\Delta$[Fe/H] =
(-2.4$\pm$0.3) $\times 10^{-3}$ dex degree$^{-1}$. This trend (seen in
their Figure 4) also passes through the mean metallicity of [Fe/H]
$\approx$ -0.6 derived by \citet{mbb+07} in a narrow region centered
at $\Lambda_{\odot} = 100\arcdeg$. 

For consistency, all of the data included in the \citet{kyd10} study
(including those from \citealt{mbb+05,mbb+07} and \citealt{ccm+10})
were derived from M giants, which are, however, biased toward
metal-rich, and therefore relatively younger, stars. In the current
study, we explore the metallicity in fields between $75 <
\Lambda_{\sun} < 130\arcdeg$ from the Sgr core along the trailing tail
using predominantly main sequence stars. Such stars near the main
sequence turnoff are much less prone to metallicity biases than M
giants, because MSTO stars are present in all stellar populations. An
additional advantage of focusing on MSTO stars is that the number
density of turnoff stars is much higher than both young, M giant
tracers and older horizontal-branch stars; this provides us a much
larger sample with which to characterize the Sgr trailing tail
metallicity. Older trailing debris populations have recently been
studied by \citet{sig+10}, who used SDSS Stripe 82 data to develop a
new technique for estimating metallicity from photometric data where
both RR Lyrae variables and main-sequence stars from the same
structure can be identified. Their work found a constant [Fe/H] =
-1.20$\pm$0.1 for Sgr debris along much of the same region of the
trailing tail we are studying. In Section~5 we explore the Sgr
trailing tail metallicity based on our samples of predominantly MSTO
stars.

An important diagnostic of the star-formation timescale in a system is
the $\alpha$-element abundance; the $\alpha$ elements (e.g., Mg, Ca,
Ti) are produced mainly in Type II supernovae (SNe), which are the
evolutionary endpoints of massive stars that dominate the chemical
evolution at early times. Once Type Ia SNe begin to occur, the
[$\alpha$/Fe] ratio will decrease, because $\alpha$-elements are less
effectively produced by these supernova progenitors, while the overall
metallicity, [Fe/H], will continue to increase. This produces a
``knee'' in the [$\alpha$/Fe] vs. [Fe/H] diagram, which acts
essentially as a chronometer for a given system, since the [Fe/H] of
the knee indicates the transition from SNII-dominated evolution to
SNIa contributions. This phenomenon has been seen in a number of dSph
systems, which typically show lower [$\alpha$/Fe] at a given [Fe/H]
than Galactic populations because of a slower enrichment (e.g., \citealt{scs01,svt+03,vis+04,tvs+03,gsw+05}; see also a recent review
by \citealt{tht09}). However, at the lowest metallicities, the
[$\alpha$/Fe] of dSphs more closely resemble those of the MW
halo. Studies by \citet{sbb+07} and \citet{mbb+05} found the same underabundance of $\alpha$-elements
relative to the Milky Way for the core of the Sagittarius dSph. This
finding has been extended into the Sgr trailing stream by \citet{mbb+07}, and into the leading arm by \citet{ccm+10}, with both studies using
M-giants from the catalog of \citet{msw+03}. However, M giants are
biased to higher metallicity and more recently star-forming
population(s) of Sgr, so a natural next step in understanding the
evolution of the original, pre-disruption Sgr dSph is to derive
detailed abundances (especially for s-process and $\alpha$-elements)
for a significant sample of the more metal-poor, older stars
populating the core or, more accessibly, in Sgr's more nearby
streams. In Section~5 we present relative Mg abundances derived from
our spectra. We show that the majority of confirmed old, metal-poor
Sgr stream members appear to have distinct Mg abundances from those of
the Milky Way stellar populations along the lines of sight probed.

\subsection{Goals of This Paper}

Here, we present data in six of the Kapteyn's Selected Areas from our
deep proper-motion survey \citep{cmg+06}. In these six fields
intersecting the trailing tidal tail of the Sgr system, we have
augmented our proper-motion catalogs with follow-up spectroscopy. Sgr
debris has been identified from among the stars with measured radial
velocities, and these Sgr candidates are used to derive the mean
three-dimensional kinematics and chemistry of the Sgr trailing stream.

In Section~\ref{photom_pm.sec}, we briefly introduce the proper motion
survey (a more detailed discussion of the survey appears in
\citealt{cmg+06}), and discuss in depth the spectroscopic observations
with the WIYN+Hydra and MMT+Hectospec multifiber instruments that
yielded a total of $>1500$ radial velocities among proper
motion-selected stars within the six fields of
view. Section~\ref{rv_meas.sec} will detail the final selection of
candidate Sgr debris in each of our fields based on RV, proper motion,
and color-magnitude selection. Section~\ref{sgr_kinematics.sec}
presents maximum-likelihood estimates of the most precise proper
motions ($\sim1-2$ mas yr$^{-1}$ per star, or $\sim 0.2-0.7$ mas
yr$^{-1}$ mean for each field) yet measured for Sagittarius
debris. These measured kinematics are compared to the models of
\citet{lm10a}, and found to agree rather well with the predictions for
Sgr debris motions. However, we follow in
Section~\ref{mw_constraints.sec} with an analysis of the residual {\it
disagreement} between our measurements and the models, or more
accurately, we use the discrepancy to reassess the magnitude of the
solar reflex motion, which is the dominant contributor to the proper
motions in the direction of Galactic longitude.  We show that our
proper motion data (specifically, $\mu_l \cos $b) are inconsistent
with the standard IAU value of 220 km s$^{-1}$ for the Local Standard
of Rest motion at the $\sim1-2\sigma$ level and favor a significantly
higher value, consistent with several of the most recent $\Theta_{\rm
LSR}$ studies using radio techniques. In Section~\ref{sgr_abund.sec}
we apply a software pipeline designed to derive stellar abundances
from low-resolution spectra to the numerous spectra we have obtained
for this project. While the metallicities show a hint of a gradient
among the metal-poor stars in our study consistent with previous work,
we cannot rule out a constant [Fe/H] over the range of stream
longitude covered. We also examine the relative magnesium and iron
index strengths for information on $\alpha$-abundance patterns of Sgr
debris. We find Mg abundances of Sgr members are typically lower at a
given [Fe/H] than field stars, consistent with the behavior seen in
most MW dSphs. Finally, Section~\ref{summary.sec} concludes with a
brief summary of our work, and future avenues these data can be used
to explore.

\section{The Data}

\subsection{Field Locations}

The data discussed here are part of our ongoing deep proper-motion
survey in a subset of Kapteyn's Selected Areas (\citealt{m92};
\citealt{cmg+06}) at declinations of $\pm15\arcdeg$ and $0\arcdeg$.  The survey as designed by \citet{k06} consists of $\sim1\arcdeg$ fields
evenly spaced at $\sim15\arcdeg$ intervals along strips of constant
declination (see Fig.~1 in \citealt{cmg+06}). A handful of our
near-equatorial survey fields (see the left panel of
Figure~\ref{fig:radec}) fall on or near the location of Sgr trailing
tidal debris as mapped by \citet{msw+03} using M-giants from
2MASS. The location of our fields relative to the models (constrained
by the Majewski et al. data, among others) of \citet{lm10a} can be
seen in the right panel of Figure~\ref{fig:radec}, which suggests that
we can expect a significant contribution from Sgr debris to the
stellar populations along these lines of sight. In
\citet{cmg+06} and \citet{ccg+08} we showed that faint ($V$ or $g \gtrsim 20$), blue ($B-V < 0.8$ or $g-r < 0.6$) overdensities (at colors and magnitudes consistent with the expected Sgr main sequence turnoff) in the
color-magnitude diagrams of those SA fields intersecting the Sgr
orbital path also show clumping in the distributions of their proper
motions; these excesses and their clumping in proper motion space
suggest that a distant, common-motion population, likely to be Sgr
tidal debris, is present in these fields.  In this work, we focus on
six fields: SAs 116, 117, 92, 93, 94, and 71 (listed in order of
increasing $\Lambda_{\sun}$).  Coordinates for these fields are given
in Table~\ref{tab:sa_tab}, which includes equatorial and Galactic
positions as well as the longitude and latitude in the Sagittarius
coordinate system.  In this study, we are discussing fields that are
$\sim74\arcdeg - 128\arcdeg$ from the core of the Sgr dSph, along its
trailing stream.

\begin{figure}[!t]
\plotone{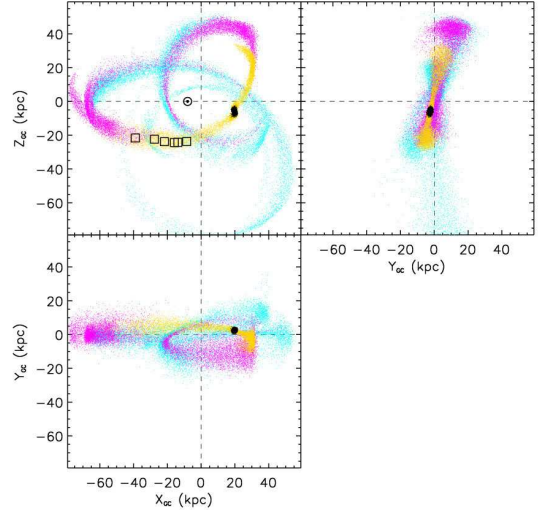}

\caption{Sagittarius debris from the best-fit triaxial halo model of
\citet{lm10a}, shown in Galactic $(X,Y,Z)_{\rm GC}$ (right-handed)
coordinates. Colors represent debris stripped on successive orbits, as
in Figure~\ref{fig:radec}, with 2 passages of additional (earlier)
debris included as cyan points. The three panels represent the
projection of Sgr debris onto the Galactic $X_{\rm GC}-Z_{\rm GC}$,
$Y_{\rm GC}-Z_{\rm GC}$, and $X_{\rm GC}-Y_{\rm GC}$ planes. In the
upper left panel (the $X_{\rm GC}-Z_{\rm GC}$ plane) the Sgr orbital
plane is nearly face-on; open black squares in this panel denote the
positions of the Kapteyn Selected Areas in this study along the Sgr
trailing tail. The Sun is represented by the circle at ($X_{\rm
GC},Z_{\rm GC}$)=(-8.0,0.0) kpc, with the Sgr core (black dots) beyond
the Galactic center as viewed from our position, and slightly below
the plane. The upper right and lower left panels (i.e., the $Y_{\rm
GC}-Z_{\rm GC}$ and $X_{\rm GC}-Y_{\rm GC}$ planes) illustrate the
near-coincidence of the Sgr orbital plane with the Galactic $X_{\rm
GC}-Z_{\rm GC}$ plane. Note that very few Sgr debris points make
excursions of more than $\sim5-10$ kpc to either side of the Galactic
$X_{\rm GC}-Z_{\rm GC}$ plane (in the $Y_{\rm GC}$ direction).}
\label{fig:sgr_xyz}
\end{figure}

The positions of the SAs in this study in the Galactic Cartesian
$X_{\rm GC}-Z_{\rm GC}$ plane are shown in the upper left panel of
Figure~\ref{fig:sgr_xyz}, overlaid atop simulated Sgr debris from the
\citetalias{lm10a} model. Of course, to place points on this figure
for the SAs requires an estimate of heliocentric distance. Where
needed throughout this work, we use mean distances to Sgr debris in
each SA field estimated from the \citetalias{lm10a} model debris along
corresponding lines of sight. We have chosen to do this rather than
measure Sgr debris distances because (a) our data in most fields don't
reach much fainter than the main sequence turnoff of Sgr, and (b) the
expected line-of-sight depth of the Sgr stream in this portion of the
trailing tail is $\sim10$ kpc, which ``smears'' the main sequence out
by as much as $\sim0.5$ magnitudes. Both of these factors make
isochrone fitting to derive distances rather unconstrained, so for all
analysis requiring a distance estimate we adopt the mean model debris
distances from \citetalias{lm10a} at each position, with the
line-of-sight depth of the stream in each field defining the
uncertainty in the distance. These values are given in
Table~\ref{tab:sgr_kinematics} below.

\begin{table*}[colspan=2,!t]
\begin{center}

\caption{Kapteyn's Selected Areas in This Study} \label{tab:sa_tab}
\begin{tabular}{cccccccc}
\\
\tableline
\\
\multicolumn{1}{c}{SA} & \multicolumn{1}{c}{RA} & \multicolumn{1}{c}{Dec} & \multicolumn{1}{c}{$l$} & \multicolumn{1}{c}{$b$} & \multicolumn{1}{c}{$\Lambda_{\odot}$\tablenotemark{a}}  & \multicolumn{1}{c}{$B_{\odot}$} & \multicolumn{1}{c}{$E(B-V)$\tablenotemark{b}} \\ \relax
 & (J2000.0) & (J2000.0) & (degrees) & (degrees) & (degrees) & (degrees) & \\
\tableline
\\
71 & 03:17:11.5 &  15:24:57.6 & 167.1 &  -34.7 & 128.2 &  -5.6 & 0.19 \\
94 & 02:55:58.1 &  00:30:03.6 & 175.3 &  -49.3 & 116.3 &   4.8 & 0.09 \\
93 & 01:54:52.1 &  00:46:40.8 & 154.2 &  -58.2 & 103.2 &  -3.2 & 0.03 \\
92 & 00:55:03.8 &  00:47:13.2 & 124.9 &  -62.1 & 90.1  & -10.6 & 0.03 \\
117 & 01:17:04.1 & -14:11:13.2 & 149.0 &  -75.7 & 87.6 &   5.1 & 0.02 \\
116 & 00:18:08.4 & -14:19:19.2 & 90.1 &  -75.0 & 74.9  & -1.4 & 0.02 \\
\tableline
\end{tabular}
\end{center}
\tablenotetext{1}{Coordinates in the Sagittarius system as defined by
\citet{msw+03}. $\Lambda_{\odot}$ and $B_{\odot}$ are analogous to
Galactic longitude and latitude, but rotated such that the Sgr core
defines the center of the system (i.e., $\Lambda_{\odot}, B_{\odot} =
0\arcdeg, 0\arcdeg)$, with $\Lambda_{\odot}$ increasing along the
trailing tidal tail. The fields in this study sample the trailing tail
between $74-128\arcdeg$ from the Sgr core.}
\tablenotetext{2}{Interstellar reddening value estimated from the maps of \citet{sfd98}.}
\end{table*}

\subsection{Photometry and Proper Motions} 
\label{photom_pm.sec}

\begin{figure*}[!t]
\begin{center}
\includegraphics[height=4.3in,trim=0.5in 0.2in 0in 0.2in, clip]{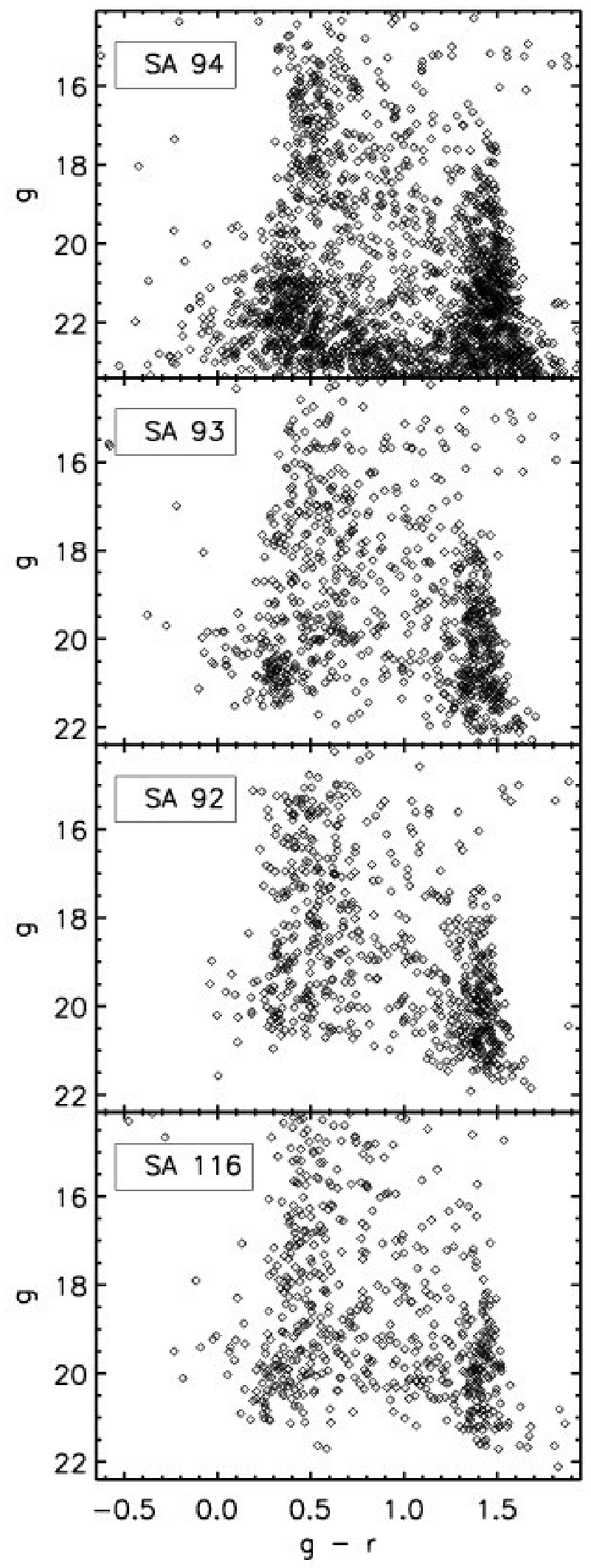}
\includegraphics[height=4.3in,trim=0.3in 0.2in 0in 0.2in, clip]{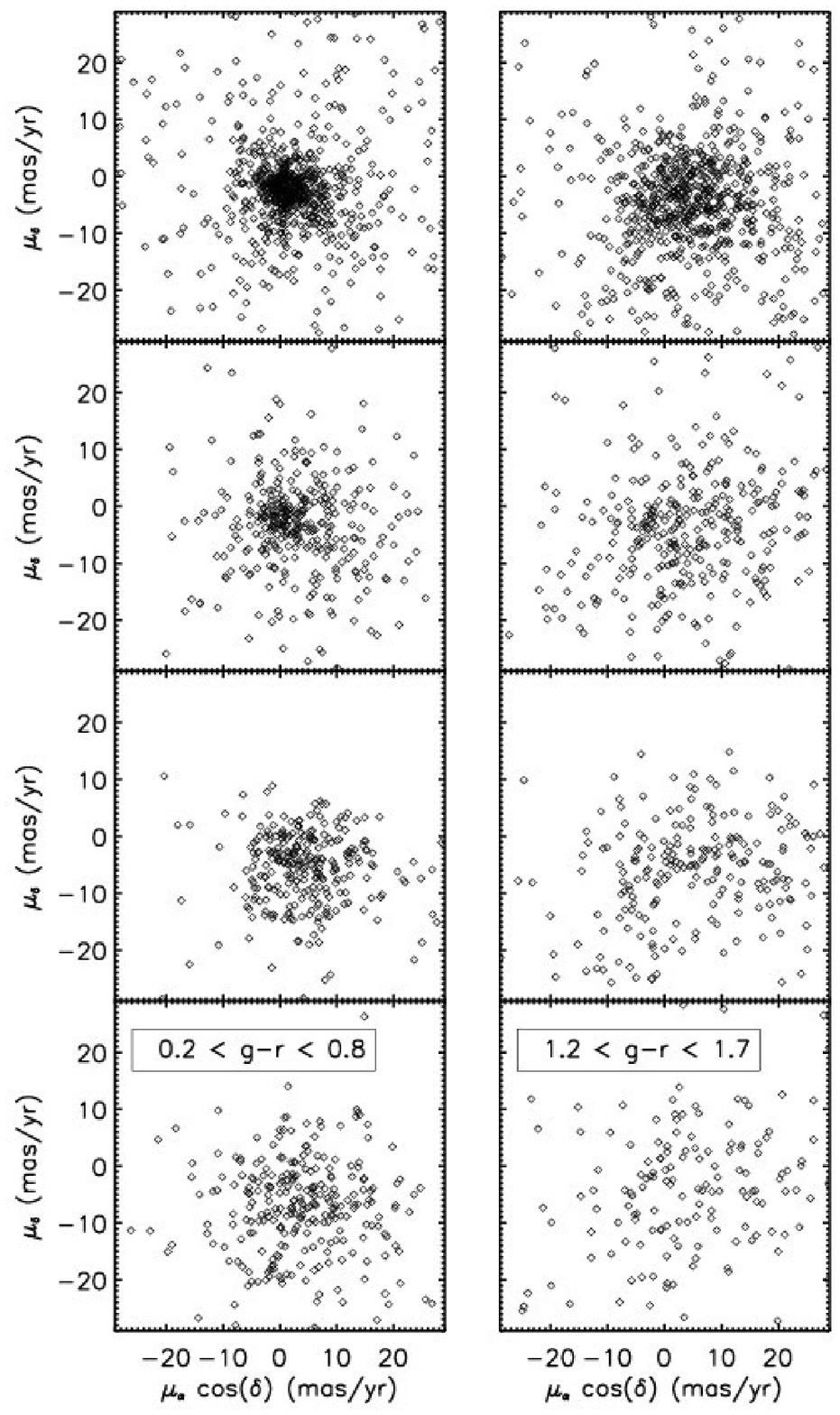}
\caption{{\it Left:} SDSS color-magnitude diagrams (CMDs) of all stars
with measured proper motions in the four fields of our survey that
overlap the SDSS footprint. Each of these CMDs shows the blue ($g-r
\sim 0.5$) swath of stars at the bright end made up of primarily Milky
Way thin- and thick-disk MSTO stars. Below this feature at similar
blue, ($g-r \lesssim 0.6$) colors, but at fainter ($g \gtrsim 19.5$)
magnitudes in each field is an apparent overdensity likely made up of
Sagittarius main-sequence stars. Note that the much deeper proper
motion catalog of SA 94 samples much more of the Sgr main sequence
than those catalogs for the other fields. {\it Right:} Proper motion
vector point diagrams (VPD) , separated into a blue ($0.2 < g-r <
0.8$) sample (middle column) and a red ($1.2 < g-r < 1.7$) subset
(right). Red stars from the prominent feature visible at red colors
($g-r \gtrsim 1.2$) in each CMD should be primarily nearby Galactic
M-dwarfs. The blue stars contain main sequence turnoff stars of MW
populations, as well as candidate Sgr MSTO stars. The proper motions
of many stars are tightly clumped in the blue samples, suggesting that
a distant, common-motion stellar population (i.e., Sgr debris) may be
present among these stars.} \label{fig:gr_cmd_vpd}
\end{center}
\end{figure*}

For those fields (SAs 92, 93, 94, and most of SA 116) that lie within
the Sloan Digital Sky Survey (SDSS) footprint, we have used photometry
(shown in Figure~\ref{fig:gr_cmd_vpd}) from SDSS Data Release 7 (DR7;
\citealt{aaa+09}) for our analyses.  The remaining photometric data for
this survey (for SAs 71 and 117) are photographic and derived from the
late epoch (du Pont 2.5-m telescope) plates from which the proper
motions were measured. Calibration of the photographic magnitudes in
the blue (IIIa-J+GG385) and visual (IIIa-F+GG495) passbands onto the
standard Johnson-Cousins system was achieved using CCD photometry
taken in 1997-1998 with the Swope 1-m at Las Campanas Observatory.
However, that UBV CCD photometry only covers a small portion
($\sim20-30\%$) of each field, and is shallower than the magnitude
limit of the photographic plates by 1-1.5 magnitudes for red stars,
and $\sim0.3$ magnitudes for blue stars, yielding poor photometric
calibration (for details, see
\citealt{cmg+06}).  Thus, for SAs 71 and 117, those aspects of our
analysis that rely on the photometry (e.g., spectroscopic target
selection, photometric parallax distance estimates, and calibration of
color-dependent systematics in the proper motions) are subject to the
uncertainties in calibration of the photographic photometry propagated
by up to 0.05 magnitudes in the $B$ and $V$ photometry.

Details of the proper motion reductions appear in
\citet{cmg+06}, so here we provide only an overview.  For all of
the near-equatorial ($-15\arcdeg \le \delta \le 15\arcdeg$) Selected
Areas in our study, proper motions are derived from plates taken with
the Mt.~Wilson 60-inch between 1909-1912, combined with deliberately
matched plates (in approximate area and plate scale) taken by
S. Majewski with the 2.5-m Las Campanas du Pont telescope between
1996-1998.  All plates were digitized with the Yale PDS
microdensitometer. Most background QSOs and galaxies are near the
limiting magnitude of the proper motion catalogs derived from solely
the Mt.~Wilson and du Pont plates; therefore if we used only these
data, the correction to an absolute proper motion frame would be
determined by only a handful of poorly-measured faint objects.  To
extend the proper motion limiting magnitude beyond the limit imposed
by the Mt.~Wilson plates, we augmented the Kapteyn survey data with
measurements of plates from the first Palomar Observatory Sky Survey
(POSS-I), which were taken in the early 1950s.  These plates, while of
much coarser plate scale ($67\farcs2$ mm$^{-1}$; compare to
$10\farcs92$ mm$^{-1}$ for the du Pont, and $27\farcs12$ mm$^{-1}$ for
the 60''), are deeper than the 60'' plates, and provide a
$\sim40$-year baseline with both the Mt.~Wilson plates {\it and} the
du Pont plates, allowing us to extend the limiting magnitude of the
survey (at least for the POSS-I/du Pont proper motion baseline) to
$V\gtrsim20.5$.

In addition, for two of the SA fields in this study (SAs 71 and 94) we
have included plates from the Kitt Peak National Observatory Mayall
4-meter telescope, taken at prime focus in the mid-1970s by A. Sandage
and in the mid-1990s by S. Majewski. This additional high-quality
plate material extends the limiting magnitude in SAs 71 and 94 to
$\sim50\%$ completeness at SDSS $r$ magnitudes of $\sim22$ (compared
to $\sim50\%$ completeness at $r\sim19.5$ in fields without 4-meter
plates; the difference in depth between SA 94 and the other fields can
be seen in Figure~\ref{fig:gr_cmd_vpd}).  The precision of the proper
motion measurements for the ``deeper" fields (i.e., those with 4-meter
plates) is improved by roughly a factor of 2 at the same magnitude
over those with no 4-m plate material; this arises because of both
more plate material with fine (18\farcs6 mm$^{-1}$) plate scale at
intermediate epochs and the increased depth provided by the 4-m
plates.

\begin{table*}[!htp]
\begin{center}
\caption{Summary of Spectroscopic Observations} \label{tab:sgr_obs_tab}
{\footnotesize
\begin{tabular}{cccccc}
\\
\tableline
\\
\multicolumn{1}{c}{SA} & \multicolumn{1}{c}{Date} & \multicolumn{1}{c}{Telescope/Instrument} & \multicolumn{1}{c}{Exposures} & \multicolumn{1}{c}{$N_{\rm stars}$} & \multicolumn{1}{c}{Mag. limit} \\
 & & & (seconds) & & \\
\tableline
\\
{\bf 71} & Dec 2002 & WIYN+Hydra\tablenotemark{a} &    4 x 1800    & 37 & 18-19\tablenotemark{c} \\
 - & Nov 2003 & WIYN+Hydra\tablenotemark{a} &    10 x 1800, 4 x 1800 & 74 & 18-19 \\
 - & Dec 2005 & WIYN+Hydra\tablenotemark{b} & 6 x 1800 & 27 & 19.5 \\
 - & Oct 2006 & WIYN+Hydra\tablenotemark{b} & 6 x 1800 & 47 & 19.0 \\
 - & Dec 2006 & WIYN+Hydra\tablenotemark{b} & 4 x 1800 & 58 & 19.3 \\ 
 - & Dec 2007 & WIYN+Hydra\tablenotemark{b} & 6 x 2400 & 53 & 19.7 \\
 - & Dec 2008 & MMT+Hectospec & 6 x 1800, 6 x 1800 & 176, 193 & 22.0 \\
\multicolumn{4}{c}{\bf ... TOTAL .................... } & {\bf 503} & \\
\\
{\bf 92} & Oct 2006 & WIYN+Hydra\tablenotemark{b} & 6 x 1800 & 39 & 19.4\tablenotemark{d} \\
 - & - & SDSS & - & 211 & 21.5 \\
\multicolumn{4}{c}{\bf ... TOTAL .................... } & {\bf 254} & \\
\\
{\bf 93} & Oct 2006 & WIYN+Hydra\tablenotemark{b} & 6 x 1800 & 55 & 19.8\tablenotemark{d} \\
 - & Oct 2007 & WIYN+Hydra\tablenotemark{b} & 6 x 1800 & 51 & 19.9 \\
 - & Dec 2008 & MMT+Hectospec & 4 x 1500 & 195 & 21.5 \\
 - & - & SDSS & - & 101 & 21.8 \\
\multicolumn{4}{c}{\bf ... TOTAL .................... } & {\bf 292} & \\
\\
{\bf 94} & Oct 2007 & WIYN+Hydra\tablenotemark{b} & 8 x 1800 & 15 & 19.0\tablenotemark{d} \\
 - & Dec 2007 & WIYN+Hydra\tablenotemark{b} & 6 x 2400 & 53 & 19.9 \\
 - & Dec 2008 & MMT+Hectospec & 8 x 1800, 6 x 1800 & 189, 165 & 22.5 \\
 - & - & SDSS & - & 88 & 22.5 \\
\multicolumn{4}{c}{\bf ... TOTAL .................... } & {\bf 432} & \\
\\
{\bf 116} & Oct 2007 & WIYN+Hydra\tablenotemark{b} & 6 x 1800 & 25 & 16.7\tablenotemark{d} \\
 - & Dec 2007 & WIYN+Hydra\tablenotemark{b} & 4 x 2700 & 59 & 17.7 \\
 - & Nov 2008 & WIYN+Hydra\tablenotemark{b} & 7 x 1800 & 60 & 19.9 \\
\multicolumn{4}{c}{\bf ... TOTAL .................... } & {\bf 122} & \\
\\
{\bf 117} & Oct 2007 & WIYN+Hydra\tablenotemark{b} & 8 x 1800 & 47 & 19.6\tablenotemark{c} \\
 - & Dec 2007 & WIYN+Hydra\tablenotemark{b} & 4 x 2400 & 49 & 19.7 \\
 - & Dec 2008 & MMT+Hectospec & (4 x 1800)+(4 x 1500) & 182 & 20.5 \\
\multicolumn{4}{c}{\bf ... TOTAL .................... } & {\bf 206} & \\
\\
\tableline
\end{tabular}
} 
\tablenotetext{1}{These WIYN+Hydra observations used the 800@30.9 grating with the red fiber cables, centered at $\sim5200$~\AA, yielding $\sim1.0$~\AA~resolution spectra.}
\tablenotetext{2}{These WIYN+Hydra observations used the 600@10.1 grating with the red fiber cables, yielding spectra covering wavelengths from $\sim4400-7200$~\AA~at $\sim3.35$~\AA~resolution.}
\tablenotetext{3}{``Roughly calibrated'' V magnitudes (see \citealt{cmg+06}).}
\tablenotetext{4}{SDSS g magnitude.}

\end{center}
\end{table*}

\subsection{Radial Velocities}
\label{rv_meas.sec}

The survey fields in which we focus this Sgr study fall on or near the
portion of the trailing stream in which \citet{mkl+04} and
\citet{mbb+07} have identified a clear Sgr radial velocity
signature. These fields can be seen relative to the orbital path of
the Sgr dSph in the left panel of Figure~\ref{fig:radec}, and with
respect to the expected location of Sgr trailing tidal debris
according to the best-fitting models of \citet{lm10a} in the right
panel of Figure~\ref{fig:radec}. We have already shown evidence
\citep{cmg+06,ccg+08} that the overdensities of faint, blue stars that
are tightly clumped in proper motions in a few of these fields are
likely made up of Sgr debris.  It was these apparent overdensities
that guided our target selection for spectroscopic follow-up.  We
began with spectroscopy from the Hydra multifiber spectrograph on the
WIYN 3.5-m telescope; in most of the shallower fields of this survey,
moderate-resolution spectra can be obtained with this instrument in a
reasonable amount of observing time.  For the deep fields (and some of
the shallower fields as well), we used another multi-object
spectrograph, the Hectospec instrument on the MMT 6.5-m, which allowed
us to observe $\gtrsim200$ Sgr stream candidates simultaneously per
setup down to faint ($g$ or $V\gtrsim21.5$) magnitudes.  We describe
the observations and data reduction for each instrument separately
below.

\subsubsection{Sample Selection}

Targets for spectroscopic follow-up were selected to lie within the
locus of the suspected Sgr main sequence turnoff (MSTO) at faint ($g$
or $V\gtrsim19.5$) magnitudes and blue ($g-r$ or $B-V \lesssim 0.8$)
colors.  \citet{cmg+06} showed that the proper motions of these Sgr
MSTO candidates clump more tightly than those of the predominantly
nearby M-dwarfs at red colors.  The tight clumping in the proper
motion vector point diagram (VPD) of stars in the MSTO feature was
used to define a selection box in proper motion space which should
contain any Sgr debris that is present in each field, and eliminate a
good fraction of unrelated stars of similar color and magnitude.  Care
was taken not to be too stringent with either the proper motion or
photometric criteria, to preserve as many potential Sgr stars in the
wings of the distributions as possible. Because the quality and depth
of the photometry and proper motions varies between fields, different
candidate selection criteria were adopted for each field.

For WIYN+Hydra observations, only stars brighter than 20th magnitude
(either $V$ or $g$, depending on whether a given field had SDSS
photometry) were included in the multifiber setups, because fainter
stars than this require rather long exposures with a 3.5-meter
telescope to achieve adequate signal-to-noise for radial velocity
measurement. After all available fibers were filled with MSTO
candidates the remaining fibers were assigned to targets at relatively
bright magnitudes ($\lesssim 18$) that also reside within the VPD
selection criteria (i.e., potential Sgr RGB stars).  Stars fainter
than 20th magnitude were targeted with the 6.5-meter MMT telescope,
which easily reaches Sgr MSTO candidates in these fields.

\subsubsection{WIYN+Hydra Observations}
\label{wiyn_obs.sec}

Spectroscopic data were obtained during a total of eight observing
runs with the WIYN 3.5-m telescope\footnote{The WIYN Observatory is a
joint facility of the University of Wisconsin-Madison, Indiana
University, Yale University, and the National Optical Astronomy
Observatory.} between December 2002 and November 2008. We used the
Hydra multi-fiber spectrograph in two different setups. The first one
(Dec. 2002, Nov. 2003 observing runs) used the 800@30.9 grating with
the red fiber cables and an order centered in the neighborhood of the
Mg triplet (5170 \AA) and covering about 980
\AA~of the spectrum.  This setup delivered a dispersion of 0.478
\AA$~$pix$^{-1}$ and a resolving power $R \sim 5400$ (resolution $\sim1$ \AA).  The second
spectrograph configuration (Dec. 2005, Oct./Dec. 2006, Oct./Dec. 2007,
and Nov. 2008) used the 600@10.1 grating with the red fiber cable to
yield a wavelength coverage $\lambda$ = 4400--7200 \AA~at a dispersion
of 1.397 \AA$~$pix$^{-1}$, for a spectral resolution of 3.35 \AA~($R
\sim 1500$ at $\lambda$ = 5200 \AA).  This spectral region was
selected to include the H$\beta$, Mg triplet, Na D, and H$\alpha$
spectral features. Typically 60-70 targets were placed on Hydra
fibers, with the remaining 15-20 fibers placed on blank sky regions to
allow for accurate sky subtraction.  Each of the 2005-6 datasets was
obtained in less than optimal conditions, including substantial
scattered moonlight in Dec. 2005 and cloudy conditions in both 2006
runs.  The majority of the 2007 and 2008 data were obtained under
favorable conditions. We further note that the Nov. 2008 observing run
occurred after the WIYN Bench Spectrograph Upgrade, which included the
implementation of a new collimator into the Bench configuration, as
well as a new CCD that delivers greatly increased throughput.

Table~\ref{tab:sgr_obs_tab} summarizes the observations.  Each Hydra
configuration was exposed multiple times (usually in sets of 30
min. exposures) to enable cosmic ray removal. Standard pre-processing
of the initial two-dimensional spectra used the CCDRED package in
IRAF.\footnote{IRAF is distributed by the National Optical Astronomy
Observatory, which is operated by the Association of Universities for
Research in Astronomy (AURA) under cooperative agreement with the
National Science Foundation.} Frames were summed, then 1-D spectra
were extracted using the DOHYDRA multifiber data reduction utilities
(also in IRAF). Dispersion solutions were fitted using 30--35 emission
lines from CuAr arc lamp exposures taken at each Hydra
configuration. On each observing run we targeted a few bright radial
velocity standards covering spectral types from F through early K
(both dwarfs and giants), each through multiple fibers, to yield
multiple individual cross-correlation template spectra.  These RV
standard spectra were cross-correlated against each other using the
IRAF tool FXCOR to determine the accuracy of the velocities and remove
any outliers (i.e., those stellar spectra that yield unreasonable
cross-correlation results due to template mismatch or some defect,
such as a poorly-removed cosmic ray, bad CCD column, or other unknown
culprit).  Measured velocities of the RV standards typically agreed
with published IAU standard values to within 1-2 km s$^{-1}$.  Radial
velocities for program stars were derived by cross-correlating all
object spectra against all of the standards taken on the same
observing run.  To maximize the $S/N$ in faint, metal-poor stars, only
$\sim200$ \AA-wide regions centered on the H$\beta$, Mg triplet, and
H$\alpha$ absorption lines were used for cross-correlation.

Radial velocity uncertainties were derived using the \citet{vmo+95}
method, as described in \citet{mcf+06} and \citet{fmp+06}.  The
Tonry-Davis ratio (TDR; \citealt{td79}) scales with $S/N$, such that
individual RV errors can be calculated directly from the TDR, provided
you have multiple observations at varying $S/N$ of some particular
standard star to map the dependence.  We have used this technique for
all datasets except those from the Dec. 2006 observing run, when only
a total of four RV standard spectra were taken.  For the SA 71
configuration observed on this run, the RV uncertainty for each stars
is derived as the standard deviation of the RV results from that
star's spectrum using cross-correlation against each of the four
standards.  Typical RV uncertainties for individual measurements for
all fields were $\sigma_{V} \approx 5-10$ km s$^{-1}$, with most
spectra having $S/N\sim$15--20 per Angstrom. The uncertainty depends
on $S/N$ (essentially magnitude) for all stars from a single Hydra
pointing; however, the varying exposure times between Hydra setups and
changing observing conditions mean that $\sigma_{V}$ is not strictly a
function of magnitude in our final catalogs.

\subsubsection{MMT+Hectospec Observations}

To obtain spectra of fainter Sgr MSTO candidates, we were granted
three nights of queue-scheduled NOAO observing time on the MMT
6.5-m.\footnote{Observations reported here were obtained at the MMT
Observatory, a joint facility of the Smithsonian Institution and the
University of Arizona. MMT telescope time was granted by NOAO
(proposal ID 2008B-0448), through the Telescope System Instrumentation
Program (TSIP). TSIP is funded by the NSF.} A total of six observing
configurations were observed with the 300-fiber Hectospec multifiber
spectrograph \citep{ffr+05} mounted at the f/5 focus of the MMT.
Targets were selected from among the apparent stellar overdensities of
blue MSTO canidates at magnitudes too faint ($g \gtrsim 20$) to be
reasonably observed with WIYN+Hydra, but using the same proper motion
criteria used to choose Hydra targets.  In each of these
configurations, a few targets previously observed with Hydra were
included for a radial velocity consistency check, and any fibers
unable to be filled with faint stars were assigned to brighter ($g <
20$) Sgr RGB candidates.

We used the 270 gpm grating, centered at $\sim$6400 \AA, to give a
working wavelength range of 3700-9150~\AA~at a dispersion of $\sim$1.21
\AA$~$pix$^{-1}$ (4.85 \AA~resolution).  This low resolution allows us
to obtain adequate signal-to-noise ($S/N \gtrsim 10$ per \AA) spectra
of stars as faint as $g=22.5$ to measure radial velocities in 3-4
hours of total exposure time.  We observed two multifiber setups each
in SAs 71 and 94, the two fields in this study that have 4-meter
plates (and thus deeper proper-motion catalogs).  The rest of the time
was devoted to one configuration each in SAs 93 and 117. Each
configuration had 200-230 targets assigned (essentially as many as the
fiber assignment software would allow), with the remainder designated
as sky fibers. The sky fibers were chosen in areas with the nearest $g
< 22$ star at least 15$\arcsec$ away, distributed throughout the field
so that a number of them would fall within each of the two CCD chips
of the Hectospec system. The number of exposures in each field, each
exposure time, the number of stars with measured radial velocities,
and the limiting magnitude of each spectroscopic field are given in
Table~\ref{tab:sgr_obs_tab}.

The Hectospec data were reduced using an external version of the SAO
``SPECROAD'' reduction pipeline \citep{mwc+07} written by Juan
Cabanela and called
ESPECROAD.\footnote{\url{http://iparrizar.mnstate.edu/~juan/research/ESPECROAD/index.php}}
The pipeline automates many reduction steps, including
bias-correction, flat-fielding, cosmic-ray rejection, fiber-to-fiber
throughput adjustments, and sky subtraction.  Wavelength calibration
was performed manually using sets of three combined HeNeAr calibration
lamp exposures from each night of observing.

We derived RVs using the IRAF task FXCOR to cross-correlate object
spectra against fourteen template RV standard spectra of nine
different stars ranging in spectral type from F through K.  We first
correlated the standard spectra against each other, and found that our
measurements agree with published RVs\footnote{From the Geneva
Radial-Velocity Standard Stars at
\url{http://obswww.unige.ch/~udry/std/}} to within 4.9 km s$^{-1}$ for
all of these stars, with zero mean offset, and $\sigma_{\Delta V} =
3.0$ km s$^{-1}$.  To minimize the effects of noise for spectra with
lower $S/N$, the cross-correlation was restricted to the regions
around the H$\alpha$, Mg triplet, and H$\beta$ lines.

Each of the object spectra was cross-correlated against all 14
standards, and the mean RV from each of these 14 measurements was
adopted as the final result. For the queue-scheduled Hectospec
observations, we relied on the queue to provide radial velocity
standards. We were thus unable to obtain repeated exposures of the
same RV standard stars to allow us to use the ``Vogt method'' (as
described in Section~\ref{wiyn_obs.sec}) to determine RV
errors. Instead, uncertainties were estimated as the standard
deviation of the 14 independent RV measurements thus derived, and vary
(essentially as a function of $S/N$) from $\sigma_{V\rm helio} \sim 3$
km s$^{-1}$ at $g = $18.0 to $\sim$15 km s$^{-1}$ at $g =$ 21.5-22.

From repeat measures of a handful of stars, including multiple Hydra
or Hectospec observations as well as many observed with both systems,
we found mean systematic offsets of $<5$ km s$^{-1}$ between observing
runs (including both Hydra and Hectospec data).  These offsets were
applied to all RVs from a given run to place all measurements on the
same system as the Dec. 2007 WIYN+Hydra velocities.

We also note that because we selected most Hectospec targets to be
faint, blue objects with miniscule proper motions, a large number of
obvious QSOs and AGN spectra appeared in our data. These were added to
the samples of QSOs and galaxies that provided the fixed absolute
proper motion frame in each field, improving the zero points in these
fields.

\subsubsection{SDSS spectra}

We supplemented our database of radial velocities by matching our
proper motion catalogs to the SDSS spectroscopic database.  The number
of additional RVs contributed by SDSS in each field is noted in
Table~\ref{tab:sgr_obs_tab}. The majority of SDSS stars are red,
nearby M-dwarfs, so very few additional Sagittarius candidates were
contributed by the addition of these spectroscopic data.  However, the
handful of Sgr stars that are present, as well as any other stars in
common with our observations, were used for a consistency check. From
the stars in common between SDSS and our observations in SAs 94 and
93, we find mean offsets of $\leq$ 5 km s$^{-1}$. The accuracies of
SDSS radial velocities are $\sim4$ km s$^{-1}$ at $g < 18$, decreasing
to $\sim15$ km s$^{-1}$ at $g \sim 20$ \citep{yrn+09}, so we choose
not to offset the RVs .

\begin{figure}[!tp]
\plotone{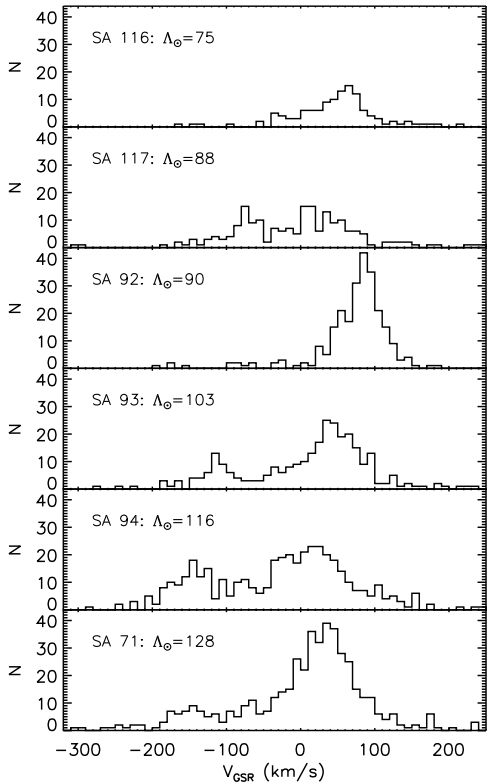}
\caption{Measured radial velocities (relative to the Galactic Standard
of Rest) in each of the six fields, displayed in order (from top to
bottom) of Sgr longitude (i.e., angular distance from the Sgr core),
$\Lambda_{\odot}$. The prominent peak in each field at $V_{\rm GSR}
\sim 0-100$ km s$^{-1}$ is made up primarily of Galactic stellar
populations. In SAs 117, 93, 94, and 71, an additional peak at lower
velocities is visible. As shown in Section 3, this peak can be attributed
to the presence of Sgr tidal debris in these
fields.} \label{fig:rvhist}
\end{figure}

\section{Sagittarius Tidal Debris Kinematics}
\label{sgr_kinematics.sec}

\subsection{Radial Velocities}
\label{sgr_rvs.sec}

Figure~\ref{fig:rvhist} shows all measured velocities in each of the
six survey fields (the total number of stellar radial velocities in
each field is given in Table~\ref{tab:sgr_obs_tab}) in the
Galactocentric ($V_{\rm GSR}$\footnote{$V_{\rm GSR} \equiv V_{\rm
helio} + 9.0 \cos b \cos l + 232.0 \cos b \sin l + 7.0 \sin b$, where
$V_{\rm helio}$ is the measured heliocentric radial velocity. This
calculation assumes a circular velocity of 220 km s$^{-1}$ at the
solar circle, and solar peculiar motion of $(U_0, V_0, W_0) = (9.0,
12.0, 7.0)$ km s$^{-1}$.}) frame.  These consist of all RVs from
WIYN+Hydra, MMT+Hectospec, and where available, SDSS.  Where multiple
measurements exist, the final catalog reflects the error-weighted mean
radial velocity.  In each of these fields, a broad peak is seen at
$V_{\rm GSR} \sim 0-100$ km s$^{-1}$ which is made up primarily of
Milky Way stellar populations in these high-latitude fields.  An
additional velocity peak is evident in SAs 71, 94, 93, and 117, well
separated (in all fields except SA 117) from the Galactic distribution
in each of these fields; it is this additional peak we shall show to
consist of mainly Sagittarius trailing tidal debris.  There is no
readily apparent peak at lower $V_{\rm GSR}$ values in SAs 92 and 116
-- this arises for different reasons in the two cases.  SA 116 is the
field in which we have the fewest measured radial velocities, and even
the stars for which we do have data are not optimally selected to find
Sgr debris. Because of limitations on exposure times due to weather,
the observed stars were all at relatively bright magnitudes ($g <
18$), where only very few Sgr red giants should be found.  More
spectra of stars at faint ($g \gtrsim 19$) magnitudes and blue colors
will be necessary to identify Sgr debris among the SA 116 data.  In SA
92, on the other hand, nearly twice as many spectra are available than
in SA 116, and mostly at relatively faint magnitudes.  The paucity of
obvious Sgr debris in this field is because of the location of SA 92
on the periphery of the stream, where Sgr stellar densities are rather
low.  There are a handful of stars at low ($V_{\rm GSR} < 0$ km
s$^{-1}$) velocities, but hardly enough candidates to assert that a
clear Sgr presence is indicated in SA 92.

To assess whether these apparent velocity overdensities are expected
among Galactic populations in each line of sight, we compared the
radial velocity distributions to those from the Besan\c{c}on Galaxy model
\citep{rrd+03}.\footnote{Model query available at
http://model.obs-besancon.fr/.}  In each SA field, the model query was
run five times to smooth out the finite sampling statistics in each
individual model run.  The five catalogs were concatenated, then for
each Kapteyn field the measured velocities were compared to the
expected radial velocities of smooth Galactic populations by scaling
the summed Besan\c{c}on model to match the total number of stars in
the broad peak in each RV histogram.  This was done separately for
"bright" and "faint" samples in each field, since stars in different
magnitude ranges preferentially sample different Galactic populations
with different velocity distributions (e.g., ``faint'' blue Galactic
stars in the region of the CMD where Sgr MSTO stars reside will be
predominantly halo stars, and thus have a much higher velocity
dispersion than stars of similar color, but much brighter magnitude,
where thin/thick disk MSTO stars predominate).

\begin{figure*}[!t]
\begin{center}
\includegraphics[width=2.25in,trim=0.5in 0 0.2in 0,clip]{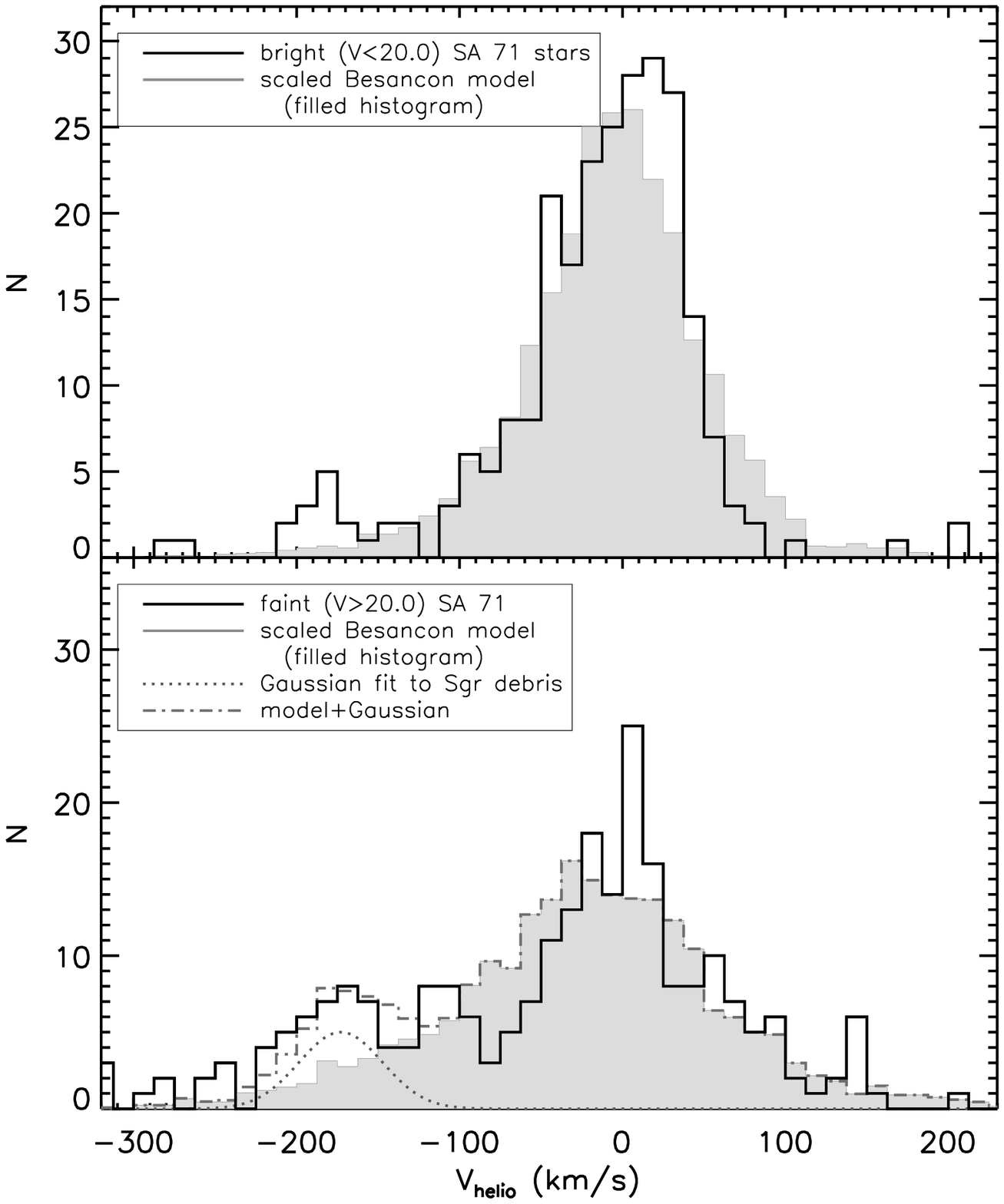}
\includegraphics[width=2.25in,trim=0.5in 0 0.2in 0,clip]{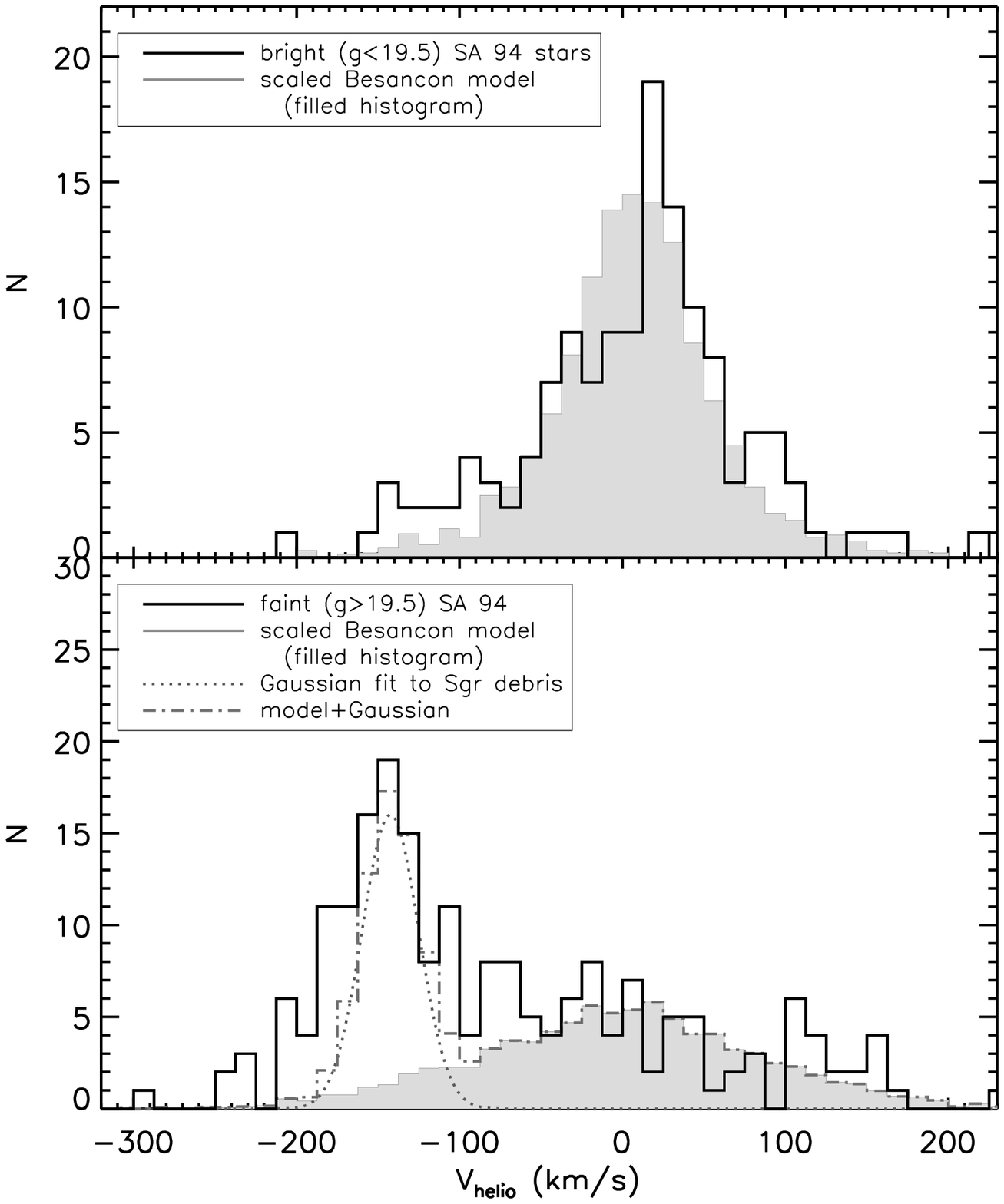}
\caption{Measured heliocentric radial velocities (solid-line
histograms) in SAs 71 (left) and 94 (right), divided into a bright (SA
71: $V<20.0$; SA 94: $g<19.5$) sample (upper panels) and a faint (SA
71: $V>20.0$; SA 94: $g>19.5$) group (lower panels). The filled gray
histogram in each panel is made up of Besan\c{c}on model points along
the corresponding line of sight in the same magnitude ranges as the
data histograms, scaled to contain the same total number of stars
between $-50 < V_{\rm helio} < 100$ km s$^{-1}$ as the observed
data. In the lower panels, a Gaussian representing the best-fitting
radial velocity and dispersion of Sgr candidates is shown (dotted
curve), along with the sum of this Gaussian and the Besan\c{c}on
distribution (dot-dashed histogram). These two fields include 4-meter
plates in the proper motion derivation, and thus contain the deepest
proper motions (and the most Sgr candidates) of any fields in the
survey. Very little, if any, Sgr debris is evident in the bright
samples.} \label{fig:sa7194_vhelhist}
\end{center}
\end{figure*}

The resulting scaled model distributions are shown as grey filled
histograms in Figure~\ref{fig:sa7194_vhelhist} for SAs 71 and 94, with
the measured heliocentric radial velocities given as solid-lined
histograms.  The broad peak is reproduced well by the model
populations, suggesting that (a) there are no large global velocity
offsets present in our data, and (b) the prominent peaks are indeed
due to foreground/background Milky Way stars.  The additional peaks at
$V_{\rm helio} \approx -170$ km s$^{-1}$ (SA 71) and $V_{\rm helio}
\approx -150$ km s$^{-1}$ (SA 94) are clearly not due to any expected
Galactic populations along these lines of sight.  

\begin{figure*}[!t]
\begin{center}
\includegraphics[width=2.25in,trim=0.5in 0 0.2in 0,clip]{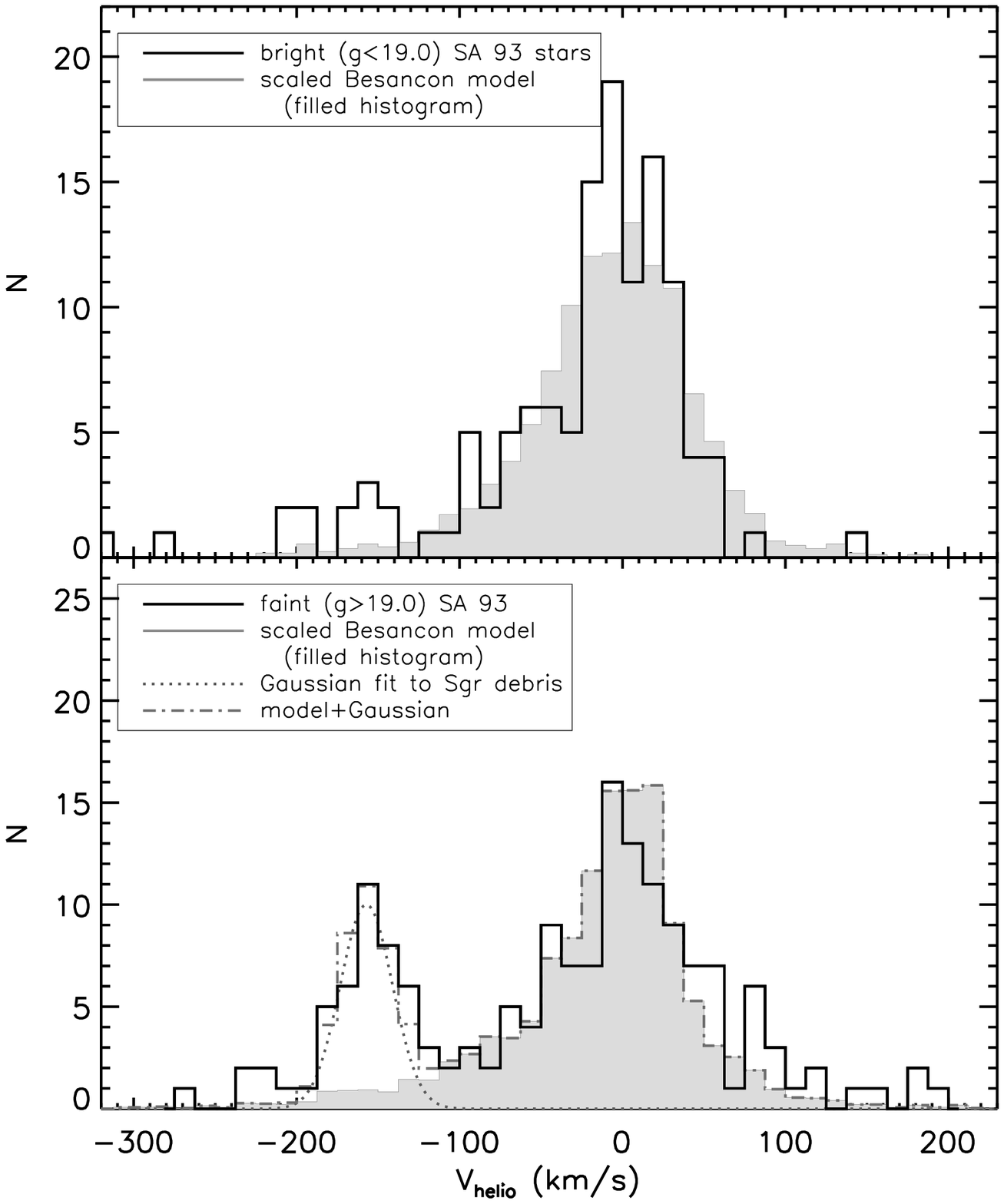}
\includegraphics[width=2.25in,trim=0.5in 0 0.2in 0,clip]{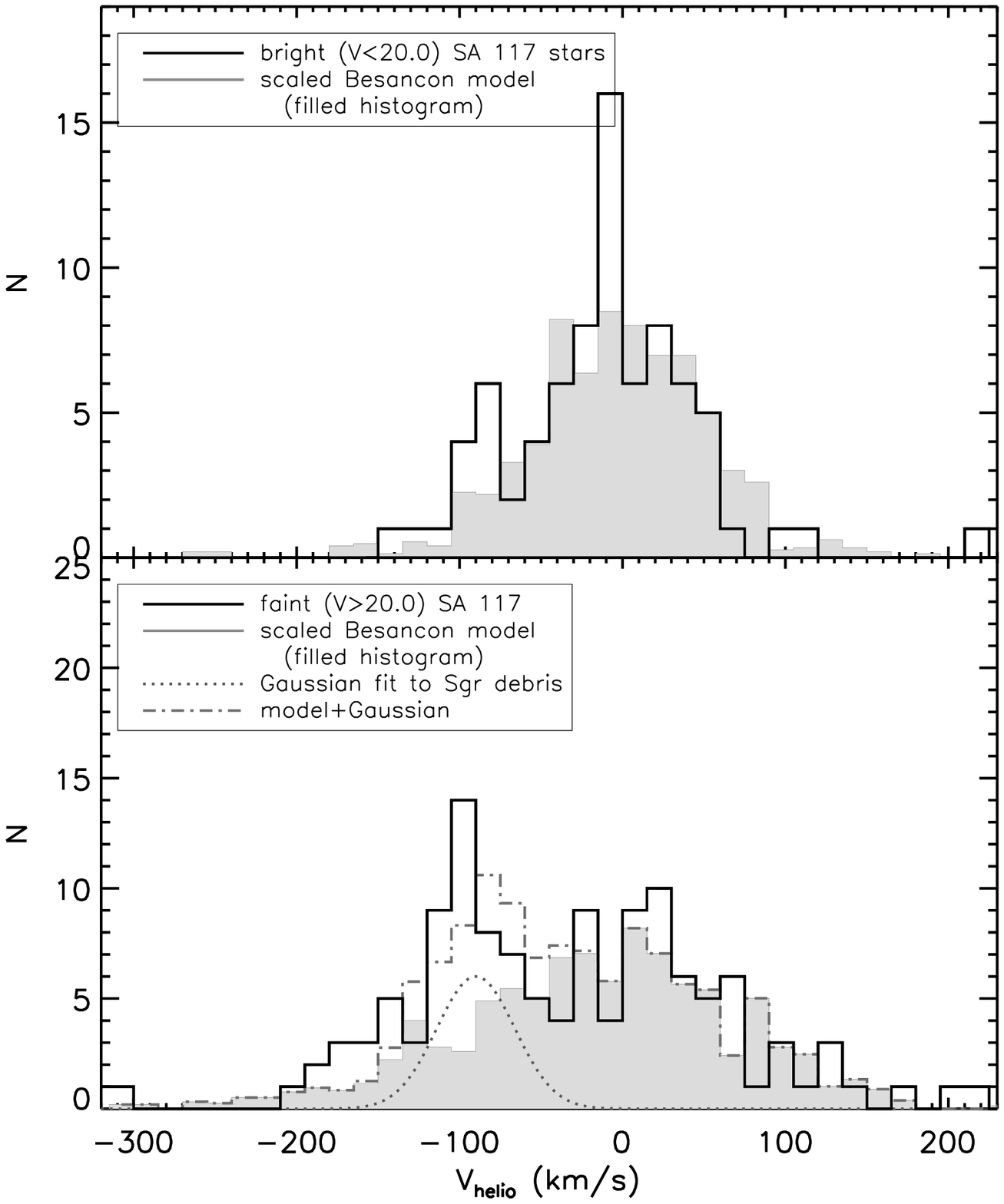}
\caption{Same as Figure~\ref{fig:sa7194_vhelhist}, but for SAs 93 (left) and 117 (right), and with slightly different definitions of the bright (SA 93: $g<19.0$; SA 117: $V<20.0$) sample (upper panels) and faint (SA 93: $g>19.0$; SA 117: $V>20.0$) group (lower panels). In both of these fields, there is a hint of a peak at Sgr-like velocities in the bright samples,
suggesting that a few Sgr red giants may have been identified in these
fields.} \label{fig:sa93117_vhelhist}
\end{center}
\end{figure*}

Similar histograms are shown for SAs 93 and 117 in
Figure~\ref{fig:sa93117_vhelhist}, which again clearly show that the
broad, prominent peak in each field is made up of Galactic
populations, and the peaks at $V_{\rm helio}
\approx -160$ km s$^{-1}$ (SA 93) and $V_{\rm helio} \approx -100$ km
s$^{-1}$ (SA 117) are inconsistent with expected Milky Way
velocities. Note that the peak in SA 117 overlaps the wings of
the Galactic distribution, making it slightly more difficult to
isolate bona fide Sgr members in this field on the basis of radial
velocities alone. 

\begin{figure*}[!t]
\begin{center}
\includegraphics[width=2.25in,trim=0.5in 0 0.2in 0,clip]{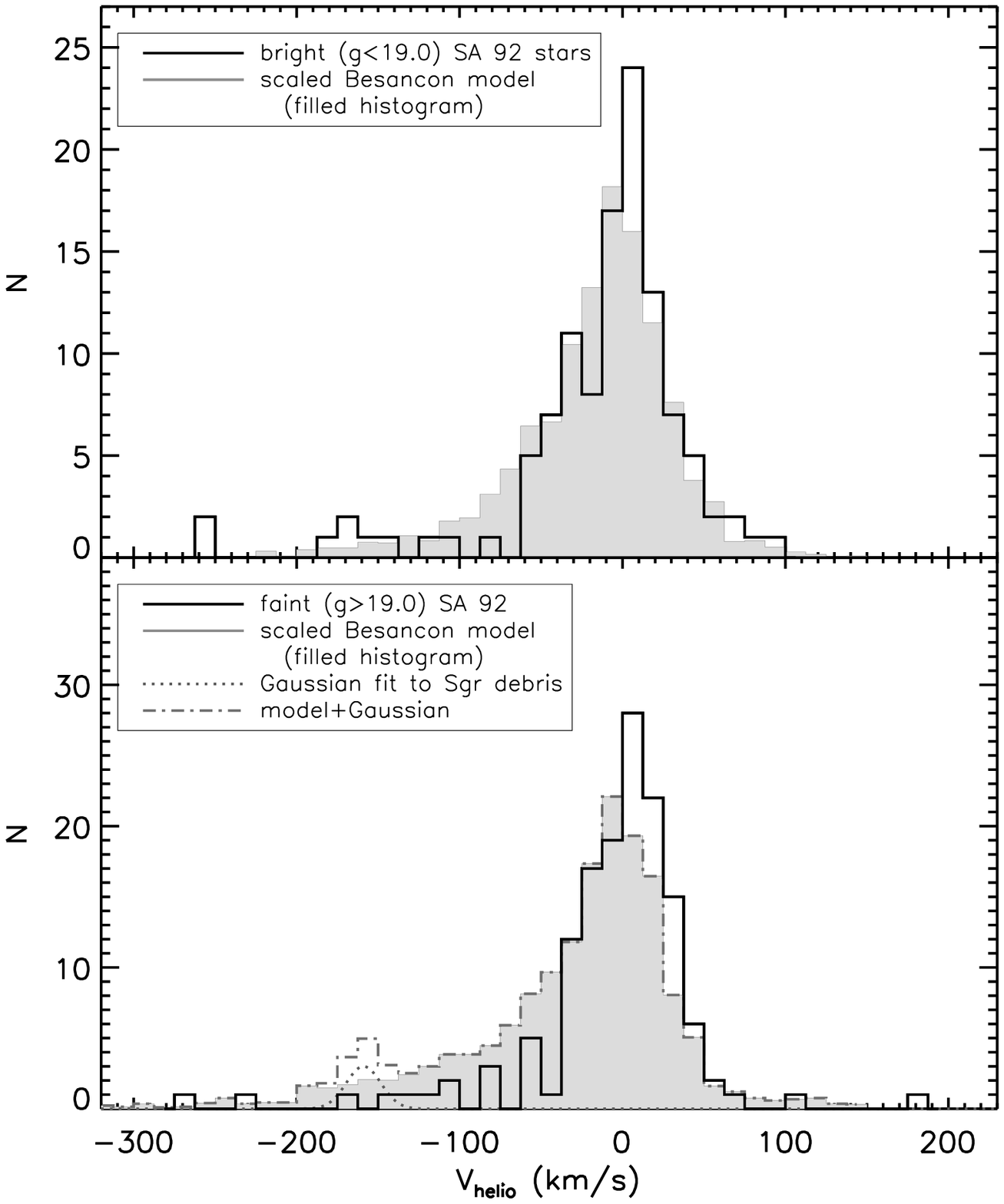}
\includegraphics[width=2.25in,trim=0.5in 0 0.2in 0,clip]{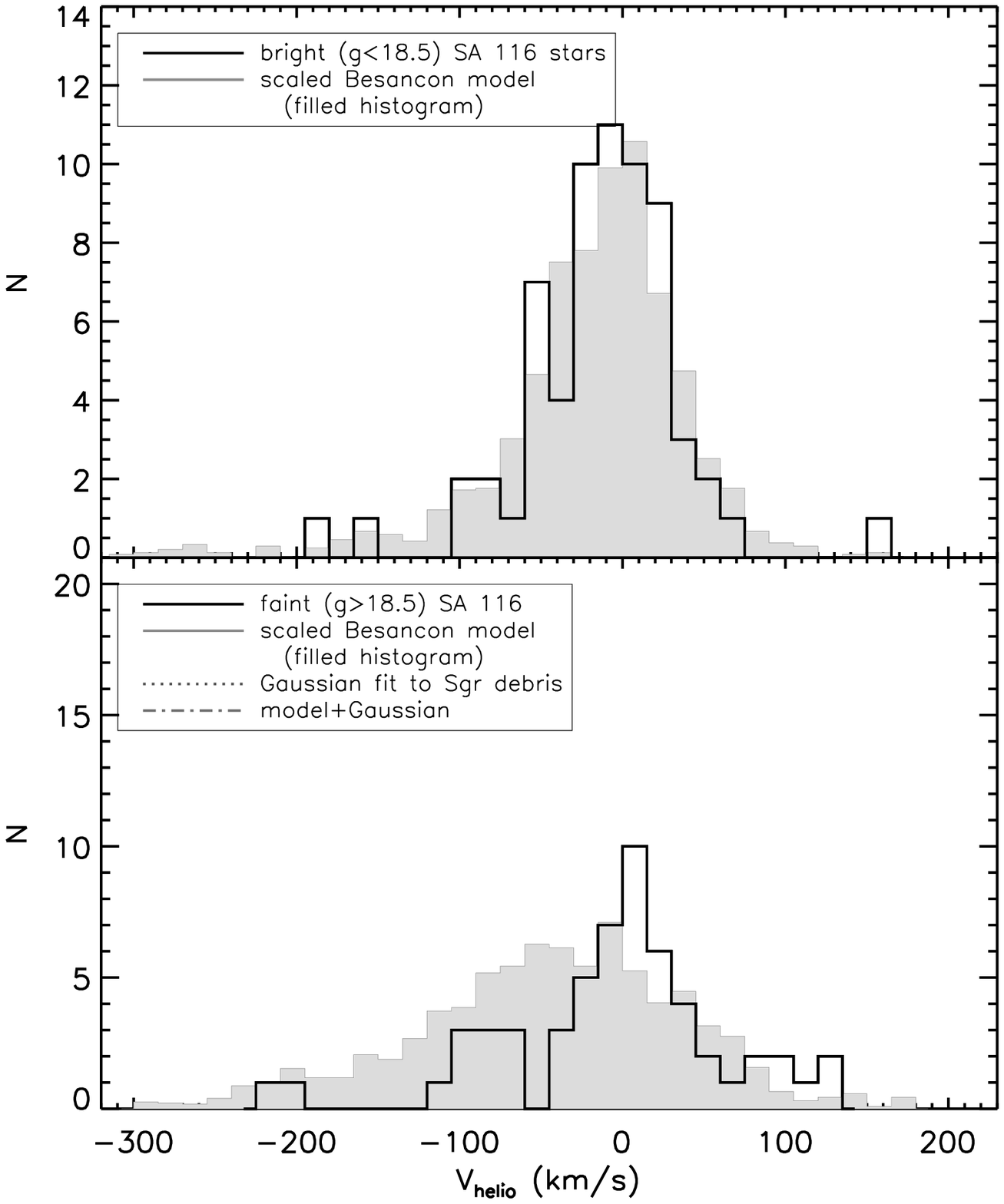}
\caption{Same as Figure~\ref{fig:sa7194_vhelhist}, but for SAs 92 (left) and 116 (right), and with slightly different definitions of the bright (SA 92: $g<19.0$; SA 116: $g<18.5$) sample (upper panels) and faint (SA 93: $g>19.0$; SA 117: $g>18.5$) group (lower panels). No excess peak is evident (relative to the Besan\c{c}on
predictions) in either of these fields, meaning we have likely sampled
very few Sgr radial velocity members along these lines of
sight.} \label{fig:sa92116_vhelhist}
\end{center}
\end{figure*}

Finally, we performed the same examination in SAs 92 and 116, with the
results shown in Figure~\ref{fig:sa92116_vhelhist}. Nearly all of the
velocities shown in SA 92 are from the SDSS database, and are
predominantly very red M-dwarfs. For this reason, the long tail of the
RV distribution at negative velocities, which is due to thick disk and
halo MSTO stars, is not well reproduced by our data set. There are a
small number of stars at Sgr-like velocities in
Figure~\ref{fig:sa92116_vhelhist}, but these fall within the expected
locus of MW stars, so we cannot definitively say that Sgr members are
present among our SA 92 sample.  In SA 116, no excess peak of measured
RVs relative to the model predictions is apparent.  This is not
surprising given (a) the caveats in the first paragraph of this
section regarding the data in SA 116, and (b) the fact that the
\citet{lm10a} model predicts Sgr debris in this field to have RVs of
$-100 \lesssim V_{\rm helio} \lesssim -50$ km s$^{-1}$, overlapping
the wings of the Galactic distribution in this field.  A handful of
Sgr members may thus be present among our velocities, but they are
difficult to distinguish from the Milky Way halo stars by their RVs.

With RVs in hand, a next culling for Sgr stream candidates was
obtained by simply taking all stars within a generously defined range
around the evident associated radial velocity peak. Such a selection
will include a few Milky Way interlopers, so we examined the samples
thus selected to remove non-Sgr stars. We first removed all stars with
proper motions $|\mu| > 10$ mas yr$^{-1}$ in either dimension; such
stars, if actually at the distance of the Sgr trailing tail in this
region of sky ($\sim$25-40 kpc), would have unreasonably large ($>
1000$ km s$^{-1}$) tangential velocities ($V_{\rm tan} = 4.74d\mu$ km
s$^{-1}$, where d is the distance in kpc and $\mu$ the proper motion
in mas yr$^{-1}$). Faint, blue stars with proper motions of this
magnitude must therefore be nearby (foreground) MW white dwarfs or
metal-poor subdwarfs. After removing these stars, we then examine the
positions of all selected candidates in the color-magnitude diagram.
We reject faint stars that are well redward of the readily apparent
Sgr main sequence, and at brighter magnitudes, we remove only stars at
positions obviously inconsistent with being Sgr red giants or
horizontal branch stars.  In SA 93, a clear offset was visible between
mean proper motions of bright ($g < 19.0$) Sgr candidates and fainter
ones, so we chose to keep only candidates at $g > 19.0$, on the
assumption that the density of stream stars should be much greater at
fainter magnitudes near the lower RGB and MSTO than along the upper
RGB.  This yields fewer total Sgr candidates, but the ones that remain
have much higher probability of being Sgr members than do the brighter
candidates.

\begin{figure}[!ht]
\includegraphics[width=3.2in,trim=0.15in 0.15in 0in 0.15in,clip]{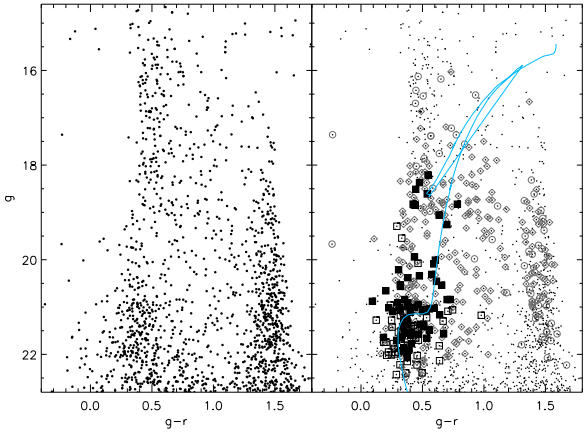}
\caption{SDSS $g$ vs. $g-r$ CMDs of SA 94; the left panel shows all
objects in our proper motion catalog that were flagged as stars by the
SDSS star/galaxy separator. The right panel overplots all stars
observed spectroscopically as larger symbols: candidates within the
initial RV selection are black squares (filled squares: final Sgr
candidates; open squares: in RV selection, but removed by other
criteria), open diamonds are stars with RVs outside the Sgr RV
selection, and open circles are stars that only have RVs in SDSS (note
that none of these ended up being selected as Sgr candidates). The
final Sgr candidates that we selected by RV, proper motion, and
color-magnitude position (filled squares) are concentrated around a
likely MSTO of Sgr debris. The blue curve is a \citet{ggo+04}
isochrone for a 10 Gyr population at [Fe/H] = -1.3 and a distance of
30 kpc.} \label{fig:sa94_cmd}
\end{figure}

\begin{figure}[!ht]
\includegraphics[width=3.2in,trim=0.15in 0.15in 0in 0.15in,clip]{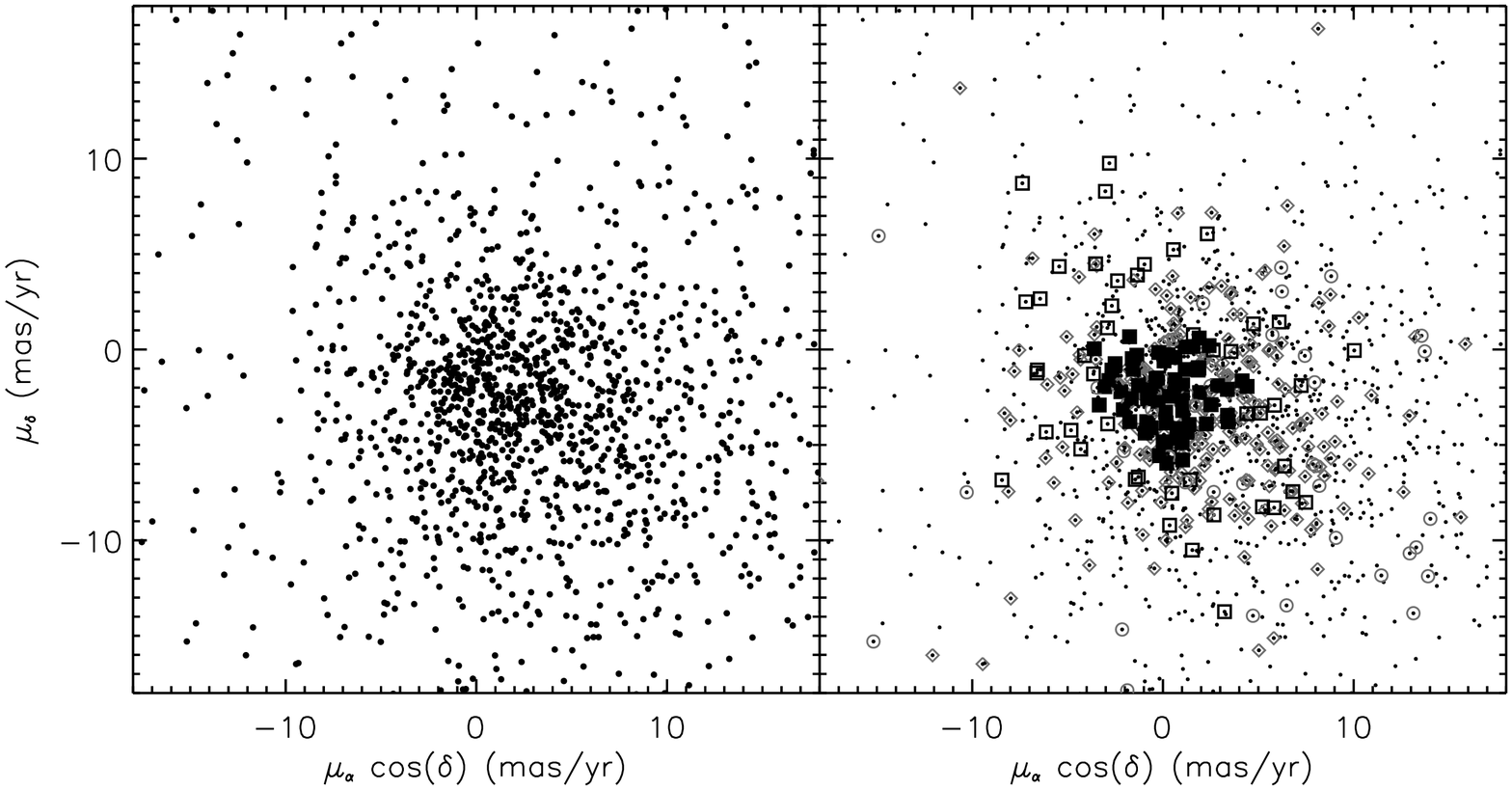}
\caption{Proper motion vector point diagram of SA 94, with panels and
symbols as in Figure~\ref{fig:sa94_cmd}.} \label{fig:sa94_vpd}
\end{figure}

\begin{figure}[!ht]
\includegraphics[width=3.2in,trim=0.15in 0.15in 0in 0.15in,clip]{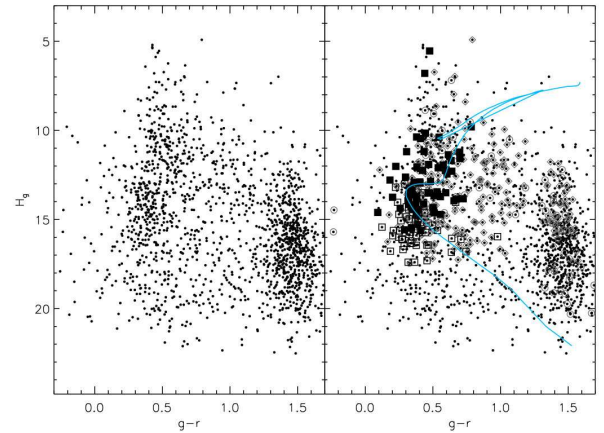}
\caption{Reduced proper motion diagram (RPMD) for SA 94, where $H_g
\equiv g + 5$ log $\mu + 5$, with $\mu$ in arcsec yr$^{-1}$. Panels
and symbols are as in Figure~\ref{fig:sa94_cmd}. The blue curve is a
\citet{ggo+04} isochrone for a 10 Gyr population at [Fe/H] = -1.3 and
a distance of 30 kpc, with the measured mean proper motion of Sgr
debris in SA 94 ($\mu_{\rm tot}$=2.34 mas yr$^{-1}$) used to convert
to reduced proper motion.} \label{fig:sa94_rpmd}
\end{figure}

\subsection{Selecting Final Sgr Candidates}

We now discuss how we pared down the samples of Sgr candidates from
the initial broad RV and proper motion selections in each field to the
final, more securely-identified samples used for analysis. For
brevity, we will show detailed examples for only two of the six fields
in the study. The first of these is SA 94, which is the "best-case"
field in our study, because it has deep proper motions due to the
availability of 4-meter plates in its data set. For comparison, we
follow the SA 94 discussion with details of SA 93, which has a much
shallower proper motion catalog than SA 94, but also has high-quality
SDSS photometry. These two fields were chosen simply to give the
reader an idea of the type and quality of the data included in this
study, and the process we followed to select Sgr candidates.

\subsubsection{Selecting Sgr Candidates in SA 94}

Figure~\ref{fig:sa94_cmd} shows the SDSS $g - r$ vs. $g$
color-magnitude diagram (CMD) for SA 94. The left panel shows all
stars for which we have measured proper motions, with SDSS-classified
galaxies removed from the sample.  On the right-hand side, open
squares depict all stars observed spectroscopically to illustrate the
candidate selection. The majority of spectroscopic targets in this
deep proper-motion field were selected from the Sgr MSTO feature of
faint, blue stars, and observed with the large-aperture MMT 6.5-m
telescope.  The remaining targets are either (a) bright stars observed
with WIYN+Hydra, or (b) targets chosen to fill unused fibers after all
possible MSTO candidates had been assigned. The final culled sample of
Sgr candidates (based initially on RV selection, with further
interactive proper-motion and CMD selection performed as discussed
below) is shown by the large, filled black squares.  As expected,
these concentrate at the Sgr MSTO locus, with a handful of brighter
stars having properties consistent with Sgr membership as RGB or red
clump stars. We have overlaid an isochrone from \citet{ggo+04} for an
old (10 Gyr), metal-poor ([Fe/H] = -1.3) population at the expected
distance (d = 29.5 kpc; \citealt{lm10a}) of Sgr debris in SA 94; the
age and metallicity of this isochrone is chosen to match the Sgr
metal-poor population identified by \citet{sdm+07}, which should be
the dominant contributor to debris in this portion of the trailing
tail. The final set of Sgr candidates in this field concentrate near
this ridgeline; most of the scatter about the isochrone is likely due
to the $\pm$5 kpc line-of-sight depth of the Sgr stream as well as the
intrinsic population dispersion in Sgr. The proper motion vector point
diagram (VPD) is displayed as Figure~\ref{fig:sa94_vpd}, with panels
and symbols the same as in Figure~\ref{fig:sa94_cmd}. Sgr candidates
clump more tightly in the VPD than the overall population of
spectroscopic targets, which is consistent with the notion that we are
indeed measuring a distant, common-motion grouping of stars.

To further cull the sample, we turn to the reduced proper motion
diagram (RPMD). Reduced proper motion, defined as $H_g \equiv g + 5$
log $\mu + 5$, where $g$ is the apparent magnitude and $\mu$ is the
total proper motion in arcsec yr$^{-1}$, compresses stars with common
tangential velocity into coherent features in the RPMD. Thus, a
common-motion population should form a sequence in the RPMD, even if
the population has a significant line-of-sight depth (see
\citealt{m99}, especially Figure 4). In Figure~\ref{fig:sa94_rpmd} we
show such a diagram for SA 94, with the same isochrone as in
Figure~\ref{fig:sa94_cmd}, shifted to the measured tangential velocity
of SA 94 Sgr debris (to be discussed in
Section~\ref{pm_meas.sec}). After an initial calculation of the mean
motion, candidates that were obviously inconsistent with a
broadly-defined region ($\gtrsim$0.3 magnitudes in $H_g$ and/or $g-r$)
about the ridgeline were manually removed from the sample.

\begin{figure}[!ht]
\includegraphics[width=3.2in,trim=0.15in 0.15in 0in 0.15in,clip]{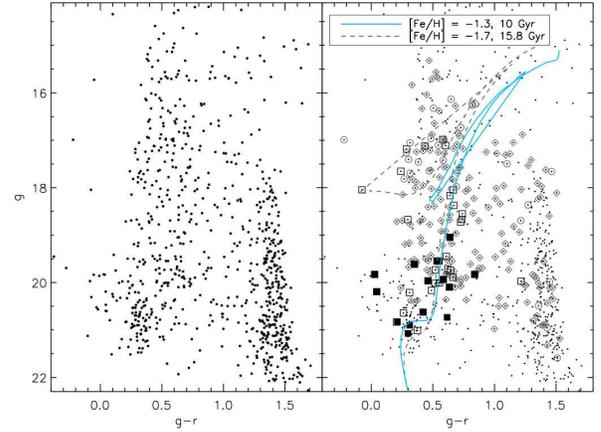}
\caption{SDSS $g$ vs. $g-r$ CMDs of SA 93, with panels and symbols the
same as in Figure~\ref{fig:sa94_cmd}. Because the proper motion
catalog in this field doesn't reach nearly as faint stars as in SA 94,
far fewer Sgr candidates have been identified among our radial
velocities. Note also that for reasons discussed in the text, only
stars with $g > 19.0$ were kept in the final sample. The blue curve is
a \citet{ggo+04} isochrone for a 10 Gyr population at [Fe/H] = -1.3
and a distance of 28 kpc. To illustrate the possible presence of very
old, evolved stars among the Sgr candidates, we also overlay a grey
dashed curve for the same distance, but [Fe/H] = -1.7 and an older
($\sim$15.8 Gyr) population. Note that many of the open squares (RV
candidates not included in the final sample) lie along the asymptotic
giant branch of this old, metal-poor population.} \label{fig:sa93_cmd}
\end{figure}

\begin{figure}[!ht]
\includegraphics[width=3.2in,trim=0.15in 0.15in 0in 0.15in,clip]{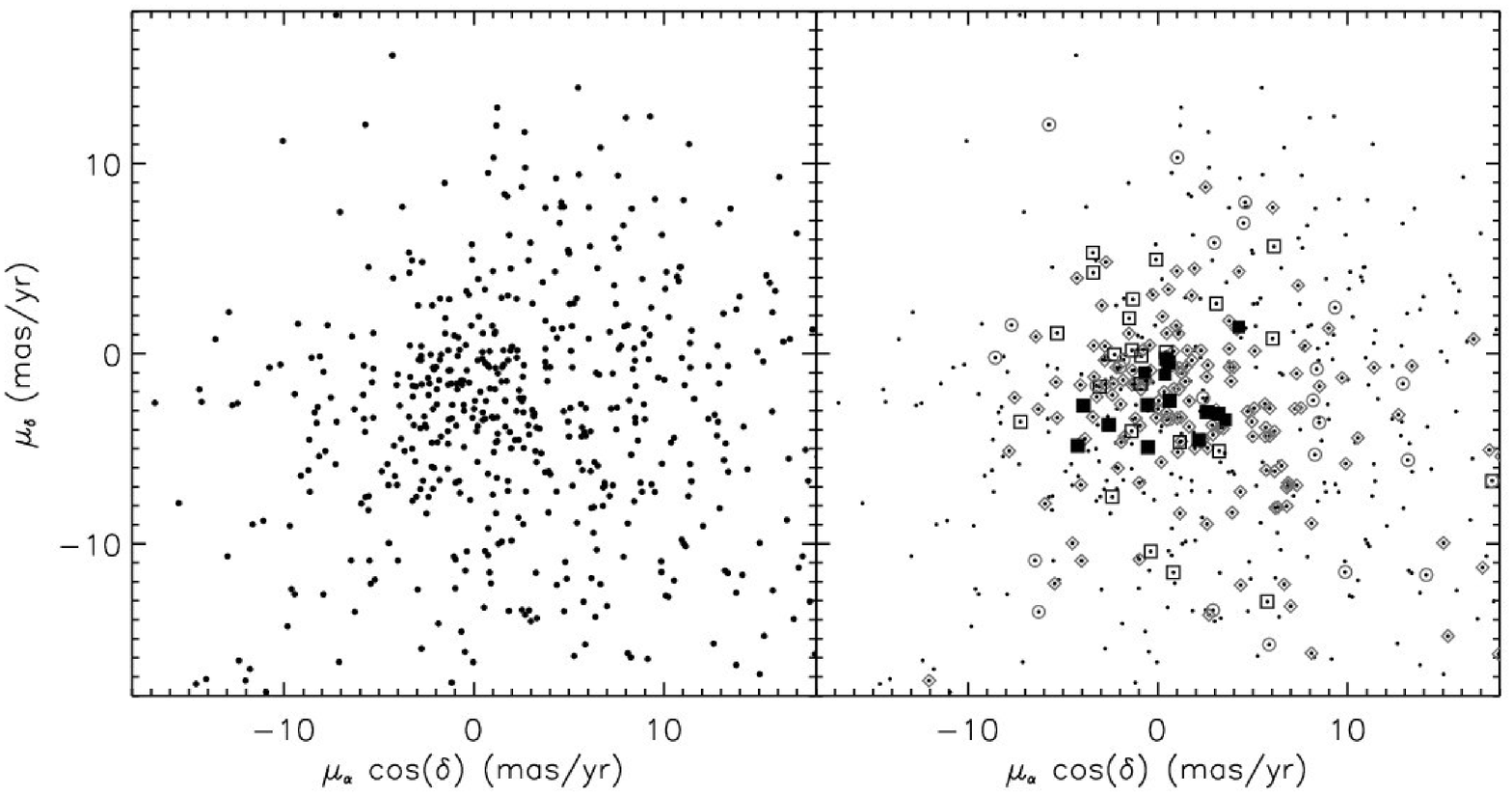}
\caption{Proper motion vector point diagram of SA 93, with panels and
symbols as in Figure~\ref{fig:sa94_cmd}.} \label{fig:sa93_vpd}
\end{figure}

\begin{figure}[!ht]
\includegraphics[width=3.2in,trim=0.15in 0.15in 0in 0.15in,clip]{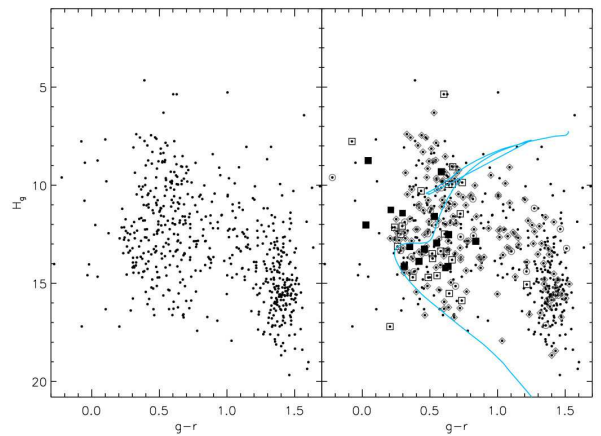}
\caption{Reduced proper motion diagram (RPMD) for SA 93, where $H_g
\equiv g + 5$ log $\mu + 5$, with $\mu$ in arcsec yr$^{-1}$. Panels
and symbols are as in Figure~\ref{fig:sa94_cmd}. The blue curve is a
\citet{ggo+04} isochrone for a 10 Gyr population at [Fe/H] = -1.3 and
a distance of 28 kpc, with the measured mean proper motion of Sgr
debris in SA 93 ($\mu_{\rm tot}$=2.69 mas yr$^{-1}$) used to convert
to reduced proper motion.} \label{fig:sa93_rpmd}
\end{figure}

\subsubsection{Selecting Sgr Candidates in SA 93}

\begin{figure}[!t]
\includegraphics[width=1.65in,trim=0.5in 0 0.2in 0,clip]{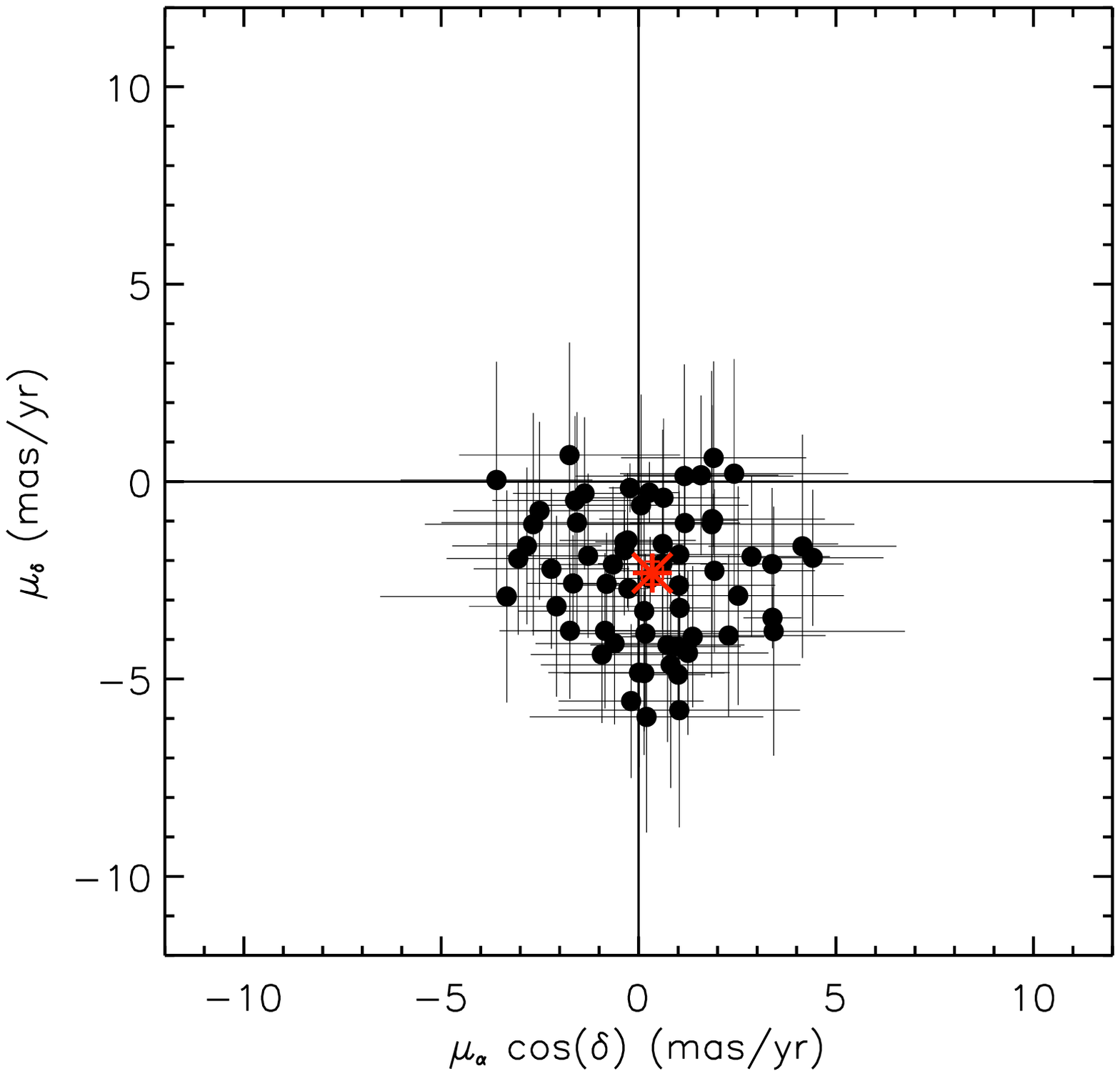}
\includegraphics[width=1.65in,trim=0.5in 0 0.2in 0,clip]{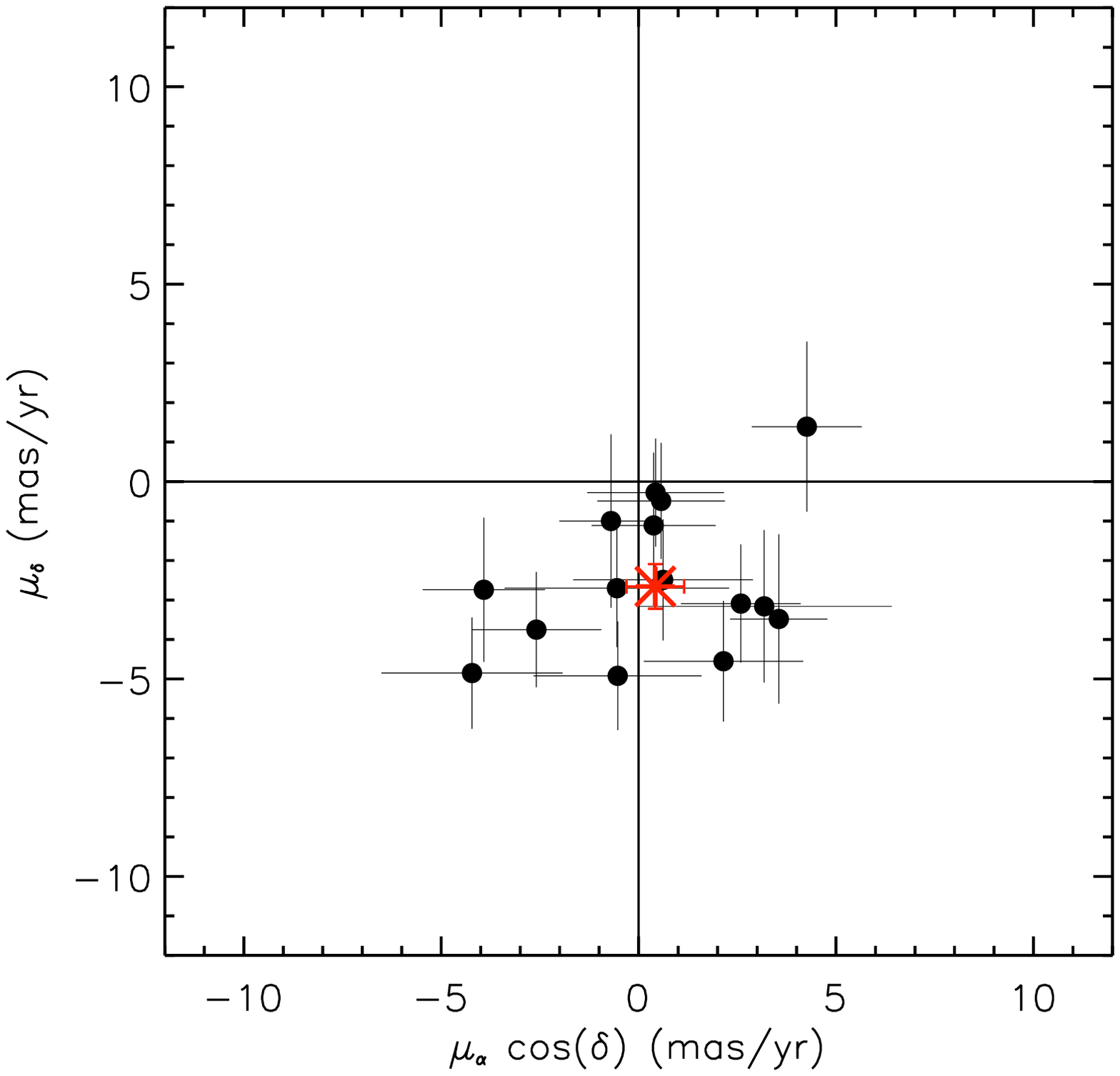}
\caption{Proper motions of only the final Sgr candidates in SAs 94 (left panel) and
93 (right panel), with individual error bars. In SA 93, a proper
motion offset was apparent between faint ($g > 19.0$) candidates and
brighter Sgr candidates; for this reason, only the faint candidates
were retained. The red asterisks represent the final maximum
likelihood estimate of the mean Sgr debris motion in each
field.} \label{fig:sa9493_vpd_members}
\end{figure}

SA 94 is one of only two fields (with SA 71) of the six in this study
that have KPNO 4-m plates, and thus have 1-1.5 magnitude deeper proper
motions.  Furthermore, of those two, SA 94 has higher-quality
photometry (from SDSS) than the photographic magnitudes used for SA
71.  Thus SA 94 is our best field in terms of overall data quality. To
illustrate how a more typical field compares to the highest-quality SA
94 data, we show in Figures~\ref{fig:sa93_cmd}, \ref{fig:sa93_vpd},
and \ref{fig:sa93_rpmd} the same type of plots as in
Figures~\ref{fig:sa94_cmd}, \ref{fig:sa94_vpd}, and
\ref{fig:sa94_rpmd}, but for SA 93. This field is located in a high
stellar density region of the stream, as is SA 94, but has shallower
proper motion data (see Fig.~\ref{fig:gr_cmd_vpd}), providing far
fewer Sgr MSTO candidates for follow-up spectroscopy. Because the
available proper-motion data do not sample the MSTO as robustly as in
SA 94, this field was given lower priority for spectroscopy, with only
one relatively short MMT+Hectospec configuration observed (see
Table~\ref{tab:sgr_obs_tab}). However, in this one Hectospec setup,
nearly all available $g \lesssim 20.2$ Sgr candidates satisfying the
(somewhat relaxed relative to SA 94) target selection criteria were
observed. The overlaid \citet{ggo+04} isochrone for an old (10 Gyr),
metal-poor ([Fe/H] = -1.3) population at the expected distance ($d$ =
28 kpc) of Sgr debris in SA 93 (blue curve in
Figure~\ref{fig:sa93_cmd}) is seemingly consistent with all but the
brightest of the identified candidates. We have also overplotted
(dashed gray curve) an older ($\sim$15.8 Gyr), more metal-poor ([Fe/H]
= -1.7) isochrone with the same distance as the blue curve; this curve
passes through the positions of the bright, blue stars, suggesting
that they may also be Sgr members of an old
horizontal-branch/asymptotic giant branch population. However, the
MSTO of such an old population is beyond the magnitude limit of our
proper-motion survey.\footnote{Note that we are not suggesting that
Sgr has stars 15.8 Gyr old -- this isochrone is simply meant as a
guide to show that these stars could plausibly be BHB stars associated
with Sgr. A blue horizontal branch is only seen in the oldest
(log(age) > 10.15) of the \citealt{ggo+04} isochrones at this
metallicity.} The RPMD (Figure~\ref{fig:sa93_rpmd}) for SA 93 shows
the same ridgeline as in the CMD, shifted by the final measured
tangential velocity for the Sgr candidates. Because the uncertainty in
the mean proper motion is much larger in this field than in SA 94 (on
the order of $\sim$100 km s$^{-1}$ in tangential velocity for SA 93,
compared to $\sim$50 km s$^{-1}$ in SA 94), it is difficult to
conclude much from the RPMD. The benefits of the additional 4-meter
plates in SAs 94 and 71 are clearly illustrated by the relatively
fewer identified Sgr stream members in the shallower SA 93 field
compared to the deeper data sets. Zoomed-in versions of the VPDs for
SAs 94 and 93 are given in Figure~\ref{fig:sa9493_vpd_members}, with
error bars shown on all points to illustrate the quality of the
proper-motion data (note that the error bars on individual stars in
each of the panels of Fig.~\ref{fig:sa9493_vpd_members} are of
comparable size, in spite of the higher-quality proper motion data in
SA 94 than in SA 93, because the majority of Sgr candidates in SA 94
are faint [$g > 21$] MSTO candidates, while SA 93 candidates are
mostly $>1$ magnitude brighter than this). As previously mentioned, a
proper motion offset was seen between faint ($g > 19.0$) stars in SA
93 and brighter candidates; we included only the faint ($g > 19.0$)
stars in our proper motion measurement, because these are more likely
to be true Sgr members. The maximum likelihood estimate of the
absolute proper motion of Sgr debris in each field is represented by
the large red asterisks, which have $1\sigma$ uncertainty smaller than
the size of the point.

\subsubsection{Summary of Sgr Candidate Selection}

In summary, the basic Sgr candidate selection in each Selected Area
began with a broad RV selection centered on the apparent Sgr velocity
peak (e.g., $-250 < V_{\rm hel} < -100$ km s$^{-1}$ in SAs 94 and
93). This was followed by removing high proper motion stars ($|\mu| >
10$ mas yr$^{-1}$ in either dimension), which would have tangential
velocities much greater than the Milky Way escape velocity if those
stars were at the 25-40 kpc distances of Sgr debris along the SA lines
of sight. We then used our knowledge of the distance and metallicity
expected for Sgr debris in these fields to remove stars at faint
magnitudes that are more than $\sim0.3$ magnitudes from the expected
color-magnitude and RPMD loci in each field (note that in some cases
we did not apply this at bright magnitudes, since there is
considerable uncertainty about the exact CMD locus for Sgr debris
above the turnoff). We calculated mean proper motions from the
selected stars, then iteratively removed outliers ($>3\sigma$) in
proper motion. Finally, we manually inspected the remaining candidates
to remove any stars that were $\sim1-2\sigma$ outliers from the
identified Sgr locus in {\it all observables} (i.e., RV, $g-r$ color,
magnitude, {\it and} proper motion).

\begin{figure*}[!t]
\begin{center}
\includegraphics[width=5.5in, trim=0.1in 0.1in 0.1in 0.1in, clip]{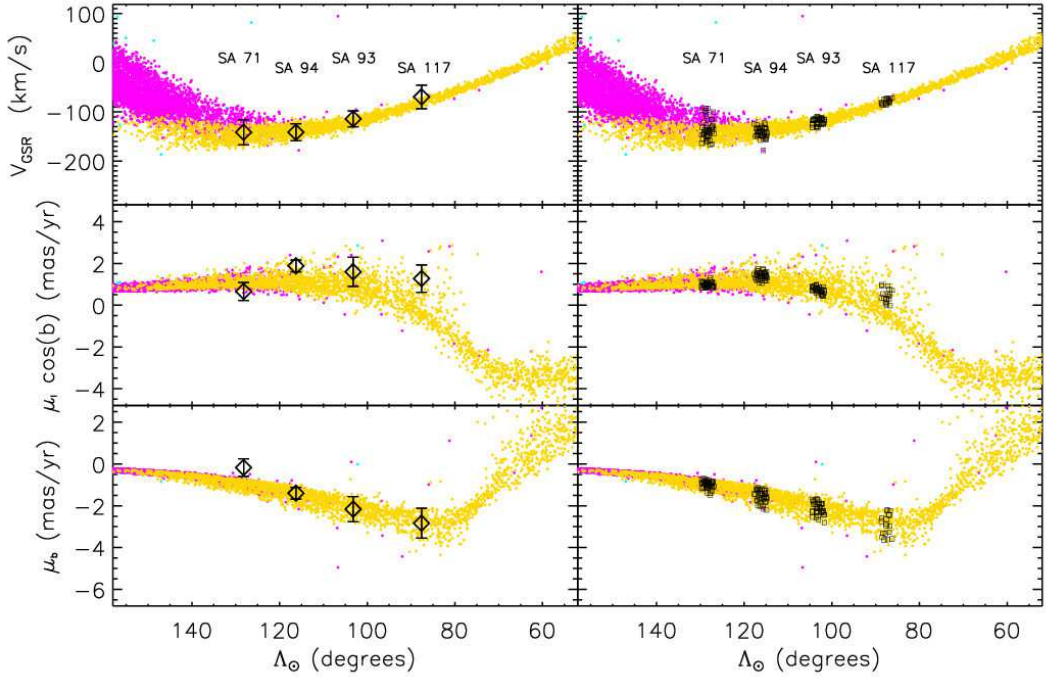}
\caption{Kinematics of Sgr candidate stars measured in four SA fields (SAs 71, 94, 93, and 117, from left to right) as a function of longitude in the Sgr
coordinate system defined by \citet{msw+03}. From top to bottom, the
panels depict GSR-frame radial velocity ($V_{\rm GSR}$), proper motion
along Galactic longitude ($\mu_l \cos$ b), and proper motion along
latitude ($\mu_b$).  Colored points depict trailing tail debris from
the best-fit Sagittarius model of
\citet{lm10a}, with different colors representing debris stripped on
successive orbits, as in \citet{lm10a}.  In each row, the right-hand
panels depict stars selected from the model to be within $\pm3\arcdeg$
of each of the SA fields as small black squares; these are the points
to which we compare the measured kinematics in each field. In the left
panels, large open diamonds with error bars represent the maximum
likelihood estimates of the mean kinematics of Sgr debris.  Note that
it is important to compare measured kinematics (on the left) to the
"clouds" of small black squares in corresponding fields in the right
panels, rather than comparing to the trend defined by all of the
colored points. This is necessary because at a given
$\Lambda_{\odot}$, the model debris plotted here can come from a large
area on the sky. We are interested in comparing only what the model
predicts "should be" seen in each pencil-beam field of view, and thus
we select only model debris in corresponding regions of the sky. The
measured $V_{\rm GSR}$ for Sgr debris matches the model very well in
all four fields, and the $\mu_b$ proper motions agree in three of the
four fields (see the text for discussion of the difficulties in
selecting Sgr debris in SA 71, the $\sim1\sigma$ discrepant point
furthest to the left). In $\mu_l \cos$ b, however, the measurements
for all but one field (again, SA 71) are systematically offset to
higher proper motions than predicted by the model. We show that this
offset can be accounted for by an upward revision of $\Theta_{\rm
LSR}$, the rotation speed at the Solar circle.} \label{fig:lambda_3dkinematics}

\end{center}
\end{figure*}

Once Sgr candidates were selected using color, magnitude, RV, proper
motion, and RPMD criteria, the kinematical properties of Sgr debris in
each SA field were estimated using a maximum likelihood method (e.g.,
\citealt{pm93,hgi+94,kwe+02}). The final measured radial
velocities and velocity dispersions, along with uncertainties in these
values, are given in Table~\ref{tab:sgr_kinematics} in both the
heliocentric and Galactocentric (GSR) frames. The estimates of the
mean radial velocity (in the GSR frame) of Sgr debris can be seen in
the top panels of Figure~\ref{fig:lambda_3dkinematics}, which overlays
our measurements atop the best-fitting debris model of
\citet{lm10a}. The colors in the figure were chosen to match those
used by \citet{lm10a}, who color-coded points along the stream
according to the orbital passage on which they became unbound. Gold
points correspond to debris stripped during the two most recent
perigalactic passages, and magenta during the previous two
passages. This portion of the trailing tail has a well-constrained
radial velocity trend, which was measured by
\citet{mkl+04} and \citet{mbb+07}, and was one of the constraints on
the \citet{lm10a} models. Sgr model trailing tail debris within
$\pm3.0\arcdeg$ of the position of each Kapteyn line of sight is shown
in the right-hand panels of Figure~\ref{fig:lambda_3dkinematics} as
small open squares. This illustrates that not only do our measured
radial velocities agree quite well with the model, but that the
dispersion in each field appears to match the distribution of model
points. However, we caution that the velocity dispersions we find (see
Table~\ref{tab:sgr_kinematics}) are higher than those derived by
\citet{mkl+04} and \citet{mbb+07} for the trailing tail. 
It is unclear whether the dispersion is truly intrinsically higher for
Sgr trailing tail MSTO populations we sampled than for the M giants
previously studied, or whether our derived dispersions are inflated by
the large measurement uncertainties for our RVs, velocity zero-point
offsets between the many data sets we have combined, or Milky Way
foreground/background contamination in our Sgr candidate
samples. While we have endeavored to account for all of these factors,
a robust conclusion likely requires high-resolution spectra of
trailing-tail MSTO stars.

\begin{deluxetable*}{ccccccccccccc}
\setlength{\tabcolsep}{0.03in}
\tablecaption{Sagittarius Stream Kinematics in Kapteyn Selected Areas \label{tab:sgr_kinematics}}
\tablewidth{0pt}
\tablehead{
\colhead{SA} & \colhead{$N$} & \colhead{$V_{\rm helio}$} & \colhead{$V_{\rm GSR}$} & \colhead{$\sigma_0$} & \colhead{$\mu_{\alpha} \cos(\delta$)} & \colhead{$\mu_{\delta}$} & \colhead{$\mu_{l}$ $\cos($b)} & \colhead{$\mu_{b}$} & \colhead{$U_{GC}$} & \colhead{$V_{GC}$} & \colhead{$W_{GC}$} & \colhead{distance} \\
& & \colhead{km s$^{-1}$} & \colhead{km s$^{-1}$} & \colhead{km s$^{-1}$} & \colhead{mas yr$^{-1}$} & \colhead{mas yr$^{-1}$} & \colhead{mas yr$^{-1}$} & \colhead{mas yr$^{-1}$} & \colhead{km s$^{-1}$} & \colhead{km s$^{-1}$} & \colhead{km s$^{-1}$} & \colhead{kpc}
}
\startdata
 71 &   33 &  -172.9$\pm$4.9 &  -141.4$\pm$4.9 &    25.6$\pm$3.5 &    0.39$\pm$0.44 &   -0.55$\pm$0.42 &    0.66$\pm$0.43 &   -0.17$\pm$0.43 &   138.1$\pm$ 47.1 &    80.7$\pm$ 77.1 &    80.1$\pm$ 63.3 &  38.0$\pm$4.5 \\ 
 94 &   64 &  -142.4$\pm$2.3 &  -141.3$\pm$2.3 &    17.4$\pm$1.8 &    0.35$\pm$0.27 &   -2.32$\pm$0.29 &    1.88$\pm$0.28 &   -1.40$\pm$0.28 &   227.9$\pm$ 34.1 &   -49.5$\pm$ 54.0 &   -12.6$\pm$ 30.8 &  29.5$\pm$4.0 \\ 
 93 &   15 &  -157.2$\pm$4.7 &  -114.1$\pm$4.7 &    16.1$\pm$3.3 &    0.42$\pm$0.73 &   -2.66$\pm$0.56 &    1.60$\pm$0.70 &   -2.16$\pm$0.60 &   210.6$\pm$ 81.3 &  -101.6$\pm$ 88.4 &   -10.7$\pm$ 45.4 &  28.0$\pm$3.0 \\ 
  117 &   43 &   -90.0$\pm$3.9 &   -69.3$\pm$3.9 &    24.0$\pm$2.8 &    0.02$\pm$0.64 &   -3.11$\pm$0.72 &    1.28$\pm$0.65 &   -2.84$\pm$0.71 &   229.3$\pm$ 83.7 &   -76.8$\pm$ 94.6 &    11.5$\pm$ 24.9 &  25.0$\pm$4.0 \\ 
\\
 92 &    6 &  -158.9$\pm$6.3 &   -78.4$\pm$6.3 &    10.6$\pm$8.7 &    1.34$\pm$1.07 &   -3.84$\pm$0.66 &    1.46$\pm$1.06 &   -3.80$\pm$0.66 &   145.6$\pm$123.3 &  -296.9$\pm$107.9 &   -84.6$\pm$ 45.6 &  27.5$\pm$2.5 \\ 
116 &   10 &   -75.4$\pm$6.7 &   -22.3$\pm$6.7 &    20.3$\pm$4.9 &    1.12$\pm$1.04 &   -3.30$\pm$1.15 &   -0.67$\pm$1.07 &   -3.42$\pm$1.12 &    87.3$\pm$124.3 &  -170.8$\pm$129.6 &   -22.7$\pm$ 35.3 &  24.5$\pm$2.0 \\ 
 \enddata
\tablecomments{All calculations assume $V_{\rm circ}$ = 220 km s$^{-1}$ at $R_{0}$ = 8.0 kpc. We used the solar peculiar motion of \citet{mb81}: ($U_0$, $V_0$, $W_0$) = (9.0, 12.0, 7.0) km s$^{-1}$ (in a right-handed frame).}

\end{deluxetable*}

\subsection{Proper Motions}
\label{pm_meas.sec}

With the sample of Sgr debris candidates identified in each field, we
used a maximum-likelihood method to estimate the Sgr absolute proper
motions in both spatial directions ($\mu_{\alpha} \cos (\delta$),
$\mu_{\delta}$). These results are seen in
Table~\ref{tab:sgr_kinematics} with their uncertainties, which include
the uncertainty in the proper motion zero point in each field added in
quadrature to the maximum likelihood error estimate. The uncertainties
for each field depend on many factors, including the depth and quality
of the plates, the number of available reference objects (i.e.,
background QSOs and galaxies) used to convert from relative to
absolute proper motions, the number and depth of spectroscopic targets
obtained (and thus the number of Sgr candidates identified), and the
position of each field relative to the highest-density regions of the
stream (i.e., the number of Sgr candidates expected in each field).

Mean proper motions along Galactic coordinates (i.e., $\mu_l \cos b,
\mu_b$) in each field are compared to the model of \citet{lm10a} as a
function of $\Lambda_{\odot}$ in the middle and lower panels of
Figure~\ref{fig:lambda_3dkinematics}, with points once again
color-coded by the orbital passage in which they became unbound, and
model points within $\pm3.0\arcdeg$ of each SA line of sight
highlighted for guidance. The results for SAs 94, 93, and 117 agree
nicely with the model predictions within the uncertainties in $\mu_b$,
but show a $\sim1-2\sigma$ offset from the main trend in $\mu_l \cos
(b$) (left middle panel).  The mean proper motions along both
directions for SA 71 (the leftmost data points in
Figure~\ref{fig:lambda_3dkinematics}) are slightly shifted (by $\sim
1-1.5\sigma$) from the mean of the model prediction for this field.  A
number of factors contribute to the difficulty in selecting a "pure"
sample of Sgr debris in SA 71, and there is thus an additional
uncertainty (besides the formal errors) in the mean Sgr debris proper
motions in this field. First, this field is at somewhat low latitude
($b=-34.7\arcdeg$), and thus suffers greater contamination from
Galactic populations, as evidenced by the extended tail of
low-velocity stars in the lower panel of
Figure~\ref{fig:sa7194_vhelhist}. This lower latitude also means that
SA 71 suffers significantly more reddening than higher-latitude fields
-- nearly 0.6 magnitudes of extinction using the \citet{sfd98}
maps. Secondly, the distance of Sgr debris increases with
$\Lambda_{\odot}$ along the portion of the trailing tail in this
study, so that stream members in SA 71 are nearly 40 kpc away, making
them fainter than in the other fields. Finally, the poor-quality
photographic photometry we are limited to in this field renders
inscrutable the typical CMD features such as the blue edge of the
Galactic disk MSTO and the Sgr upper main sequence.

Assuming distances to Sgr debris in each field as given in
Table~\ref{tab:sgr_kinematics} and values of $V_{\rm circ}$ = 220 km
s$^{-1}$ and $R_0$ = 8.0 kpc, we converted the measured Sgr debris
motions to Galactocentric $UVW_{\rm GC}$ velocities (i.e., Cartesian
velocities such that the Sun is moving at 9.0, 232.0, 7.0 km s$^{-1}$
assuming $V_{\rm circ}$ = 220 km s$^{-1}$ and ($U_0$, $V_0$, $W_0$) =
(9.0, 12.0, 7.0) km s$^{-1}$ \citep{mb81} for the solar motion
relative to the Local Standard of Rest; to facilitate direct
comparison, values used for these constants are the same as those in
the \citetalias{lm10a} model), which are shown in
Table~\ref{tab:sgr_kinematics} (note that we placed SAs 92 and 116 --
the two fields with no securely identified Sgr members -- in a
separate section in Table~\ref{tab:sgr_kinematics}. These data are
given for completeness, but are not used for subsequent
analysis.). The $U_{\rm GC}$ component should dominate the total space
velocity of Sgr debris in each of these trailing arm fields. This can
be surmised from Figure~\ref{fig:sgr_xyz}, which shows that the motion
of the Sgr trailing tail is oriented almost parallel to the Galactic
$X$-axis in the $X_{\rm GC}-Z_{\rm GC}$ plane. This, in addition to
the fact that the Sagittarius orbital plane is only slightly
misaligned with the Galactic $X_{\rm GC}-Z_{\rm GC}$ plane
\citep{msw+03,mlp+06}, suggests that most of the motion in this part
of the trailing tail is inward toward the Galactic center and roughly
parallel to the Galactic disk (i.e., the $X_{\rm GC}-Y_{\rm GC}$
plane) at a distance of $\sim$20-25 kpc below the disk.  This is borne
out by the measured $UVW_{\rm GC}$ velocities in our SA fields -- the
$U_{\rm GC}$ component in the four fields with quality measurements is
by far the largest component of the 3-D motion.  This can be seen even
more clearly by considering the proper motions along Galactic
coordinates, but in a Galactic rest frame (designated as $\mu'_{l}
\cos($b) and $\mu'_{b}$). These proper motions, given in
Table~\ref{tab:sgr_pms_gsr}, show $\mu'_{l}
\cos($b) proper motions of nearly zero in each field -- as expected
for streaming motions confined to the $X_{\rm GC}-Z_{\rm GC}$-plane.
As we will show in Section~4, the offset of these derived longitudinal
proper motions (reflected in the $V_{\rm GC}$ Galactic velocity
component) from zero can be used to reevaluate the velocity of the
Local Standard of Rest (under the assumption that the longitudinal
motions {\it should be} zero).

\begin{table}[colspan=2,!t]
\begin{center}
\caption{Galactic Frame-of-Rest Proper Motions of Sagittarius Debris in Selected Areas \label{tab:sgr_pms_gsr}}
\begin{tabular}{cccccc}
\\
\tableline
\\
\multicolumn{1}{c}{SA} & \multicolumn{1}{c}{$V_{\rm GSR}$} & \multicolumn{1}{c}{$\mu'_{\alpha} \cos(\delta$)} & \multicolumn{1}{c}{$\mu'_{\delta}$} & \multicolumn{1}{c}{$\mu'_{l} \cos($b)} & \multicolumn{1}{c}{$\mu'_{b}$} \\
 & \multicolumn{1}{c}{(km s$^{-1}$)} & \multicolumn{1}{c}{(mas yr$^{-1}$)} & \multicolumn{1}{c}{(mas yr$^{-1}$)} & \multicolumn{1}{c}{(mas yr$^{-1}$)} & \multicolumn{1}{c}{(mas yr$^{-1}$)} \\
\tableline
\\
  71 & -141.4 & -0.47 & 0.39 & -0.61 &  0.00 \\
 94 & -141.3 & -0.77 & -1.09 & 0.22 & -1.31 \\
 93 & -114.1 & -0.71 & -1.36 & 0.00 & -1.54 \\
117 & -69.3 & -1.17 & -1.57 & -0.44 & -1.91 \\
\tableline
\end{tabular}
\end{center}
\tablenotetext{1}{All calculations assume $V_{circ}$ = 220 km s$^{-1}$ at $R_{0}$ = 8.0 kpc, and solar peculiar motion of ($U_0$, $V_0$, $W_0$) = (9.0, 12.0, 7.0) km s$^{-1}$ (in a right-handed frame).}
\end{table}

\section{Constraints on Milky Way Stucture}
\label{mw_constraints.sec}

As discussed in Section~1.1, the opportune orientation of the Sgr
trailing tidal tail means that the observed motion of tidal stream
stars in the Galactic $Y$ direction (i.e., towards $[l,b] =
[90^{\circ}, 0^{\circ}]$) is dominated by the solar reflex motion,
which consists of the solar peculiar motion and the Galactic
rotational motion at the solar circle (i.e., the Local Standard of
Rest $\Theta_{\rm LSR}$).  As shown by \citet{mlp+06}, the intrinsic
motion of Sgr debris along the $Y$ direction ($V_{\rm GC}$, contained
primarily in the $\mu_l \cos($b) component of proper motion) varies
only slowly across the region of the trailing tail between $70\arcdeg
\le \Lambda_{\odot} \le 130\arcdeg$, making the fields of view in
which we have deep proper motion data ideal for constraining
$\Theta_{\rm LSR}$. It can be seen in Table~\ref{tab:sgr_kinematics}
that $V_{\rm GC}$ for Sgr debris in each of the four fields (SAs 71,
94, 93, and 117) with reliable data is non-zero at the $\sim1\sigma$
level. Setting aside SA 71, in which it is difficult to securely
identify Sgr debris, the remaining three fields exhibit $V_{\rm GC}$
systematically offset to negative values. If indeed the expected
$V_{\rm GC}$ for Sgr debris in these fields is zero, this suggests
that the value of $\Theta_{\rm LSR}$ that was subtracted from the
$V$-component of these velocities was {\it lower} than it should be --
i.e., $\Theta_{\rm LSR}$ should be greater than the canonical 220 km
s$^{-1}$.

In this section, we use variations on the \citetalias{lm10a} numerical
model of the Sgr tidal stream to isolate the contribution to $\mu_l
\cos($b) from the solar reflex motion and identify the value of
$\Theta_{\rm LSR}$ favored by our proper motion data.

\subsection{$N$-body models}
\label{nbody.sec}

Though the measured Galactic Cartesian $V-$velocity (i.e., motion
along the Galactic $Y$-component) of Sgr trailing tidal debris is
dominated by Solar reflex motion, there is some contribution of
intrinsic Sgr motion to the $V-$component of debris velocities. In
particular, we must consider the following effects when trying to back
out $\Theta_{\rm LSR}$ from measured $V_{\rm GC}$ for Sgr debris: (1)
the slight inclination of the Sgr debris plane to the Galactic
$XZ_{\rm GC}$ plane means a small fraction of Sgr space motion is
projected onto the measured motions (i.e., the $V$ velocity is in fact
a function of both $\Theta_{\rm LSR}$ {\it and} intrinsic Sgr motion);
(2) the Galactic Standard of Rest (GSR) frame radial velocities used
to constrain the Sgr model were derived assuming a value of
$\Theta_{\rm LSR}$; (3) changing $\Theta_{\rm LSR}$ correspondingly
changes the Milky Way mass scale, which thus affects the space
velocity of the Sgr dSph in the models. Therefore, taking into account
these dependencies on the assumed value of $\Theta_{\rm LSR}$, we
repeat the \citetalias{lm10a} analysis, changing $\Theta_{\rm LSR}$ to
construct self-consistent models for the Sgr tidal stream in each of
four choices for the Local Standard of Rest speed, namely $\Theta_{\rm
LSR} = 190, 250, 280, 310$ km s$^{-1}$ (in addition to the original
\citetalias{lm10a} value of $\Theta_{\rm LSR} = 220$ km s$^{-1}$ used
in earlier sections of this paper).

Our methodology is described in detail by \citetalias{lm10a}.  In
brief, we constrain the model Sgr dwarf to lie at the observed
location $(l,b)=(5.6^{\circ}, -14.2^{\circ})$, distance $D_{\rm Sgr} =
28$ kpc (\citealt{siegel2011}, \citealt{sdm+07}), and radial velocity
$v_{\rm Sgr} = 142.1$ km s$^{-1}$ in the heliocentric frame.  The
orbital plane is constrained to be that defined by the trailing arm
tidal debris, which has experienced minimal angular precession
\citep{jlm05}, and the speed of Sgr tangential to the line of sight
($v_{\rm tan}$; see Section 3.3 of \citetalias{lm10a}) is constrained
by a $\chi^2$ minimization fit to the radial velocities of trailing
arm tidal debris.  We fix the mass and radial scalelength of the Sgr
progenitor so that the fractional mass loss history of the dwarf is
similar in all models to that of \citetalias{lm10a}.

The adopted Milky Way Galactic mass model consists of three
components: a Hernquist spheroid (representing the Galactic bulge), a
\citet{mn75} disk, and a logarithmic dark matter halo.  The Local
Standard of Rest in this model is given by: \begin{equation}
\Theta_{\rm LSR} =
\sqrt{R_{\odot} (a_{\rm bulge} + a_{\rm disk} + a_{\rm
halo})}\end{equation} where $a_{\rm bulge}$, $a_{\rm disk}$, and
$a_{\rm halo}$ respectively represent the gravitational acceleration
exerted on a unit-mass at the location of the Sun due to the Galactic
bulge, disk, and halo components.  In the \citetalias{lm10a} model
(for which $\Theta_{\rm LSR} = 220$ km s$^{-1}$), the bulge/disk/halo
respectively contribute 32\%/49\%/19\% of the total centripetal
acceleration at the position of the Sun, corresponding to bulge/disk
masses $M_{\rm bulge} = 3.4 \times 10^{10} M_{\odot}$ and $M_{\rm
disk} = 1.0 \times 10^{11} M_{\odot}$, and a total mass within 50 kpc
of $4.5 \times 10^{11} M_{\odot}$.

\begin{table}
\begin{center}
\caption{Masses of Galactic Bulge and Disk Components in the Sagittarius Models \label{tab:bulgediskmasses}}
\begin{tabular}{cccc}
\\
\tableline
\\
\multicolumn{1}{c}{$\Theta_{\rm LSR}$} & \multicolumn{1}{c}{$M_{\rm disk}$} & \multicolumn{1}{c}{$M_{\rm bulge}$} & \multicolumn{1}{c}{$\alpha$} \\
\multicolumn{1}{c}{(km s$^{-1}$)} & \multicolumn{1}{c}{($M_{\odot}$)} & \multicolumn{1}{c}{($M_{\odot}$)} &  \multicolumn{1}{c}{-} \\
\tableline
\\
190 & $6.8\times10^{10}$ & $2.3\times10^{10}$ & 0.68 \\
220 & $1.0\times10^{11}$ & $3.4\times10^{10}$ & 1.00 \\
250 & $1.4\times10^{11}$ & $4.6\times10^{10}$ & 1.35 \\
280 & $1.8\times10^{11}$ & $6.0\times10^{10}$ & 1.76 \\
310 & $2.2\times10^{11}$ & $7.5\times10^{10}$ & 2.21 \\
\tableline
264 & $1.5\times10^{11}$ & $5.2\times10^{10}$ & 1.53 \\
232 & $1.1\times10^{11}$ & $3.9\times10^{10}$ & 1.14 \\
\tableline
\end{tabular}
\end{center}
\tablenotetext{1}{The mass of the disk and bulge components in each of the models of the Sgr stream. Each model is specified by the value of $\Theta_{\rm LSR}$ that constrained the fit; the 220 km s$^{-1}$ model is that of \citetalias{lm10a}. The ratio of disk to bulge mass is constant throughout -- the constant $\alpha$ is the scaling factor, such that $M_{\rm disk} = \alpha*M_{\rm disk, 220 km/s}$ and $M_{\rm bulge} = \alpha*M_{\rm bulge, 220 km/s}$. The total mass, axis ratios, and scale length of the Galactic dark matter halo were fixed to the best-fit values of \citetalias{lm10a}.}
\end{table}

Since the baryonic Galactic disk and bulge components are the dominant
factors in determining $\Theta_{\rm LSR}$ (together comprising $>
80$\% of the total centripetal force), we therefore scale the total
bulge $+$ disk mass as necessary to normalize the rotation curve at
the solar circle ($R_{\odot} = 8$ kpc) to the chosen value of
$\Theta_{\rm LSR}$.  The masses of the disk and bulge components in
each of the models are given in Table~\ref{tab:bulgediskmasses}. We
leave the {\it ratio} of bulge/disk mass fixed in order to preserve
the shape of the rotation curve interior to the solar circle.  In
addition, we fix the Galactic dark matter halo parameters (mass, axis
ratios, and scalelength) to the best-fit values derived by
\citetalias{lm10a} since these authors found that these values were
relatively insensitive to factors of $\sim 2$ variation in the mass
scale of the baryonic Galactic components.

\subsection{Results}
\label{nbodyresults.sec}

Constraints on $\Theta_{\rm LSR}$ were derived in two ways. In the
first of these methods, we assumed (as argued previously in this
paper, as well as in \citetalias{mlp+06}) that the dominant
contribution to the measured $\mu_l \cos($b) component of Sgr motion
is due to the solar rotation, and that the largest component of
$\Theta_{\rm LSR}$ is along $\mu_l \cos($b). We have shown that these
are reasonable first-order assumptions, and thus use only the
longitudinal proper motions as constraints on fitting $\Theta_{\rm
LSR}$ in our first attempt.  After doing so, however, we performed a
similar analysis, but using all three dimensions of Sgr debris motions
as constraints to determine $\Theta_{\rm LSR}$.  In the following, we
present both results, which come out somewhat different from each
other (though consistent within 1$\sigma$).  Fits using only $\mu_l
\cos($b) tend to prefer relatively high values of $\Theta_{\rm LSR}$,
while those constrained by full 3-D kinematics tend toward lower
values more in line with the IAU standard of 220 km s$^{-1}$.

\subsubsection{$\Theta_{\rm LSR}$ Constraints Using Only $\mu_l \cos($b) Motions of Sgr Debris}

\begin{figure*}
\begin{center}
\includegraphics[height=6.0in]{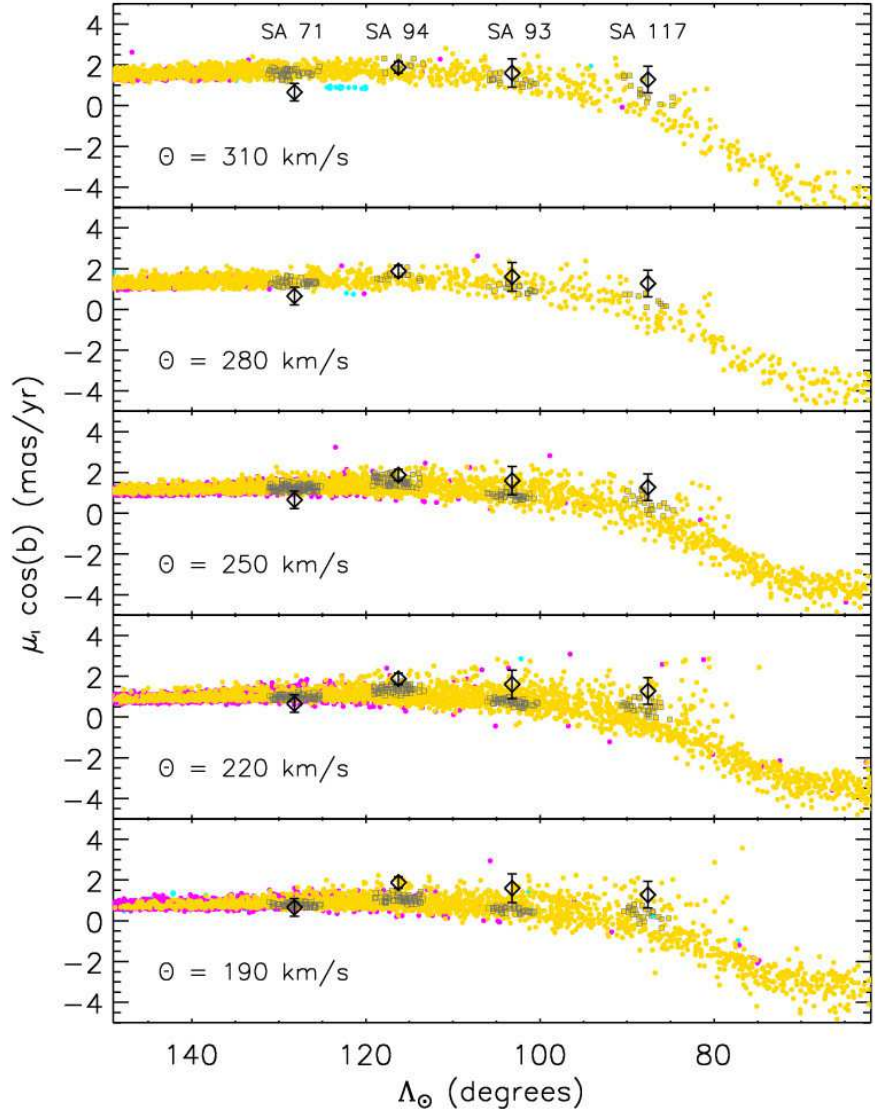}
\caption{Mean longitudinal proper motions, $\mu_l \cos($b) (large open diamonds with error bars),  in (from left to right) SAs 71, 94, 93, and 117 as a function of $\Lambda_{\odot}$. Model debris from the Sgr simulations are shown as the colored points; from top to bottom, these represent models with $\Theta_{\rm LSR}$ = 310, 280, 250, 220, and 190 km s$^{-1}$. The sudden drop in $\mu_l \cos($b) for $\Lambda_{\odot} \lesssim 90\arcdeg$ is due to the inversion in sign that occurs as the debris sweeps past the South Galactic pole. Small open gray squares denote the model debris corresponding (within $\pm3\arcdeg$ in RA, Dec, and $\Lambda_{\odot}$) to each SA field. It is clear that higher values of $\Theta_{\rm LSR}$ provide a better match of the small grey squares to the observed $\mu_l \cos($b) values in the Selected Areas (diamonds).} \label{fig:mul_varyvcirc}
\end{center}
\end{figure*}

We quantify the agreement of our proper motions with those of
simulated Sgr tidal debris from each of our grid of models using a
$\chi^2$ statistic. The $\chi^2$ fitting was performed using mean Sgr
debris proper motions in only SAs 71, 94, 93, and 117 -- as discussed
in Section~\ref{sgr_rvs.sec}, the results in SAs 92 and 116 are
unreliable for a variety of reasons. For our model comparison, we
first select all \citetalias{lm10a} Sgr model points within
$\pm3.0\arcdeg$ in both $\Lambda_{\odot}$ and $(\alpha,\delta)$ of
each SA field. The large area (relative to the $40\arcmin \times
40\arcmin$ coverage of each SA field) used to select model debris
corresponding to each SA position ensures that enough $N-$body
particles are selected for robust measurement of model debris motions
at each position. This also makes the fitting less sensitive to
small-scale differences in positions and densities of debris stars
between the models and the actual stream that arise due to the
vagaries of the modeling and our incomplete knowledge of the Sgr
trailing tail properties. Figure~\ref{fig:mul_varyvcirc} shows the
model debris $\mu_l \cos($b) as a function of $\Lambda_{\odot}$ for
each of the five Sgr simulations, with points corresponding to each SA
field shown as small open gray squares. It is clear that $\mu_l
\cos($b) changes very little over the $6\arcdeg$ ranges in
$(\alpha,\delta)$ used. Furthermore, the small number of selected
model points, even in such a large selection region, shows that these
broad selection criteria are necessary to have sufficient model points
for comparison. The maximum likelihood proper motion results in SAs
71, 94, 93, and 117 are shown in Figure~\ref{fig:mul_varyvcirc} as
open black diamonds, with error bars reflecting $1\sigma$
uncertainties. It can be seen in the figure that the models with
$\Theta_{\rm LSR} > 220$ km s$^{-1}$ tend to reproduce the
longitudinal proper motions for most of the fields better than the
standard 220 km s$^{-1}$ value of this fundamental constant. Note that
SA 71, at $\Lambda_{\odot} = 128\arcdeg$, is the exception to this
trend -- as discussed in Section~3.3, identifying bona fide Sgr debris
in this field is more difficult than the others and so these data are
more suspect.

\begin{figure}
\begin{center}
\includegraphics[width=2.4in,trim=0.15in 0.15in 0in 0.15in,clip]{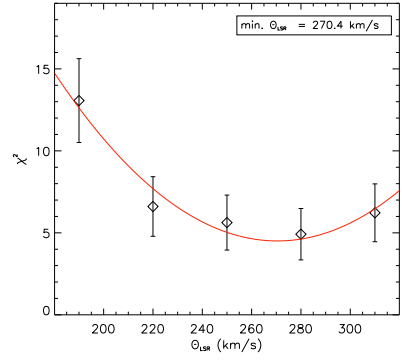}
\caption{Total $\chi^2$ residuals for the final mean proper motions relative to corresponding model debris. Each point represents a $\chi^2$ for one of the five models in which we vary $\Theta_{\rm LSR}$ from 190-310 km s$^{-1}$. A parabola fit to the results (the red curve) yields a minimum $\chi^2$ at $\Theta_{\rm LSR} = 270.4$ km s$^{-1}$.} \label{fig:chisq_vs_theta}

\end{center}
\end{figure}

Each of the Sgr model simulations provides predicted kinematics of Sgr
debris for a Galactic potential constrained by a given value of
$\Theta_{\rm LSR}$. As discussed previously, the Galactic
$V$-component of the Sgr debris space velocity along the trailing tail
contains little contribution due to the Sgr motion; nearly all the $V$
velocity (as measured by the $\mu_l \cos($b) component of the proper
motion) is reflected Solar motion. Thus, to first order, we can simply
compare our mean Sgr proper motions along Galactic longitude in the
trailing tidal tail to our models of Sgr debris for different values
of $\Theta_{\rm LSR}$ and determine the value of $\Theta_{\rm LSR}$
that best reproduces the measured PMs. To do so, we defined a $\chi^2$
residual:

\begin{equation}
\chi^2_\mu = \sum_i \frac{\mu_{l,\rm SA} [\rm i] - \mu_{l,\rm mod} [\rm i]}{\sigma_{\mu_{l,\rm SA}[\rm i]}}
\end{equation}

\noindent where $\mu_{l,\rm SA} [\rm i]$ represents the mean $\mu_l \cos($b)
proper motion in each of the four SA fields, and $\mu_{l,\rm mod} [\rm
i]$ the mean proper motion of the corresponding model debris for each
field. The residuals are weighted by the uncertainty,
$\sigma_{\mu_l,\rm SA [\rm i]}$, in each SA proper motion. This
$\chi^2$ statistic was initially calculated for the proper motions
presented in Table~\ref{tab:sgr_kinematics} relative to each of the
five Sgr debris models (corresponding to $\Theta_{\rm LSR} = 190, 220,
250, 280, 310$ km s$^{-1}$). Rather than running new (and laborious)
$N$-body simulations for many intermediate values of $\Theta_{\rm
LSR}$ and calculating $\chi^2$ for each of them, we choose to find the
minimum $\chi^2$ by fitting a parabola to the $\chi^2$ results for
each of the five modeled values of $\Theta_{\rm LSR}$. The results of
the $\chi^2$ calculation and the parabolic fit are seen in
Figure~\ref{fig:chisq_vs_theta} for the proper motions of the ``best''
SA samples given in Table~\ref{tab:sgr_kinematics}. The minimum of the
parabola yields $\Theta_{\rm LSR,min} = 270.4$ km s$^{-1}$.

To estimate the uncertainty in $\Theta_{\rm LSR}$, we choose a
bootstrap (resampling with replacement) method (see \citealt{a10} and
references therein). This technique uses the entire sample of
individual Sgr candidate star proper motions, and thus yields an
estimate of the errors in $\Theta_{\rm LSR}$ including the effects of
proper motion measurement errors and "contamination" of the
proper-motion samples by Milky Way stars.  Our selected samples of Sgr
candidates in SAs 71, 94, 93, and 117 should contain mostly Sgr
debris, plus some amount of contamination by MW stars that will vary
depending on the depth of the proper-motion and radial-velocity
catalogs, the local Sgr stream density, and the Galactic latitude of
each field. From the Sgr samples in each field, we performed 100,000
bootstrap resamplings, wherein $N$ random selections were made from
the $N$ original stars in each field (i.e., the catalogs of candidates
were resampled with replacement).  Iteratively $3\sigma$-clipped mean
proper motions of these resampled Sgr candidates were measured, and
the mean proper motions used in an identical $\chi^2$ fitting routine
to that described above. Assuming that the contaminants in each sample
are somewhat uniformly distributed in their kinematical quantities,
this method should yield a statistically robust result for
$\Theta_{\rm LSR}$ and its uncertainty (due to both the intrinsic
measurement errors and the MW contamination). The best-fitting values
of $\Theta_{\rm LSR}$ for these 100,000 samples are given as a
histogram in the left panel of Figure~\ref{fig:theta_hist}, along with
a Gaussian fit (red curve) to the results. From this Gaussian, we
derive a final value of $\Theta_{\rm LSR} = 264 \pm 23$ km s$^{-1}$.

\subsubsection{$\Theta_{\rm LSR}$ Constraints Using Three-Dimensional Motions of Sgr Debris}

\begin{figure}
\includegraphics[height=1.35in,trim=0.15in 0.15in 0in 0.15in,clip]{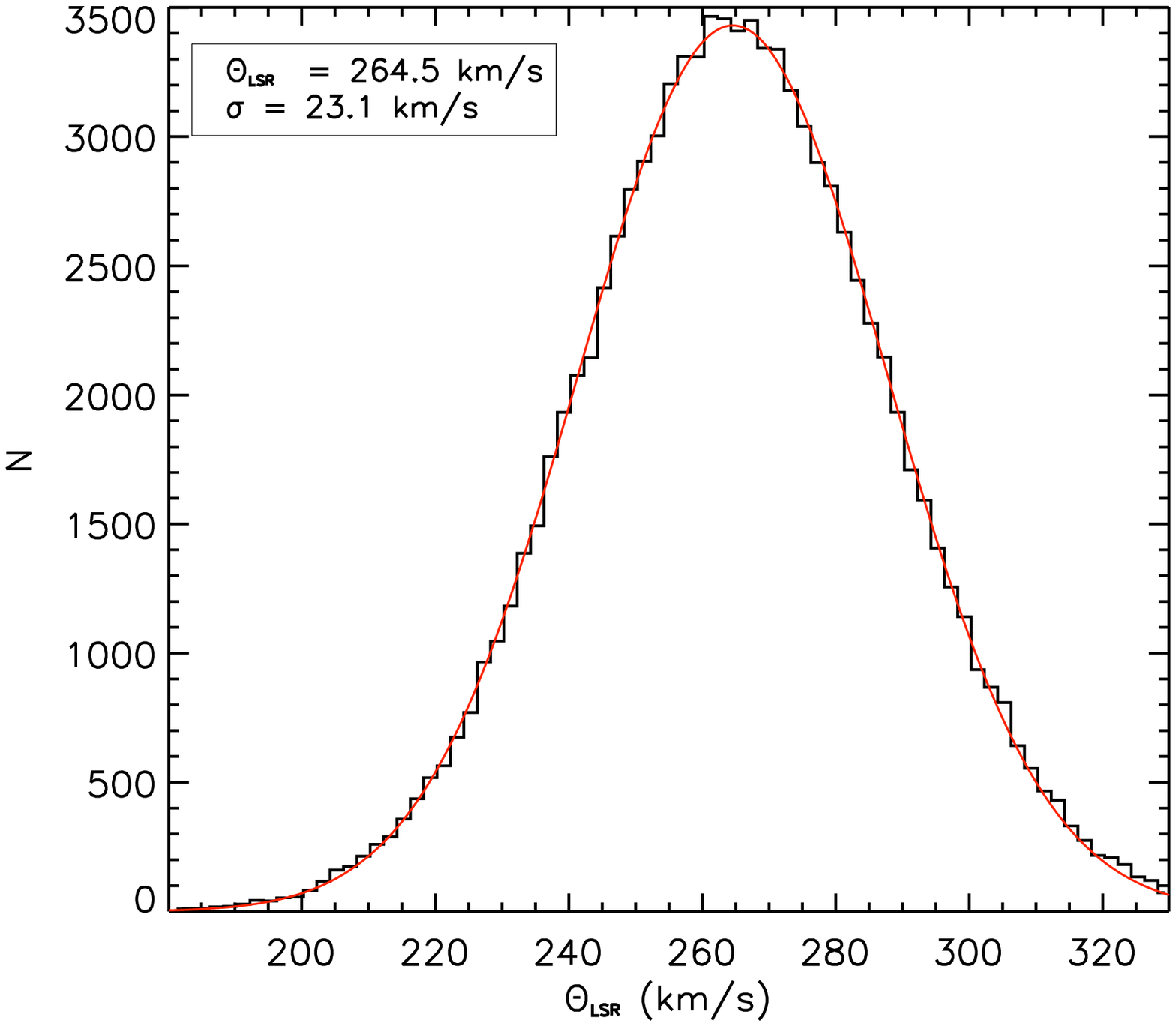}
\includegraphics[height=1.35in,trim=0.15in 0.15in 0in 0.15in,clip]{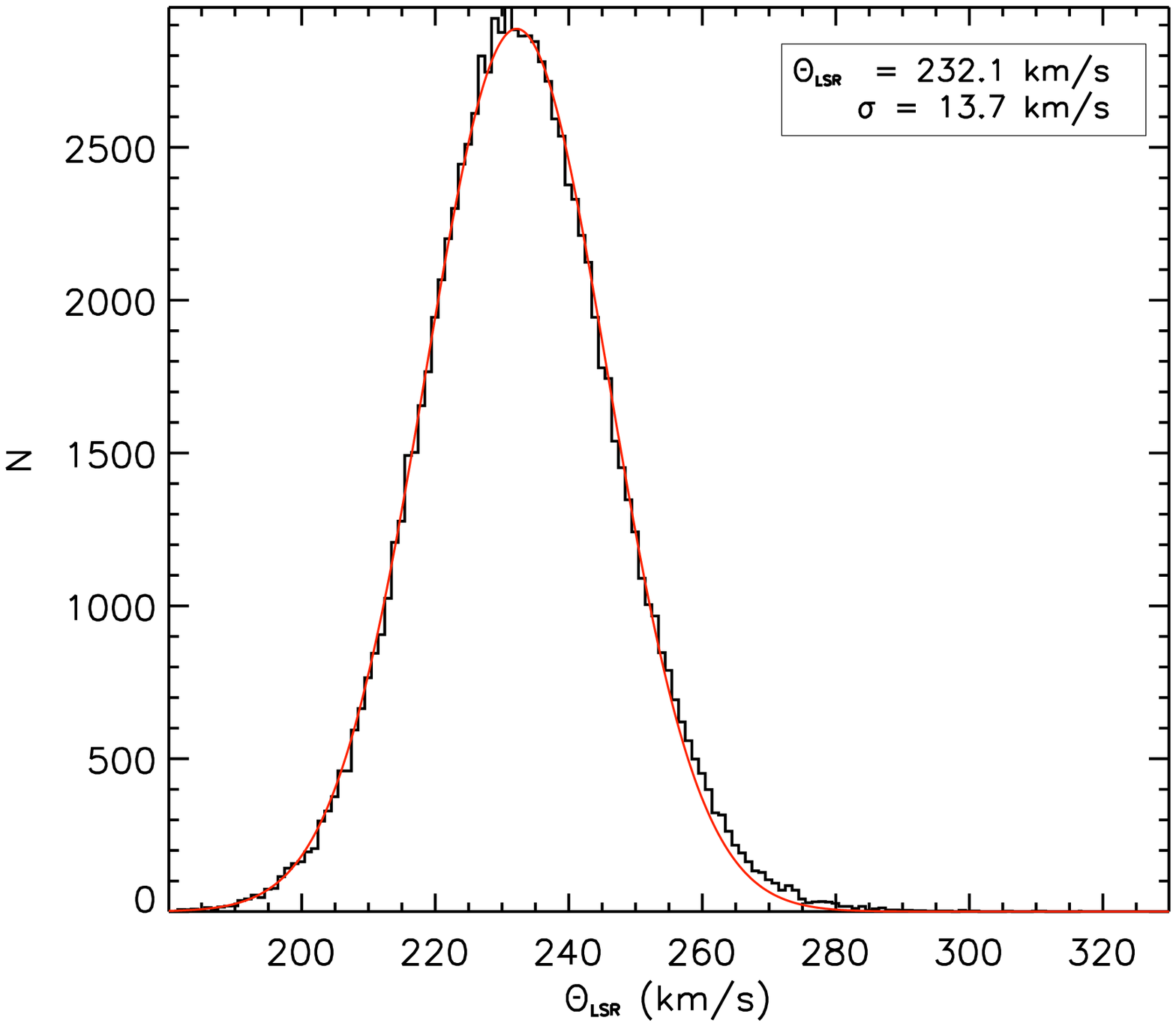}
\caption{Resulting $\Theta_{\rm LSR}$ corresponding to the minimum $\chi^2$ of 100,000 bootstrap resamplings of the individual stellar proper motions in each SA field. The left panel shows fits using only the $\mu_l \cos($b) component of Sgr debris motions in SAs 71, 94, 93, and 117. A Gaussian fit (overlaid as the red curve) to the results yields $\Theta_{\rm LSR} = 264 \pm 23$ km s$^{-1}$. The right panel shows the results using all three dimensions of Sgr debris motions (i.e., $\mu_l \cos($b), $\mu_b$, and $V_{\rm GSR}$) as constraints, which yields $\Theta_{\rm LSR} = 232 \pm 14$ km s$^{-1}$.} \label{fig:theta_hist}
\end{figure}

The constraints we derived on $\Theta_{\rm LSR}$ using only the
longitudinal proper motions assume that the contributions of
$\Theta_{\rm LSR}$ to $\mu_b$ and $V_{\rm GSR}$ are negligible. If the
Sgr orbital plane was exactly coincident with the Galactic $XZ_{\rm
GC}$ plane, then indeed the rotation velocity at the solar circle
would only be reflected in the $\mu_l \cos($b) motions. In reality,
the Sgr orbital plane is {\it not} perfectly aligned with the Milky
Way $XZ$-plane, so there is some projection of $V_{\rm circ}$ onto
$\mu_b$ and $V_{\rm GSR}$. In fact, for Sgr debris in the four fields
of view comprising this study (SAs 71, 94, 93, and 117), only (75\%,
87\%, 61\%, 58\%, respectively) of the total value of $\Theta_{\rm
LSR}$ is projected onto $\mu_l \cos($b). We ran the $\chi^2$ fitting
again, but this time including all three dimensions of the motion as
constraints. The bootstrap analysis gave a result of $\Theta_{\rm LSR}
= 232 \pm 14$ km s$^{-1}$ -- a histogram of the bootstrap results is
seen in the right panel of Figure~\ref{fig:theta_hist}. This mean
value is lower by $\sim1.4\sigma$ than the result using only $\mu_l
\cos($b). Formally, this is a better fit than the one-dimensional
result (with uncertainty of 14 km s$^{-1}$ compared to an uncertainty
of 23 km s$^{-1}$ from the fits using only longitudinal proper
motions), but the two are consistent within their 1$\sigma$
uncertainties.

Finally, we performed the same exercise using all three dimensions of
Sgr debris motions, but excluding the less reliable SA 71 field. The
uncertain identification of Sgr debris in SA 71 is likely the reason
this field (at $\Lambda_{\odot} = 128\arcdeg$) is an outlier from the
predicted kinematical trends in Figures~\ref{fig:lambda_3dkinematics}
and \ref{fig:mul_varyvcirc}. The bootstrap fit using only SAs 94, 93,
and 117 yields $\Theta_{\rm LSR} = 244 \pm 17$ km s$^{-1}$. This
slightly higher value for $\Theta_{\rm LSR}$ suggests that (as is
evident in Figures~\ref{fig:lambda_3dkinematics} and
\ref{fig:mul_varyvcirc}) the Sgr candidates in SA 71 skew our results
toward lower $\Theta_{\rm LSR}$.

Ultimately, we have derived three estimates of $\Theta_{\rm LSR}$ --
one based on a simple one-dimensional analysis (using all four fields)
that gave $264 \pm 23$ km s$^{-1}$, another based on three-dimensional
data yielding $232 \pm 14$ km s$^{-1}$, and a final 3-D result with SA
71 excluded, which gave $244 \pm 17$ km s$^{-1}$. It is likely that
the true result is somewhere between the two extremes (264 km s$^{-1}$
and 232 km s$^{-1}$) from our methods.

\subsubsection{Sgr Disruption Models For Best-Fitting $\Theta_{\rm LSR}$}

\begin{figure}
\plottwo{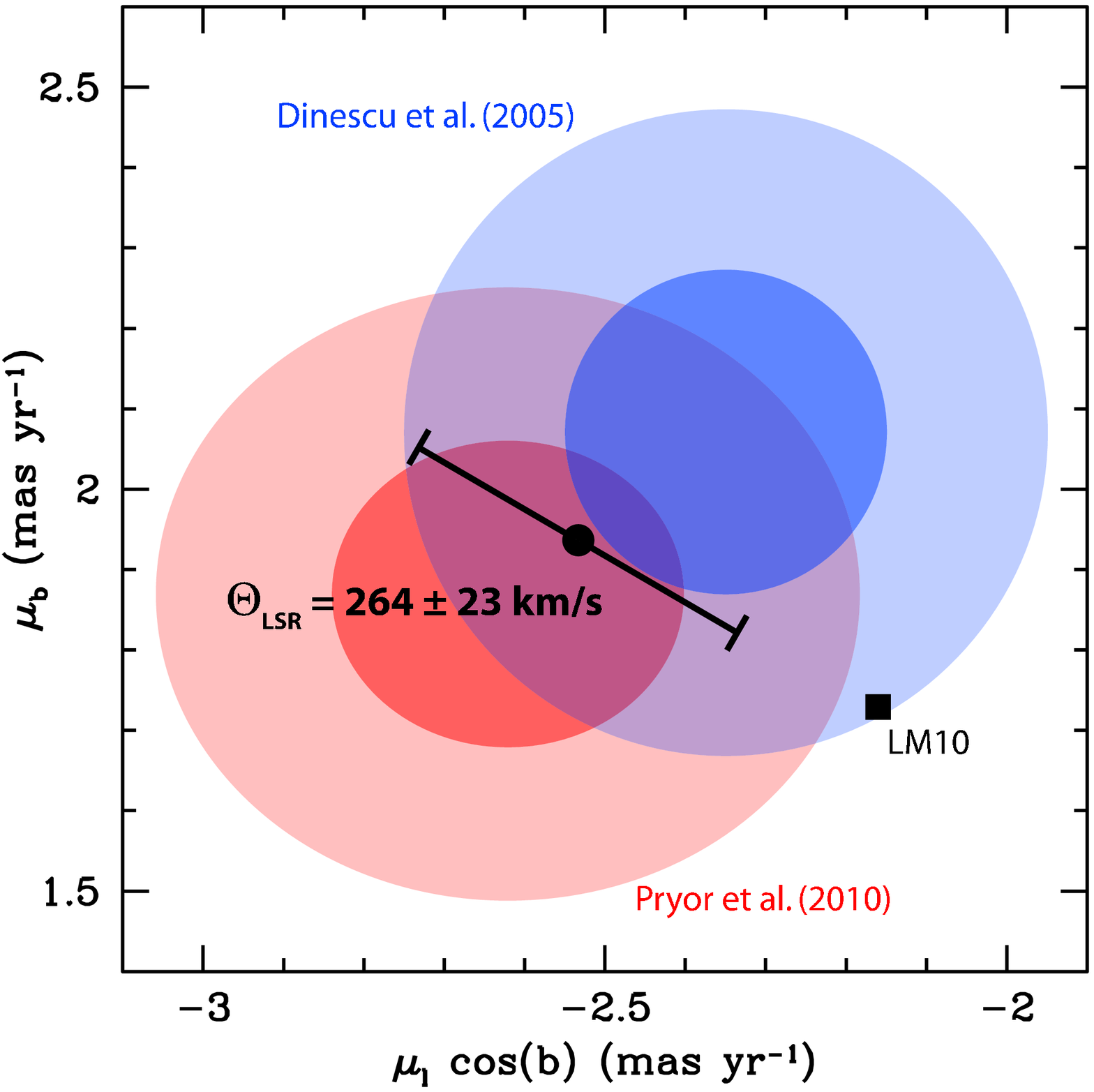}{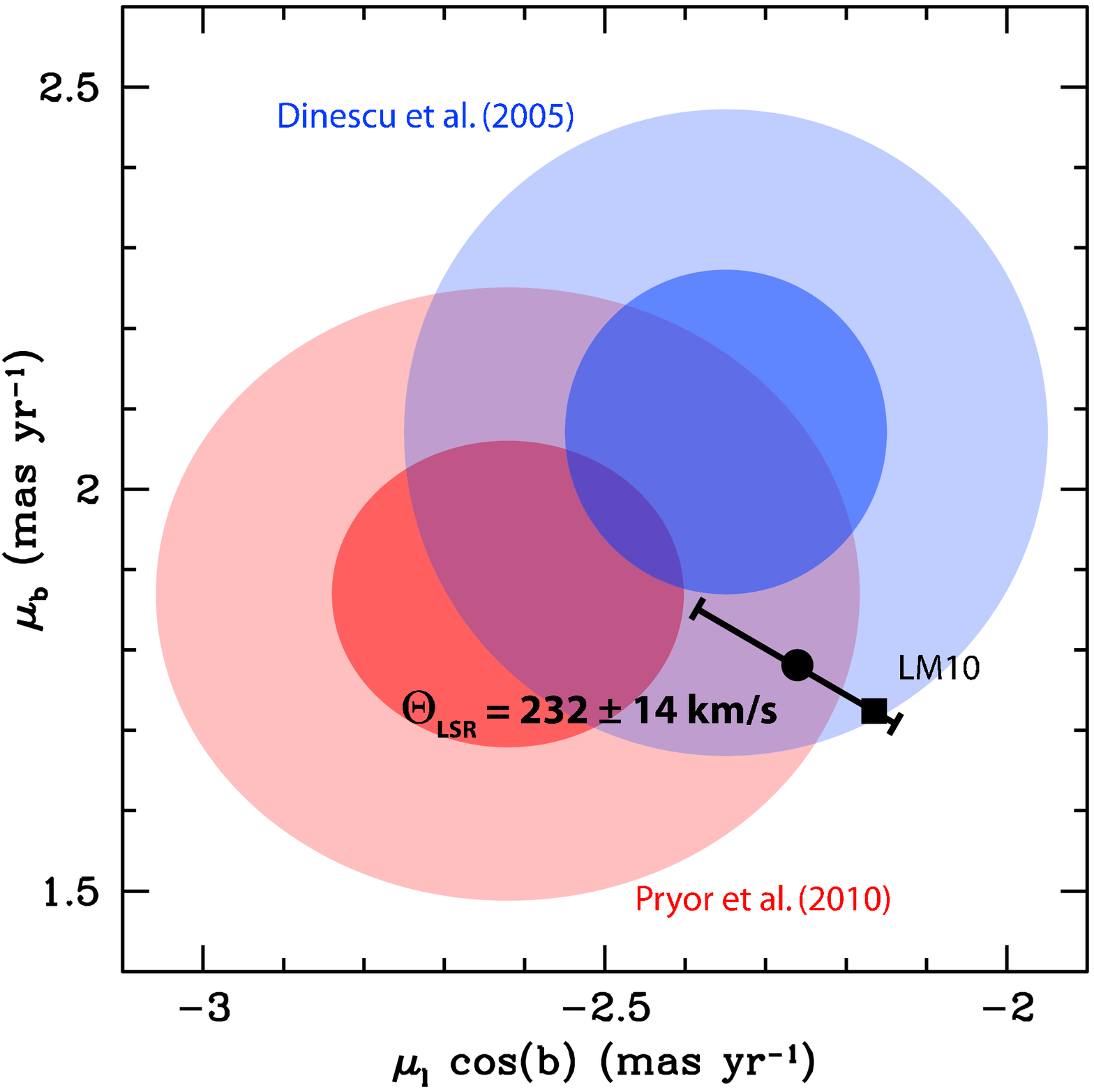} 
\caption{Proper motion estimates  for the Sgr core in Galactic coordinates.  Blue and red shaded ellipses show 1$\sigma$ and 2$\sigma$ uncertainty regions around the measurements
of \citet{dgv+05} and \citet{ppo10} respectively.  The proper motion
of the Sgr model dwarf described by \citetalias{lm10a} ($\Theta_{\rm
LSR} = 220$ km s$^{-1}$) is indicated by a filled black square.  In
the left panel the $1\sigma$ range of proper motions corresponding to
the value $\Theta_{\rm LSR} = 264\pm23$ km s$^{-1}$ we found using
only the $\mu_l \cos($b) component of Sgr debris motions is indicated
by the error bars surrounding the filled black circle. The right panel
is similar, but for the result ($\Theta_{\rm LSR} = 232\pm14$ km
s$^{-1}$) using all three dimensions of Sgr debris kinematics. The
orientation of the error bars represents the direction to which
changes in $\Theta_{\rm LSR}$ correspond in this diagram; the proper
motion of the Sgr core is constrained to lie along this line because
the orbital plane is fixed by the observed position of the tidal
debris leading and trailing Sgr, which trace its
orbit.} \label{sgrpm.fig}
\end{figure}

We now repeat the $N$-body analysis described in
Section~\ref{nbody.sec} two times, first taking $\Theta_{\rm LSR} =
264$ km s$^{-1}$ as a constraint on the models, then again using
$\Theta_{\rm LSR} = 232$ km s$^{-1}$.  The resulting $N$-body model
for the 264 km s$^{-1}$ case matches the angular position, distance,
and radial velocity trends of the observed Sgr tidal streams (using
all of the observational constraints included in the original
\citetalias{lm10a} model) almost equally as well as did the LM10 model
(formally, $\chi = 3.1$ for the $\Theta_{\rm LSR} = 264$ km s$^{-1}$
model, compared to $\chi = 3.4$ for the \citetalias{lm10a} model; see
discussion in Section 4.3 of \citetalias{lm10a}).  In addition, as
demonstrated in the left panel of Figure~\ref{sgrpm.fig} the proper
motion of the remnant core of the Sgr dwarf in this revised model
($\mu_l \cos($b)$ = -2.54$ mas yr$^{-1}$, $\mu_b = 1.92$ mas
yr$^{-1}$) is a substantially better match to observations (e.g.,
\citealt{dgv+05,ppo10}) than was the LM10 model.

However, the $N$-body model in a Milky Way halo with $\Theta_{\rm LSR}
= 232$ km s$^{-1}$ fits equally well as does the 264 km s$^{-1}$ case,
$\chi = 3.1$. For this model, the Sgr core proper motion (seen in the
right panel of Figure~\ref{sgrpm.fig}) is intermediate between those
of the \citetalias{lm10a} model and the 264 km s$^{-1}$ result, as
might be expected. In this case, the proper motions are discrepant
with both the \citet{dgv+05} and \citet{ppo10} results at the
$\sim1.5\sigma$ level. In the following subsection we discuss the
ramifications of these $\Theta_{\rm LSR}$ results for the Milky Way
halo.

\subsection{Robustness of the Galactic Mass Models}

The results for $\Theta_{\rm LSR}$ from our analysis in the previous
subsection suggest that the true value of $\Theta_{\rm LSR}$ as
constrained by Sgr trailing tail debris likely lies between 232-264 km
s$^{-1}$, but that more (or more sensitive) proper motion measurements
of Sgr trailing debris are needed to resolve this issue.  In this
subsection, we will discuss the implications of the $\Theta_{\rm LSR}$
constraints resulting from our two methods; we remind the reader that
the upper end of the range (i.e., the 264 km s$^{-1}$ result) is less
robustly determined than the results that produced lower values of the
circular velocity. However, we include this value in our discussion to
present the reader with the range of possible ramifications of what,
in either case, represents an upward revision of $\Theta_{\rm LSR}$
from the accepted value.

\begin{figure}
\begin{center}
\includegraphics[width=2.25in]{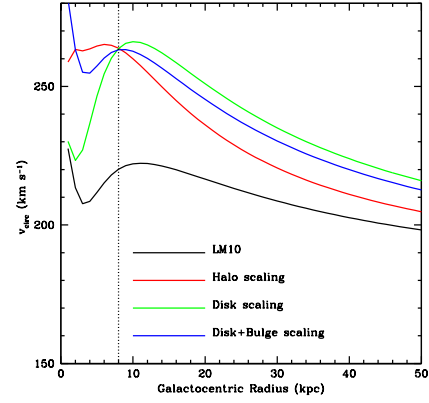}
\caption{Model Milky Way rotation curves as a function of radius from the Galactic center.  Solid blue/green/red lines respectively
represent rotation curves with $\Theta_{\rm LSR} = 264$ km s$^{-1}$
achieved via scaling the Galactic bulge$+$disk, Galactic disk alone,
and Galactic halo alone.  Included for comparison is the rotation
curve of the original LM10 model (solid black line) normalized to
$\Theta_{\rm LSR} = 220$ km s$^{-1}$.  The vertical dotted line
represents the location of the Sun at $R_{\odot} = 8$ kpc.}
\label{RotCurve.fig}
\end{center}
\end{figure}

The value of $\Theta_{\rm LSR} = 232 \pm 14$ km s$^{-1}$ we found
using all three dimensions of Sgr debris kinematics is consistent with
the canonical 220 km s$^{-1}$ value at the roughly 1$\sigma$
level. However, this value is also consistent (within the 1$\sigma$
uncertainties) with recent determinations of $\Theta_{\rm LSR}$ that
have found the rotation speed to be higher than the standard 220 km
s$^{-1}$ value [e.g., \citet{rmz+09} -- $\Theta_{\rm LSR} = (254\pm16)
(R_0/8.4)$ km s$^{-1}$; \citet{bhr09} -- $\Theta_{\rm LSR} = 244\pm13$
km s$^{-1}$]. For a change in $\Theta_{\rm LSR}$ of only about 10 km
s$^{-1}$, it is difficult to make any conclusions about whether the
additional mass required to increase the rotation speed must reside in
the Galactic halo or the disk/bulge. We note that placing the
additional mass in the disk and bulge (with the halo fixed) yields
$M_{\rm bulge} = 3.9 \times 10^{10} M_{\odot}$ and $M_{\rm disk} = 1.1
\times 10^{11} M_{\odot}$ for the 232 km s$^{-1}$ model -- an increase
of $\sim10\%$ over the disk and bulge mass from the model of
\citetalias{lm10a}.  The relatively high value of $\Theta_{\rm LSR} =
264 \pm 23$ km s$^{-1}$ found by our analysis using only the $\mu_l
\cos($b) motions would require that the mass of the Galactic bulge and
disk components be increased by $\sim50\%$ from the values assumed by
\citetalias{lm10a} to $M_{\rm bulge} = 5.2 \times 10^{10} M_{\odot}$
and $M_{\rm disk} = 1.53 \times 10^{11} M_{\odot}$.  This disk mass is
near the peak of the probability distribution ($M_{\rm disk} = 1.35
\times 10^{11} M_{\odot}$) found by \citet{krh10} based on
fitting the GD-1 stream in a three-component gravitational potential
similar to our own.  We caution however that the orbit of Sgr is
largely insensitive to the {\it distribution} of the excess mass
between the two baryonic components, and solutions that yield similar
$\chi^2$ can be found by ascribing all or part of the needed
adjustment in $\Theta_{\rm LSR}$ to changes in the mass of either the
disk or bulge components alone.  We do note, however, that the
relative fraction of the total disk+bulge mass in each component is
constrained by the need to reproduce the shape of the observed Milky
Way rotation curve interior to $R_{\odot}$ (see Figure
\ref{RotCurve.fig}).

The total mass of the Milky Way interior to 50 kpc in the 264 km
s$^{-1}$ model is $5.2 \times 10^{11} M_{\odot}$, similar to the value
of $4.5 \times 10^{11} M_{\odot}$ in the \citetalias{lm10a} model.
Since we have accounted for the increased $\Theta_{\rm LSR} = 264$ km
s$^{-1}$ by increasing the disk+bulge mass (which is a relatively
small component of the total virial mass), the mass of the Milky Way
interior to 200 kpc is $M_{\rm vir} = 1.6
\times 10^{12} M_{\odot}$, similar to the value of $1.5 \times 10^{12}
M_{\odot}$ derived by \citetalias{lm10a} assuming that $\Theta_{\rm LSR} = 220$
km s$^{-1}$.

We note that it was not possible to obtain a satisfactory model
(within our parameterization of the Milky Way components; exploration
of different dark halo models is beyond the scope of this work) for
the Sgr stream by leaving both the bulge and disk masses fixed at
their \citetalias{lm10a} values and accounting for changes in
$\Theta_{\rm LSR}$ by scaling the dark matter halo.  The dark matter
halo profile is characterized by the parameters $v_{\rm halo}$ and
$r_{\rm halo}$ (see Eqn. 3 of \citetalias{lm10a})\footnote{The halo
triaxiality is an added complication that has little bearing on the
present discussion.}, which describe the total mass normalization and
radial scalelength of the halo respectively.  Since dark matter in the
\citetalias{lm10a} model contributes only 19\% of the total
centripetal acceleration in the solar neighborhood, $v_{\rm halo}$
must be scaled up drastically (by a factor of $\sim 3$ in total halo
mass) to increase $\Theta_{\rm LSR}$ from 220 km s$^{-1}$ to 264 km
s$^{-1}$, and necessitates a large increase in the space velocity of
Sgr along its orbit (to $\sim 400$ km s$^{-1}$) to produce a leading
arm debris stream at an observed peak distance of $\sim 50$ kpc (see
Figure 6 of \citetalias{lm10a}).  However, such a rapidly moving
satellite model yields radial velocities along the tidal debris
streams that are systematically discrepant from observations by $\sim
75$ km s$^{-1}$.  Similarly, it is neither possible to obtain a
satisfactory fit for larger $\Theta_{\rm LSR}$ values ($\Theta_{\rm
LSR} \gtrsim 280$ km s$^{-1}$, for which the halo mass scaling problem
is even more extreme), nor for much lower values ($\Theta_{\rm LSR}
\sim 190$ km s$^{-1}$, because the baryonic bulge+disk mass component
alone require $\Theta_{\rm LSR} > 190$ km s$^{-1}$).

Another possibility we considered was to again fix the baryonic mass
(i.e., the bulge+disk component), but to a smaller value than the
\citetalias{lm10a} model, and allow the halo mass to vary. In
particular, we attempted to fit a model with $\Theta_{\rm LSR} = 232$
km s$^{-1}$, but with the bulge+disk mass decreased by 10\% from the
\citetalias{lm10a} values. Even this small change in the baryonic mass
required scaling up the dark matter halo mass by $\sim50$\% to keep
$\Theta_{\rm LSR} = 232$ km s$^{-1}$. The best-fit $N$-body model in
this case fit the Sgr trailing-tail velocities, but was a poor fit to
the leading arm SDSS distances because of the increased speed of the
Sgr core necessitated by the much larger halo. This illustration
highlights the large changes in the halo mass effected by even small
changes in the baryonic mass or the LSR velocity when fitting to
observational data on the Sgr system. In fact, it is a testament to
how well-constrained the Sgr system is by the current observational
data that we are unable to fit the data if we change the dark halo
model substantially.

One possible way to construct an $N$-body model of the Sgr dwarf that
fits the observational data relatively well while dramatically
changing $\Theta_{\rm LSR}$ is to adopt a Galactic halo whose
scalelength is a factor of $\sim 10$ shorter than commonly adopted
(from $r_{\rm halo} = 12$ kpc in the LM10 model to $r_{\rm halo} = 1$
kpc).  However, the Galactic rotation curve implied by such a short
halo scalelength declines steeply outside the solar circle (Figure
\ref{RotCurve.fig}), in conflict with observations (e.g.,
\citealt{sho09}).  We therefore conclude that it is not possible to
satisfactorily model the Sgr dwarf in a Milky Way model with the bulge
and disk masses fixed at the \citetalias{lm10a} values of $M_{\rm
bulge} = 3.4 \times 10^{10} M_{\odot}$, $M_{\rm disk} = 1.0 \times
10^{11} M_{\odot}$, and the Milky Way halo scaled to produce
$\Theta_{\rm LSR}$ much higher than 220 km s$^{-1}$. Thus our (and
other recent) suggestions that $\Theta_{\rm LSR}$ is due an upward
revision implies that the disk and/or bulge components -- but {\it
not} the halo -- of the Milky Way are more massive than previously
thought.

\section{Abundances}
\label{sgr_abund.sec}

While the spectra in the Selected Areas were obtained primarily with
kinematics in mind, they have sufficient resolution and, for a large
fraction of stars, sufficient $S/N$, to obtain information not only on
metallicity but also abundance patterns. This allows us an independent
estimate of abundance distributions that, while of lower precision
than the echelle work of \citet{mbb+05,mbb+07} and
\citet{cmc+07,ccm+10}, is derived for many Sgr stars, and is less
biased than those M-giant studies.

\subsection{Sgr Metallicity}

Metallicities were measured for all stars using a software pipeline
entitled "EZ\_SPAM" (Easy Stellar Parameters and Metallicities),
details of which will appear in a forthcoming paper (Carlin et
al. 2011, {\it in prep.}). EZ\_SPAM relies on the well-understood and
calibrated Lick spectral indices (see, e.g., \citealt{wfg+94,f87}) to
measure stellar properties from low-resolution spectra. In particular,
estimates of [Fe/H] are derived for target stars using eight Fe
indices combined with the H$\beta$ index. Calibration of these
multi-dimensional data comes from fits of known [Fe/H] values as a
function of the Lick Fe and H$\beta$ indices for stars in the atlas of
\citet[based on the spectra of \citealt{j98}]{s07}. The EZ\_SPAM code
yields [Fe/H] measurements with $1\sigma$ precision of $\sim0.3$ dex
at $S/N \approx 20$, decreasing to $\sim0.1$ dex at higher
signal-to-noise ($S/N \gtrsim 50$).

\renewcommand{\thefootnote}{\alph{footnote}}

\begin{table}[!ht]
\begin{center}
\caption{Mean [Fe/H] for Sgr Debris in Kapteyn Selected Areas of This Study \label{tab:sa_feh_tab}}
\begin{tabular}{ccccc}
\\
\tableline
\\
\multicolumn{1}{c}{SA} & \multicolumn{1}{c}{$<$[Fe/H]$>$} & \multicolumn{1}{c}{$\sigma_{\rm [Fe/H]}$} & \multicolumn{1}{c}{N$_{\rm [Fe/H]}$\tablenotemark{a}} & \multicolumn{1}{c}{$\Lambda_{\odot}$}  \\
 & & & & (degrees) \\
\tableline
\\
71 & -1.14$\pm$0.19 &  0.97$\pm$0.14 & 24 & 128.2 \\
94 & -1.13$\pm$0.08 &  0.61$\pm$0.06 & 57 & 116.3  \\
93 &  -1.25$\pm$0.11 &  0.47$\pm$0.08 & 23 & 103.2 \\
92 & -0.97$\pm$0.16 &  ...\tablenotemark{b} & 2 &  90.1 \\
117 & -1.08$\pm$0.08 & 0.52$\pm$0.06 & 43 &  87.6 \\
116 & -1.21$\pm$0.21 & 0.64$\pm$0.15 & 10 & 74.9 \\
\tableline
\tablenotetext{1}{Number of spectra with $S/N > 20$ providing reliably-measured [Fe/H].}
\tablenotetext{2}{Cannot be measured for this field -- too few spectra.}
\end{tabular}
\end{center}
\end{table}

\begin{figure}[!th]
\includegraphics[width=3.0in, trim=0.1in 0.1in 0.1in 0.1in, clip]{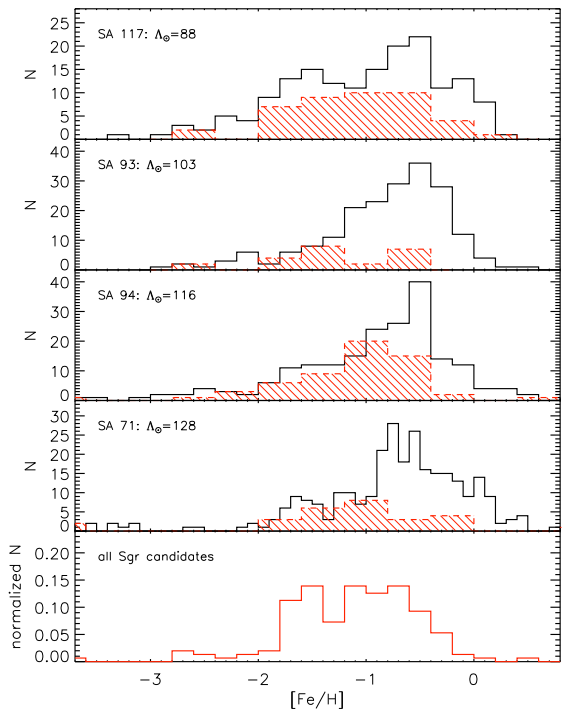}
\caption{Measured values of [Fe/H] for all stars having spectra with $S/N > 20$ in each SA field. The (red) hashed histogram is made up of Sgr candidates from our final samples in each field, and the solid black line represents all other stars (i.e., mostly Milky Way field stars, with perhaps some unidentified Sgr debris included) for which we obtained spectra. The distribution of Sgr metallicities is clearly different from that of the field stars in all of these regions (except perhaps SA 117, which is somewhat ambiguous), peaking at a more metal-poor mean value in each field. The bottom panel shows the metallicity distribution function (MDF) for all four fields in the $88\arcdeg < \Lambda_{\sun} < 128\arcdeg$ portion of the trailing tail in our study. This fractional MDF consists of all the Sgr debris metallicities (red histograms) from the previous four panels, normalized by the total number of stars (147) in the sample.}  \label{fig:feh_hist_sas}
\end{figure}

The metallicity distribution for all well-measured stars (i.e., those
with spectra having $S/N > 20$) in each of the four fields (SAs 71,
94, 93, and 117) in which Sgr debris are reliably identified is given
in Figure~\ref{fig:feh_hist_sas}. For each field, the solid (black)
histogram shows [Fe/H] of non-Sgr stars, and the hashed (red)
histogram gives the distribution of [Fe/H] for stars selected to be
Sgr members. The bottom panel represents the distribution of
metallicities for all Sgr members from the four trailing-tail fields,
normalized by the total number of stars (147) in the sample to produce
a fractional distribution. In each field, Sgr members are typically
more metal-poor than the field stars, with the possible exception of
those in SA 117.

\begin{figure}[!t]
\includegraphics[width=3.0in, trim=0.1in 0.1in 0.1in 0.1in, clip]{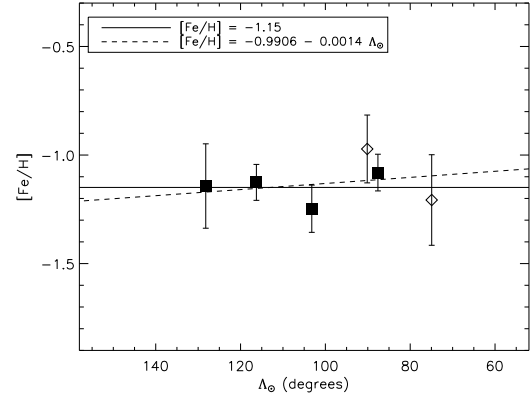}
\caption{Measured values of [Fe/H] for Sgr candidates in each SA field
  as a function of Sgr longitude, $\Lambda_{\odot}$.  Filled squares
  and diamonds (and associated error bars) show the maximum likelihood
  estimate from the individual Sgr candidates in each field (diamonds
  are the two fields lacking secure identification of candidates). The
  mean value of the four well-measured fields (the filled
  squares), [Fe/H] = -1.15, is represented by the solid line that
  reproduces the measurements well. A linear fit to the same four
  fields is shown as a dashed line, and is suggestive of a slight metallicity
  gradient of 1.4$\times10^{-3}$ dex degree$^{-1}$ along the stream
  (though the fit is consistent with zero slope within the
  uncertainties).} \label{fig:feh_lambda}
\end{figure}

For each SA field in the survey, a maximum likelihood estimate for
[Fe/H] was derived from all well-measured stars in the final Sgr
candidate sample. The resulting values for Sgr debris metallicities in
each field are given in Table~\ref{tab:sa_feh_tab} along with
$\sigma_{\rm [Fe/H]}$, the dispersion in [Fe/H] about the mean. As was
the case for the kinematics in SAs 92 and 116, we regard the [Fe/H]
results in these fields (and, to a lesser degree, those in SA 71) with
some skepticism, because the identification of Sgr debris in these
fields is rather unreliable. The mean metallicities for Sgr stars are
displayed in Figure~\ref{fig:feh_lambda} as a function of
$\Lambda_{\odot}$; solid squares depict SAs 71, 94, 93, and 117 (i.e.,
the ``well-measured" fields), with open symbols included for SAs 92
and 116. Error bars represent the uncertainties in the mean value from
the maximum likelihood estimator; however, the scatter of [Fe/H] for
Sgr candidates in each field is rather large. Typical fields have
$\sigma_{\rm [Fe/H]} = 0.5-0.6$ dex about the quoted mean values,
similar to the broad metallicity distribution function for Sgr stars
seen by, e.g., \citet{sm02}, \citet{zbb+04}, \citet{sdm+07}, and
\citet{mbb+05}. The scatter is even larger in SA 71 (at
$\Lambda_{\odot} = 128\arcdeg$); this may arise for a number of
reasons. As can be discerned from Figure~\ref{fig:radec}, SA 71 may be
sampling Sgr debris stripped on multiple pericentric passages (i.e.,
both gold and magenta debris may be present in this field).
Furthermore, this is the lowest-latitude field among those in this
study, and thus may be also suffering more contamination from Galactic
thick disk stars. Finally, we note that SA 71 has been shown by
\citet{ccg+08} to contain a significant number of stars from the
"Monoceros stream" overdensity, which could contribute to the
inflation of the metallicity dispersion in this field, though it is
unlikely that many Monoceros stars would lie within our Sgr radial
velocity criteria for this field. Also shown in
Figure~\ref{fig:feh_lambda} is a solid line at constant [Fe/H] =
-1.15, which is the mean value from the four well-measured fields; the
tight correspondence of the mean values of each field to this line is
consistent with the notion that debris along this narrow stretch of
the trailing tail has constant metallicity. However, there is a hint
of a shallow gradient, which we confirm by fitting a linear trend to
the four good data points.  This fit, overlaid as a dashed line in
Figure~\ref{fig:feh_lambda}, is [Fe/H] = -0.991$\pm$0.003 -
(0.0014$\pm$0.0036) $\Lambda_{\odot}$. While suggestive of a slight
gradient, the slope given is also consistent with zero within the
errors of the fit. This is not surprising considering that nearly all
debris in the portion of the stream contained within this study is
expected to have been stripped on the same pericentric passage of the
Sgr core, as evidenced by the fact that all of our fields overlap
gold-colored debris in Figure~\ref{fig:radec} (i.e., debris that
became unbound during the last two perigalactic passages; see
\citealt{lm10a} for more detail about the color scheme used).

Our measured metallicity of [Fe/H] $\sim$ -1.2 for Sgr trailing debris
is $\sim0.6$ dex more metal-poor than the result obtained by
\citet{kyd10} for trailing-tail M-giants. This is not surprising, as M
giants are biased toward relatively younger, more enriched stellar
populations. In spite of this difference between the older, metal-poor
Sgr stars in our SA fields and the more metal-rich M-giants, we
measure a shallow gradient in [Fe/H] as a function of
$\Lambda_{\odot}$, with a slope consistent with the \citet{kyd10}
measurement, and just a simple offset in the zero-point metallicity. A
more apt comparison for the mean metallicity of Sgr debris in our SA
fields is the work of
\citet{sig+10}, who used SDSS Stripe 82 data to develop a new
technique for estimating metallicity from photometric data, which relies on
combined information from both RR Lyrae variables and main-sequence
stars from the same structure. Because SAs 94, 93, and 92 are within
Stripe 82 (and, in fact, we have used those SDSS data in our analysis),
the Sesar et al. study probes identical stellar populations from the
Sgr trailing tail as our work. This is borne out by the fact that our
measured [Fe/H] = -1.15 is in excellent agreement with the value of
[Fe/H] = -1.20$\pm$0.1 derived by \citet{sig+10} for Sgr debris along
the same portion of the trailing stream.

\begin{figure}[!th]
\plotone{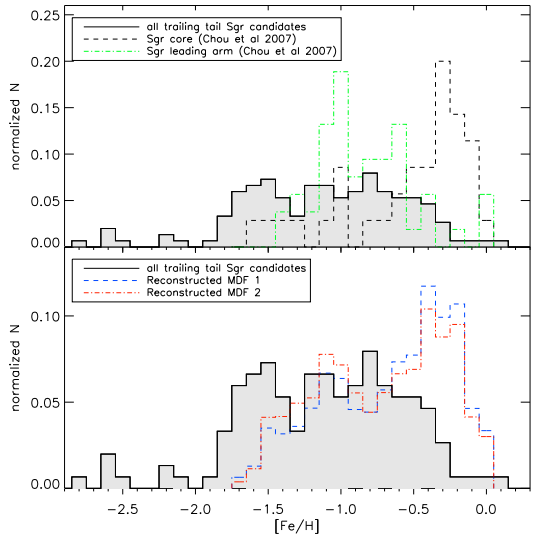}
\caption{The normalized MDF from all 147 Sgr candidates in SAs 71, 94, 93, and 117 with spectra having $S/N > 20$ is shown as a solid black histogram with grey fill. For comparison, in the upper panel we show the MDFs from \citet{cmc+07} of the Sgr core (black dashed line) and Sgr leading arm M-giants (green dot-dashed line). In the lower panel the red (dot-dashed) and blue (dashed) histograms show the approximate Sgr MDF from several Gyr ago reconstructed by Chou et al. from linear combinations of the core and leading arm samples. The first of these was created by interpolating the MDFs at different orbital longitudes, and MDF 2 was created by assigning observed MDFs to particles in the \citet{ljm05} Sgr model by the times they became unbound.} \label{fig:sgr_mdf}
\end{figure}

\subsection{Metallicity Distribution Function}

Previous attempts to measure the metallicity distribution function
(MDF) of the Sgr stream have suffered from both small number
statistics and stellar tracers that have an inherent metallicity bias.
For example, \citet{cmc+07} measured the MDF for M giant stars at
several points along the Sgr leading stream as well as its core; using
these data they attempted to reconstruct the MDF that the Sgr stream
progenitor would have had several Gyr ago.  However, because M giants
only form in metal rich stellar populations, this analysis, while able
to show that the mean metallicity of stars varies along the stream,
was inadequate to assess the MDF across the full metallicity range of
the system.  In addition, the analysis was based on a relatively small
sample (about 7 dozen stars), nearly half of which are in the core of
the Sgr dSph.  \citet{mbb+07} derived abundances of Sgr M giants along
the {\it trailing} tidal tail, from which they derived
$\langle$[Fe/H]$\rangle = -0.61 \pm 0.13$ between $80\arcdeg <
\Lambda_{\sun} < 100\arcdeg$ (from 6 M giants), and
$\langle$[Fe/H]$\rangle = -0.83 \pm 0.11$ (mean of 4 stars) for debris
even further along the trailing tail. The existence of a metallicity
gradient among Sgr stream M giants along both the trailing and leading
tails was confirmed by \citet{kyd10}, who combined their additional
measurements of 5 stars at $\Lambda_{\sun} = 66\arcdeg$
($\langle$[Fe/H]$\rangle \sim -0.5$) and 6 stars at $\Lambda_{\sun} =
132\arcdeg$ ($\langle$[Fe/H]$\rangle \sim -0.7$) with the Chou et
al. and Monaco et al. results to confirm the gradient in [Fe/H] among
Sgr stream M giants. However, all of these M-giant studies suffer an
inherent bias toward metal-rich stellar populations, and are likely
not showing the true MDF of the Sgr system.

Blue horizontal branch stars (BHBs) are another easily-identified and
rather unambiguous tracer of halo substructure that has been used to
probe the Sgr stream. However, BHB stars arise only in old, metal-poor
populations, and are thus not ideal tracers of the global MDF of a
system consisting of multiple stellar populations. \citet{ynj+09}
performed an extensive study of the Sgr tails using SDSS and SEGUE
spectroscopy of BHB stars in both the northern and southern Galactic
caps. BHB stars in the portions of both the leading ($200\arcdeg <
\Lambda_{\sun} < 300\arcdeg$) and trailing ($70\arcdeg <
\Lambda_{\sun} < 110\arcdeg$) tails in this study have MDFs peaking at
$\langle$[Fe/H]$\rangle \sim -1.7$, with significant numbers of stars
as low as [Fe/H] $\sim$ -2.5. Although this result turns up metal-poor
populations not seen in M giants, it is difficult to make conclusions
about the overall MDF of the Sgr stream or progenitor based on biased
metallicity tracers such as BHB stars and M giants.

Fortunately, because the present analysis makes use of MSTO stars, it
is far less susceptible to metallicity biases and can provide new
insights into the MDF (particularly at the intermediate to metal-poor
end) of the stream (and therefore the progenitor) MDF.  Of course, our
spectra have $\sim10\times$ worse resolution than the various echelle
resolution studies, but our sample of Sgr stream stars is
significantly larger, including 147 with good enough S/N ($>20$) for
[Fe/H] measurements to the approximately $\lesssim$ 0.3 dex level.  As
shown in Figure~\ref{fig:feh_hist_sas} and \ref{fig:feh_lambda}, the
mean [Fe/H] for that portion of the stream probed by our SA data is
about -1.1, but with a significant tail to both solar metallicity as
well as very metal poor ($< -2.0$) stars.  Indeed, our sample includes
some rather metal-poor stars associated with the Sgr system, with
stars as metal-poor as the -2.5 dex BHBs seen by \citet{ynj+09}.

Figure~\ref{fig:sgr_mdf} shows the MDF derived from our data in the
$88\arcdeg < \Lambda_{\sun} < 128\arcdeg$ portion of the Sgr trailing
tail. The MDF we derive is significantly broader and extending to much
more metal-poor stars than indicated by the biased, M giant studies
(shown in the upper panel of Figure~\ref{fig:sgr_mdf} as a black
dashed histogram for the Sgr core and green dot-dashed lines for the
leading arm), encompassing both the M-giant and BHB results.  A
comparison of our MDF to the \citet{cmc+07} reconstruction of the Sgr
M-giant MDF from several Gyr ago based on their core and leading-arm
samples is given in the lower panel of
Figure~\ref{fig:sgr_mdf}. Clearly our Sgr trailing-tail sample is
lacking the metal-rich component seen in the present-day core, but
shows a similar distribution to the metal-poor tail of the
reconstructed MDF. Additional metal-poor stars are present in our
sample that are not seen in the M giant samples; these are likely
drawn from similar populations to those in the \citet{ynj+09} study.

Obviously, as has been shown by Chou et al. and others, the total MDF
of the entire Sgr system will include a higher contribution of
metal-rich stars when the core is included, but we also expect more
metal-poor stars from those parts of the tails with larger separation
from the core than we explore.  Thus, while we cannot yet accurately
reconstruct the total MDF of the Sgr system, at least we now have a
better feel of the {\it breadth} of the Sgr MDF from the data shown in
Figures~\ref{fig:feh_hist_sas} and \ref{fig:sgr_mdf}.  Comparison of
the latter MDF with those of other MW dSphs (summarized, e.g., in
\citealt{kls+11}) shows Sgr to be more typical of other MW satellites.
In particular, the Sgr MDF resembles even more that of the LMC (as has
been previously suggested by, e.g., \citealt{mbf+03},
\citealt{ctg+05}, and \citealt{mbb+05}), which has been argued to be a
chemical analog to the Sgr progenitor by \citet{ccm+10} and a
morphological analog by \citet{lkm+10}.

\subsection{``Alpha'' Abundances}

\begin{figure*}[!t]
\begin{center}
\includegraphics[width=4.6in,trim=0.2in 0.2in 0.2in 0.2in,clip]{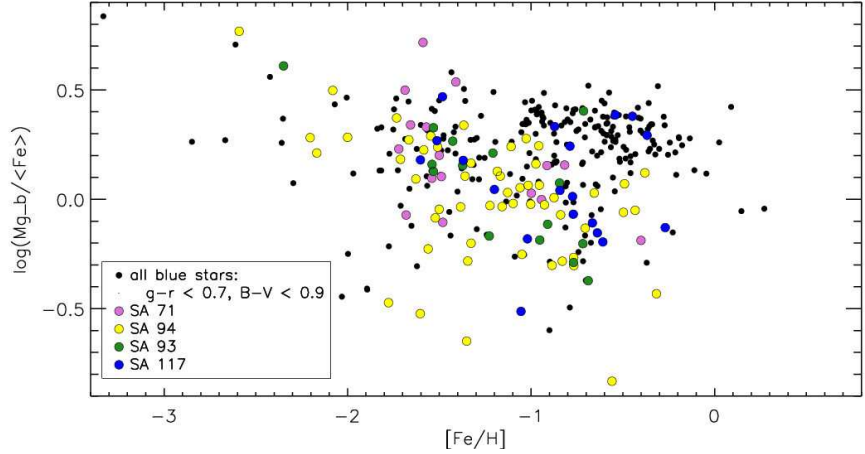}
\caption{Relative values of Lick index ratio log (Mg b/$<$Fe$>$), where the
  indices are as described in the text and Carlin et al.~2011 ({\it in prep.}), for all blue
  stars ($0.2 < g-r < 0.7$ for SAs 94 and 93, and $B-V < 0.9$ for SAs
  71 and 117) having spectra with $S/N > 30$ in the four SA fields
  with securely identified Sgr debris. When comparing only
  predominantly dwarf stars of similar, blue photometric colors, the
  log (Mg b/$<$Fe$>$) ratio can be thought of as a proxy for [Mg/Fe],
  because variations in [Mg/Fe] with log $g$ and color (i.e.,
  temperature) are then minimized. Colored points represent all stars
  within the initial Sgr candidate RV selections, with color codes as
  in the legend. Black dots are all other stars outside the Sgr
  velocity range. For [Fe/H] $\gtrsim$ -1.5, Sgr candidates (colored
  points) typically occupy a region of lower Mg abundance at a given
  [Fe/H] than the black dots that are likely Galactic foreground
  stars. This behavior is typical for stars from most Galactic dSphs
  (relative to Galactic disk populations). At lower metallicities, the
  distributions converge.} \label{fig:mgfe_feh}
\end{center}
\end{figure*}

As shown in Section~5.1, we have identified metal-poor populations in
(at least) four of the fields from our study, which explore a
different segment of the stellar populations in the Sgr stream than
previous M-giant studies. We have observed stars in these fields only
at low resolution, and thus cannot do detailed element-by-element
chemical analysis such as that enabled by high-resolution
spectroscopy. However, we can use the low-resolution Lick indices to
explore relative $\alpha$-abundances for the stars in our
study. Specifically, we explore the relative Mg abundances using the
Lick Mg b index centered at 5160-5190 \AA. Calibrating the Mg b index
to an actual [Mg/Fe] abundance is difficult, because the strength of
Mg lines is highly sensitive to surface gravity, with some additional
sensitivity to effective temperature and [Fe/H]. Disentangling these
effects is difficult with low-resolution spectra, but we can still
explore a subset of the stars in our samples in a way that is
relatively free of the effects of surface gravity and temperature of
individual stars. To do so, we select only blue ($0.2 < g - r < 0.7$,
or $B - V < 0.9$) stars, which should be mostly main-sequence dwarfs
(thus, with similar surface gravity), since no giants are found at
such blue colors.  Furthermore, the temperature sensitivity of the Mg
line strength, which is already much smaller than the log $g$
sensitivity, is mitigated by concentrating on a limited color
range. In Figure~\ref{fig:mgfe_feh} we show a ``pseudo-[Mg/Fe]" ratio,
given as the logarithm of the ratio of the Lick Mg b index to the mean
of all eight Lick Fe indices (after transforming them to a common
scale), for all of the blue stars in SAs 71, 94, 93, and 117 for which
we have high enough signal-to-noise ($>30$) to precisely measure
indices and [Fe/H]. Black points in this diagram are all stars with
non-Sgr radial velocities, while our samples of all stars with
Sgr-like RVs in each field are shown as colored points. It is readily
apparent that the black (MW) points mostly occupy a different region
of the diagram than the colored (Sgr) dots, which suggests an
intrinsic chemical difference between the populations (though some
overlap is expected, especially at low metallicities, where many MW
halo stars likely resemble dSphs in their abundance patterns). Indeed,
the behavior seen in Figure~\ref{fig:mgfe_feh} is exactly that seen
for many MW dSphs -- for more metal-rich dSph stars, the Mg (or
$\alpha$) abundance is lower (on average) at a given [Fe/H] than in
the Galactic populations, with the two populations converging at lower
metallicities (i.e., at the "knee" in the dSph's distribution). Among
the more metal-rich (and younger) M-giant populations of the Sgr
stream, there is some indication that the knee in [$\alpha$/Fe] occurs
at -1.2 $\lesssim$ [Fe/H] $\lesssim$ -1.0
\citep{ccm+10,mbb+07}, but this is difficult to assess because of the
lack of M-giants at lower metallicity. Thus the apparent convergence
of Sgr trailing tail [Mg/Fe] with the plateau seen in Galactic stars
at [Fe/H] $\lesssim$ -1.5 may be an extension of the same behavior
seen in the M-giant studies. Alternatively, since we've already shown
that the mean [Fe/H] along the trailing tail differs between the
M-giant sample of \citet{mbb+07}, who find [Fe/H] $\sim$ -0.6, and our
result of [Fe/H] $\sim$ -1.2 (which is also consistent with the
findings of \citealt{sig+10}), our study may be sampling a distinctly
older, more metal-poor population of Sgr debris than the M-giant
tracers. Further characterization of the $\alpha$-element behavior
along the Sgr trailing tail would benefit from either a calibration of
our Mg b/$<$Fe$>$ index onto [Mg/Fe] abundance or the identification
of bona fide stream giant stars that are bright enough for
echelle-resolution spectroscopic follow-up.

\section{Summary}
\label{summary.sec}

We have presented the first large-scale study of the 3-D kinematics of
the Sagittarius trailing tidal stream, with data spanning
$\sim60\arcdeg$ along the trailing tail. The data include deep,
precise proper motions derived from photographic plates with a
$\sim$90-year baseline, and radial velocities from more than 1500
low-resolution stellar spectra, of which $>150$ have been identified
as Sgr debris stars. Mean absolute proper motions of these Sgr stars
in four of the six 40$\arcmin \times 40\arcmin$ fields from our survey
have been derived with $\sim0.25-0.7$ mas yr$^{-1}$ per field
precision in each dimension (depending on the quality and depth of
plate material and the number of spectra obtained in each field). Mean
three-dimensional kinematics in each of these four fields have been
shown to agree with the predicted $V_{\rm GSR}$ and $\mu_b$ from the
Sagittarius disruption models of \citetalias{lm10a}. However, there is
a systematic disagreement in the $\mu_l \cos($b) proper motions (with
the exception of the somewhat problematical SA 71 field), which we use
to assess refinements to the mass scale of the Milky Way (particularly
its disk and bulge components).

While proper motions along the portion of the trailing tail in this
study provide constraints on Sgr tidal disruption models, the
fortuitous orientation of the Sgr plane also allows us to use the
measured proper motions to derive the circular velocity at the Solar
circle (or ``Local Standard of Rest''), $\Theta_{\rm LSR}$.  Our
first-order approximation using only the $\mu_l \cos($b) proper
motions as constraints yields $\Theta_{\rm LSR} = 264 \pm 23$ km
s$^{-1}$. From our measured 3-D kinematics, we find this fundamental
Milky Way parameter to be $\Theta_{\rm LSR} = 232 \pm 14$ km s$^{-1}$,
or $\sim1\sigma$ higher than the IAU standard value of 220 km
s$^{-1}$. When we remove SA 71, a field in which it is more difficult
to unambiguously identify Sgr debris, from the sample we find
$\Theta_{\rm LSR} = 244 \pm 17$ km s$^{-1}$. We suggest that the true
value of $\Theta_{\rm LSR}$ lies somewhere between 232-264 km
s$^{-1}$, while noting that all three of these estimates are
consistent with each other within their 1$\sigma$ uncertainties.

Our general result that the circular velocity at the Solar radius is
higher than the IAU standard of 220 km s$^{-1}$ agrees with the recent
derivation of $\Theta_{\rm LSR} = 254 \pm 16$ km s$^{-1}$ by
\citet{rmz+09} using trigonometric parallaxes of star forming regions
in the outer disk. The same maser data from the Reid et al. study were
reanalyzed by \citet{bhr09}, and yield a result of 246 $\pm$ 30 km
s$^{-1}$ (244 $\pm$ 13 km s$^{-1}$ if priors on the proper motion of
Sgr A* are included, and 236 $\pm$ 11 km s$^{-1}$ if the additional
contribution of orbital fitting to the GD-1 stellar stream is
included). Again, our independent result is consistent with these
studies, and inconsistent with the IAU accepted value of 220 km
s$^{-1}$ for this fundamental constant at the 1-2$\sigma$ level.
Identification of additional Sgr candidates could increase the
accuracy of our determination of $\Theta_{\rm LSR}$, as could the
addition of another epoch of accurate data to the proper motion
measurements.

We note that while \citet{rmz+09} argued that their measurement of 254
km s$^{-1}$ would imply an upward revision of the total Milky Way mass
by a factor of $\sim2$ (to a mass similar to that of M31), we have
shown that this is not required to produce $\Theta_{\rm LSR}$ even
higher than that of Reid et al. Scaling the Milky Way dark matter halo
up in mass to a level that yields $\Theta_{\rm LSR} = 264$ km s$^{-1}$
while simultaneously reproducing known leading arm debris requires the
Sgr core to have a high ($\sim400$ km s$^{-1}$) space velocity,
resulting in RVs in the tidal streams that are discrepant by $\sim75$
km s$^{-1}$ from measured values. Instead, we show that because $>
80\%$ of the centripetal force at the location of the Sun is
contributed by mass in the Galactic disk and bulge, an increase of
$\sim50\%$ in the mass of the disk+bulge accounts for the additional
acceleration needed to produce 264 km s$^{-1}$ rotation at the solar
circle, while contributing only a small ($\sim7\%$) increase to the
total virial mass of the Milky Way. With the additional constraint on
the disk+bulge mass provided by our measurement of $\Theta_{\rm LSR}$,
we have found a satisfactory model of Sgr disruption that matches all
of the constraints used in fitting the \citetalias{lm10a} model, while
additionally predicting a proper motion for the Sgr dwarf that is in
much better agreement with observations than the \citetalias{lm10a}
model.

Stellar metallicities have been derived from the low-resolution
spectra of Sgr candidates, and the mean metallicity of Sgr tidal
debris derived in each field. We find that a constant [Fe/H] = -1.15
is consistent with the observations of all four fields for which Sgr
members were reliably identified. However, a linear fit to these four
data points suggests that a gradient of ($1.4 \times 10^{-3}$) dex
degree$^{-1}$ is also reasonable (though this value is consistent with
zero slope within the uncertainty), in line with previous findings
(e.g.,
\citealt{cmc+07, kyd10}) of a metallicity gradient among
M-giants along both the leading and trailing tidal tails. The scatter
of [Fe/H] in each of the survey fields is $\gtrsim$0.5 dex, which is
typical of the stellar populations seen in the core of the Sgr dSph
(e.g., \citealt{sm02,zbb+04,sdm+07,mbb+05}). We show the metallicity
distribution function for the trailing tail that is free from the
biases inherent in previous studies of the Sgr MDF. We find that the
MDF of trailing debris is similar to MDFs of typical classical Milky
Way dwarf spheroidals. A ``pseudo-[Mg/Fe]'' was measured based on the
ratio of Lick Mg b and $<$Fe$>$ indices; the behavior of log (Mg
b/$<$Fe$>$) with [Fe/H] for Sgr main-sequence candidates is markedly
different from Galactic stars of similar photometric colors
(identified by radial velocity) from among the same
datasets. Furthermore, the trend is similar to that typically seen for
dSphs, in that [Mg/Fe] is deficient at a given [Fe/H] for Sgr stars
relative to the Milky Way field populations, and converges to a
``knee'' in Figure~\ref{fig:mgfe_feh} at lower ([Fe/H] $\sim$ -1.5)
metallicity. This is a lower [Fe/H] than previously reported for Sgr M
giants, and may reflect a bias intrinsic to those earlier M-giant
studies. High-resolution spectroscopic follow-up will be necessary to
confirm this trend among the old, metal-poor populations of
recently-stripped Sgr debris.

We appreciate the useful comments provided by the anonymous
referee. We thank Mei-Yin Chou for kindly sharing the Sgr MDF data
used in Figure 23, and Heidi Newberg for many useful discussions. JLC
acknowledges support from National Science Foundation grant
AST-0937523, and observing travel support from the NOAO thesis student
program for proposal ID 2008B-0448. JLC and SRM acknowledge partial
funding of this work from NSF grant AST-0807945 and NASA/JPL contract
1228235.  DIC and TG acknowledge NSF grant AST-0406884. DRL
acknowledges support provided by NASA through Hubble Fellowship grant
\# HF-51244.01 awarded by the Space Telescope Science Institute, which
is operated by the Association of Universities for Research in
Astronomy, Inc., for NASA, under contract NAS 5-26555.

SRM is grateful to Allan Sandage for alerting him to the existence of
the Mt.~Wilson Selected Areas plates, for suggesting their use as a
first epoch to expand our proper motion work in Kapteyn Selected
Areas, and for his contribution of KPNO plates to our survey.

{\it Facilities:} \facility{WIYN (Hydra)}, \facility{MMT (Hectospec)}, \facility{Sloan}, \facility{Swope}, \facility{MtW:1.5m}, \facility{Du Pont}, \facility{Mayall}

Funding for the SDSS and SDSS-II has been provided by the Alfred
P. Sloan Foundation, the Participating Institutions, the National
Science Foundation, the U.S. Department of Energy, the National
Aeronautics and Space Administration, the Japanese Monbukagakusho, the
Max Planck Society, and the Higher Education Funding Council for
England. The SDSS Web Site is http://www.sdss.org/.

The SDSS is managed by the Astrophysical Research Consortium for the
Participating Institutions. The Participating Institutions are the
American Museum of Natural History, Astrophysical Institute Potsdam,
University of Basel, University of Cambridge, Case Western Reserve
University, University of Chicago, Drexel University, Fermilab, the
Institute for Advanced Study, the Japan Participation Group, Johns
Hopkins University, the Joint Institute for Nuclear Astrophysics, the
Kavli Institute for Particle Astrophysics and Cosmology, the Korean
Scientist Group, the Chinese Academy of Sciences (LAMOST), Los Alamos
National Laboratory, the Max-Planck-Institute for Astronomy (MPIA),
the Max-Planck-Institute for Astrophysics (MPA), New Mexico State
University, Ohio State University, University of Pittsburgh,
University of Portsmouth, Princeton University, the United States
Naval Observatory, and the University of Washington.

\end{document}